\documentclass[12pt]{article}
\usepackage{amssymb,amsmath,latexsym}
\usepackage{graphicx}
\usepackage[final]{pdfpages}
\usepackage{verbatim}
\usepackage{color}
\usepackage{amsfonts,amsthm,hyperref,mathrsfs,ulem,tikz,mathtools}
\usepackage{amsmath}
\usepackage{todonotes}

\usetikzlibrary{arrows,patterns}

\title{
Limit absorption and Green function estimates
\\
for matrix-valued periodic operators
}

\author{Miguel Ballesteros$^1$, Gerardo Franco C\'ordova$^{1,2}$,  Hermann Schulz-Baldes$^2$, 
\\
{\small $^1$ IIMAS, UNAM, Mexico City, Mexico}
\\   
{\small $^2$Department Mathematik, Friedrich-Alexander-Universit\"at Erlangen-N\"urnberg, Germany}
}

\date{ }

\newtheorem{theo}{Theorem}

\newtheorem{proposi}[theo]{Proposition}
\newtheorem{lemma}[theo]{Lemma}
\newtheorem{hyp}[theo]{Hypothesis}

\newtheorem{rem}[theo]{Remark}

\newcommand{\CM}{{\mathbb C}}
\newcommand{\NM}{{\mathbb N}}
\newcommand{\RM}{{\mathbb R}}
\newcommand{\SM}{{\mathbb S}}
\newcommand{\TM}{{\mathbb T}}
\newcommand{\ZM}{{\mathbb Z}}

\newcommand{\HM}{{\mathbb H}}

\newcommand{\Ee}{{\cal E}}
\newcommand{\Pp}{{\cal P}}

\newcommand{\Ff}{{\cal F}}

\newcommand{\Ww}{{\cal W}}

\newcommand{\Oo}{{\cal O}}
\newcommand{\Tr}{\mbox{\rm Tr}}

\newcommand{\Rr}{{\cal R}}

\newcommand{\Cc}{{\cal C}}

\newcommand{\Ll}{{\cal L}}
\newcommand{\Qq}{{\cal Q}}

\newcommand{\one}{{\bf 1}}

\newcommand{\Hess}{\nabla^2}

\newcommand{\spec}{{\mbox{\rm spec}}}

\newcommand{\sig}{{\mbox{\rm sig}}}

\newcommand{\reg}{{\mbox{\rm\tiny reg}}}

\newcommand{\Sym}{{\mbox{\rm Sym}}}
\newcommand{\GL}{{\mbox{\rm GL}}}

\newcommand{\supp}{\mbox{\rm supp}}

\newcommand{\red}[1]{\textcolor{red}{#1}}

\usepackage{cancel}

\def\XXint#1#2#3{{\setbox0=\hbox{$#1{#2#3}{\int}$}
     \vcenter{\hbox{$#2#3$}}\kern-.5\wd0}}

\begin{document}

\maketitle

\begin{abstract}
The boundary value of the resolvent of a generic periodic tight-binding Hamiltonian with matrix symbols is shown to satisfy a limit absorption principle which is continuous in energy in dimensions $d=3$, and in dimension $d=2$ away from critical points of the energy bands corresponding to van Hove singularities. The analysis away from critical points of the energy bands is based on the coarea formula, while at the critical points it involves a parametric Morse lemma and stationary phase arguments. In particular,  at Weyl points a new type of oscillatory integrals is dealt with.

\vspace{.1cm}

\hspace{4.7cm} MSC2010 database: 81Q10, 35J08, 42B20




\end{abstract}

\vspace{.5cm}


\section{Introduction}

Bloch-Floquet theory shows that periodic selfadjoint operators have bands of absolutely continuous spectrum, possibly overlapping and even touching over some subsets of the Brillouin torus. Hence the norm of the resolvent explodes as one approaches these real intervals in the complex energy plane. Limit absorption principles state that by damping the resolvent with suitable decaying weights, one obtains operators that actually do have bounded limits as one approaches the bands. Such statements were first proved for the Laplacian on $\RM^d$ and other operators with constant symbols, and were then used as a crucial technical input to develop scattering theory (see the books \cite{Kur,RS3,Yaf,Yaf2} for a detailed history). Periodic operators were only addressed in very few works, most notably \cite{Ger}. Close to the band edges it is even possible to derive quantitative estimates on the off-diagonal decay of the Green function, that is, the integral kernels of the resolvents \cite{MT,KR,KKR}. This has also been proved in the probability community because it allows to analyze transient random walks \cite{Spi,LL}. In view of all these results, this work may at first sight appear as an old story, but there are several new twists explained next.

\vspace{.2cm}

The focus here is on matrix-valued periodic tight-binding Hamiltonians, notably periodic selfadjoint operators acting on the Hilbert space $\ell^2(\ZM^d,\CM^L)$ where the integer $L$ is the fiber dimension. A prototypical example for the case $L=1$ is the discrete Laplacian for which the limit absorption was proved in \cite{KSM} via a detailed analysis of the special functions representing the resolvent. In a more general and conceptional treatment, one can treat the regular points of the energy bands by an adaption of the methods in the aforementioned works, but the critical points (also called van Hove singularities, see \cite{Eco} for a discussion from the perspective of physics) and band touching points (in dimension $d=3$ called Weyl points and for $d=2$ rather Dirac points) lead to problems which have not been addressed before, except in \cite{Ger} by algebraic geometry techniques, and for one-band lattice operators in \cite{BSB} where only matrix elements of the resolvent were considered. While the critical points of definite signature can be dealt with by adapting the techniques of \cite{MT,KKR,LL}, those with indefinite signature require the use of a parametric Morse lemma \cite{Dui} in order to control oscillatory integrals. Furthermore, new types of oscillatory integrals have to be dealt with when addressing Weyl points. These two latter points are the essentially novel contributions of this work.  

\vspace{.2cm}

Let us note that the current tight-binding framework can be considered as an ultra-violet cut-off of matrix-valued periodic Schr\"odinger-type operators (see \cite{Par,PaS} for a modern presentation and novel results). While those operators are not dealt with here, we believe that the techniques can be extended to this case. 

\subsection{Mathematical framework}

Let us now start by introducing some notations which will allow us to state the hypothesis and main results. For integer dimension $d$ and fiber dimension $L$, let $H=H^*$ be a bounded selfadjoint Hamilton operator on $\ell^2(\ZM^d,\CM^L)$. It is supposed to be translation invariant, namely one has $\langle n+l|H|m+l\rangle=\langle n|H|m\rangle$ for all $n,m,l \in \mathbb{Z}^d$ . Here $|n\rangle$ is Dirac's ket notation for the partial isometry from $\CM^L$ onto the states in $\ell^2(\ZM^d,\CM^L)$ over site $n$, and $\langle n|=(|n\rangle)^*$ is the associated bra. Restricting the attention to such translation invariant (or $1$-periodic) operators in each direction is no restriction, as larger periods can be reduced to $1$-periodic operators by increasing the number $L$ of fibers. The focus in this work is on dimensions $d\geq 2$.

\vspace{.2cm}

The Fourier transform $\Ff:\ell^2(\ZM^d)\to L^2(\TM^d)$ where $\TM^d\cong (-\pi,\pi]^d$ is equipped with the standard Lebesgue measure, densely defined by
$$
(\Ff\phi) (k)
\;=\; 
(2\pi)^{-\frac{d}{2}}
\sum_{n\in \mathbb{Z}^d} 
e^{-\imath k\cdot n}\,\phi_n
\;,
$$
where $\phi=\sum_{n\in\ZM^d}\phi_n|n\rangle \in \ell^1(\ZM^d)$. Then $\Ff$ is naturally extended fiberwise to a unitary map $\Ff:\ell^2(\ZM^d,\CM^L)\to L^2(\TM^d,\CM^L)$. One finds that
$$
\Ff H\Ff^*
\;=\;
\int^\oplus \!\!\! dk\, \Ee(k)
$$
where $k\in\TM^d\mapsto \Ee(k)=\Ee(k)^*\in\CM^{L\times L}$ is a selfadjoint matrix-valued function. It will be assumed throughout that $H$ is sufficiently short range such that $\Ee$ is real analytic. In the following section, several assumptions will be made on $\Ee$. The aim of this work is to control the resolvent operator $R^z=(H-z\one)^{-1}$ as $\Im m(z)\to 0$. If $E=\Re e(z)$ lies in the spectrum of $H$, this limit can clearly not exist as a bounded operator, but it is well-known \cite{Yaf,Yaf2} that the resolvent damped by powerlaw factors of the position does have limits as $\Im m(z)\to 0$. Such results are called limit absorption principles. 

\subsection{Hypotheses}
\label{Hyp}

Let us introduce the set
$$
\Ww
\;=\;
\{k\in\TM^d\,:\,\Ee(k)\,\mbox{ has degenerate spectrum}\}
\;.
$$

\begin{hyp}
\label{hyp-Wdisc}
$\Ww$ is a discrete set.
\end{hyp}

\begin{hyp}
\label{hyp-Wdeg}
For $k^W\in \Ww$, the degeneracy of eigenvalues of $\Ee(k^W)$ is at most two.
\end{hyp}

A point $k^W$ is called a Weyl point. In even dimension $d$ and presence of a chiral symmetry, such points are often also called Dirac points. Both Hypothesis~\ref{hyp-Wdisc} and \ref{hyp-Wdeg} are generic in dimension $d=3$. Indeed, a celebrated theorem of von Neumann and Wigner \cite{NW} states that generically an analytic function $k\mapsto \Ee(k)$ with values in the selfadjoint matrices has at most double degeneracies on hypersurfaces of dimension $d-3$. For $k\not\in\Ww$, analytic perturbation theory \cite{Kat} assures that there is a neighborhood $U$ of $k$ such that the spectrum of $\Ee$ consists of $L$ so-called bands $k\in U\mapsto \Ee_1(k),\ldots,   \Ee_L(k) $ which are analytic. 

\begin{hyp}
\label{hyp-Morse}
Away from Weyl points, all energy bands are Morse functions.
\end{hyp}

Recall that this means that for each $j=1,\ldots,  L$ the critical points of $k\in U\mapsto \Ee_j(k)$, namely points $k^*$ with vanishing gradient $\nabla \mathcal{E}_j(k^*) =0$, are such that the Hessian  $\nabla^2 \mathcal{E}_j(k^*) $ is a non-degenerate selfadjoint matrix. It is well-known that also Hypothesis~\ref{hyp-Morse} holds generically. 

\vspace{.2cm}

Further conditions will be assumed to hold on the shape of the energy bands at the Weyl  points. This is tailored for dimension $d=3$, but not strictly restricted to it. To describe the conditions in detail, let us focus on the two touching bands appearing in Hypothesis~\ref{hyp-Wdeg}. Using a Riesz projection (see Section~\ref{sec-Global} for details),  these two bands are locally described by an analytic function $k\in B_r(k^W)\mapsto \Ee(k)\in\CM^{2\times 2}$ of selfadjoint $2\times 2$ matrices on a ball $B_r(k^W)$ of size $r$ around $k^W$. Then decompose 
\begin{equation}
\label{eq-2BandDecomp}
\Ee(k)
\;=\;
e(k)\,\one_2\,+\,\langle h(k),\sigma\rangle
\;,
\end{equation}
where $e(k)=\tfrac{1}{2}\,\Tr(\Ee(k))$, $h(k)=(h_1(k),h_2(k),h_3(k))\in\RM^3$ satisfies $h(k^W)=0$ and $\sigma=(\sigma_1,\sigma_2,\sigma_3)$ are the three Pauli matrices. Finally $\langle\,.\,,\,.\,\rangle$ denotes the euclidean scalar product such that $\langle h(k),\sigma\rangle= \sum_{j=1,2,3}h_j(k)\sigma_j$. With the notation \eqref{eq-2BandDecomp}, the two bands are locally given by $\Ee_\pm(k)=e(k)\pm|h(k)|$ where $|v|$ denotes the euclidean length of a vector $v\in\RM^d$. 

\begin{hyp}
\label{hyp-WeylType}
For all $k^W\in\Ww$, $\det\big(D h(k^W)\big)\not=0$ and $| \nabla (e\circ h^{-1})(0)|<\kappa$ for some $\kappa<1$. 
\end{hyp}

Here $Dh(k^W)\in\RM^{d\times d}$ denotes the Jacobian of $h$ at $k^W$ and $\nabla$ is again the gradient. The condition $\det\big(D h(k^W)\big)\not=0$ assures the linear growth of $h$, namely by Taylor's formula
\begin{equation}
\label{eq-hCond}
\tfrac{1}{\gamma}\,|k-k^W|\;\leq\;|h(k)-h(k^W)|  \;\leq\;\gamma\,|k-k^W|
\;,
\end{equation} 
uniformly in a pointed neighborhood of $k^W$ for some constant $\gamma$. It also implies that $h$ is a local diffeomorphsm so that $h$ is locally invertible and the second condition $| \nabla (e\circ h^{-1})(0)|<1$ makes sense. This second condition then guarantees weak tilting of the Weyl point, or in physics terminology \cite{AMV}, that the Weyl point is of so-called {\rm Type I}. Indeed, by the mean value theorem one has $|e(h^{-1}(x))-e(k^W)| \leq \kappa |x|$ for $x$ sufficiently small. Equivalently $\kappa |h(k)| \geq |e(k)-e(k^W)|$ and this then implies $\pm(\Ee_\pm(k)-e(k^W))\geq 0$.  Also note that $\nabla (e\circ h^{-1})(0)=(Dh(k^W)^{-1})^T\nabla e(k^W)$ so that the second condition holds if the smallest singular value of $Dh(k^W)$ dominates the length of the vector $\nabla e(k^W)$. Imposing Hypothesis~\ref{hyp-WeylType} on all Weyl points is clearly an open condition, but there are energy bands with other type of Weyl points and they are not considered in this work.

\subsection{Main results}

As already stated above, limit absorption principles concern the limit behavior of the weighted resolvent operator $R^z=(H-z\one)^{-1}$  in the limit $\Im m(z)\to 0$. To introduce the weights, let $X$ denote the (vector-valued, unbounded, selfadjoint) position operator densely defined on $\ell^2(\ZM^d)$ by $X|n\rangle=n|n\rangle$ where $|n\rangle$ is the Dirac notation for the state localized at  $n\in\ZM^d$. Then $X$ is naturally extended to $\ell^2(\ZM^d,\CM^L)$ and furthermore let us use the notation $\langle X\rangle=(\one+X^2)^{\frac{1}{2}}$. Then $\alpha$-damped resolvent is defined as
\begin{equation}
\label{eq-Ralpha}
\Rr^z_\alpha
\;=\;
\langle X\rangle^{-\alpha}
R^z
\langle X\rangle^{-\alpha}
\;.
\end{equation}

\begin{theo}
\label{theo-3dMain}
Let $d=3$ and suppose that {\rm Hypothesis~\ref{hyp-Wdisc}} to {\rm \ref{hyp-WeylType}} hold. For all $E\in\RM$ and $\alpha > \frac{5}{4}$, the limit
\begin{equation}
\label{eq-OpLimits}
\Rr^{E\pm \imath 0}_\alpha
\;=\;
\lim_{\epsilon\downarrow 0}
\Rr^{E\pm \imath \epsilon}_\alpha
\end{equation}
exists in operator norm.  Moreover, the limit operators for such $\alpha$ are H\"older continuous in $E$ in operator norm:
\begin{equation}
\label{eq-OpHoelder}
\|
\Rr^{E\pm \imath 0}_\alpha
-
\Rr^{E'\pm \imath 0}_\alpha
\|
\;\leq\;
C_\beta\,|E-E'|^\beta
\;,
\end{equation}
with  $0<\beta < 1$ with $\beta < \min\{2\alpha- \frac{5}{2},\frac{1}{2}, \alpha-1\} $.
\end{theo}

For the special case of the discrete Laplacian, the estimates of Theorem~\ref{theo-3dMain} were obtained in \cite{KSM} based on explicit formulas for the Green function in terms of Bessel functions. The main novelty of Theorem~\ref{theo-3dMain} is that it holds for generic operators, and allows the bands to touch in weakly tilted Weyl points. Let us note that Theorem~\ref{theo-3dMain} and other statements below can readily be  strengthened to obtain convergence in \eqref{eq-OpLimits} w.r.t. the Hilbert-Schmidt norm or other Schatten classes, but no details on this will be provided here.

\vspace{.2cm}

The proof of Theorem~\ref{theo-3dMain} will ultimately be given in Section~\ref{sec-Global}. It consists of many intermediate steps and results, some of which also allow to deal with higher dimensions $d> 3$ as well as $d=2$. For $d>3$, the theorem von Neumann and Wigner \cite{NW} states that bands generically touch on submanifolds of codimension $3$, a situation that is not dealt with in detail in the present work (however, we expect the techniques to transpose to such a case, see the remark at the end of Section~\ref{sec-Weyl}). Another (non-generic) situation not addressed here concerns Weyl or Dirac points with higher degenerecies which can be constructed using irreducible representations of the Clifford algebra \cite{SSt}; under a generic perturbation, these higher degenerecies will will dissolve to the above submanifolds of codimension $3$ with double degenerecies. Neither of these problems appears in the case $L=1$ of a single band, and, more generally, if one assumes absence of band touching which is an open condition because it is stable under perturbations.

\begin{theo}
\label{theo-OpNorm}
Suppose that $\Ww=\emptyset$, $d \geq 3$, and that {\rm Hypothesis \ref{hyp-Morse}} holds. For all $E\in\RM$ and $\alpha > \frac{d+2}{4}$, the limit \eqref{eq-OpLimits} exists in operator norm.  Moreover, the limit operators for such $\alpha$ satisfy the H\"older estimate \eqref{eq-OpHoelder} for positive $\beta < \min\{2\alpha- \frac{d+2}{2},\frac{d-2}{2}, \alpha-1, 1\} $.
\end{theo}

Note that, while Theorem~\ref{theo-OpNorm} excludes band touching by assumption, it does deal with critical points of arbitrary signature. Its proof is also completed in Section~\ref{sec-Global}. The proofs of Theorems~\ref{theo-3dMain} and \ref{theo-OpNorm} are both based on a suitable smooth partition of unity of the Brillouin torus $\TM^d$ allowing to separate regular, critical and band touching points of the energy bands. Such a partition will be constructed in Section~\ref{sec-Global}.  For any  smooth function $\rho:\TM^d\to[0,1]$, typically given by an element of a partition of unity, considered as a multiplication operator on $L^2(\TM^d,\CM^L)$, let us set
$$
\Pp_\rho
\;=\;
\Ff^* \rho\Ff
\;,
$$
and then
\begin{equation}
\label{eq-LocResol}
R_\rho^{z}
\;=\;
\Pp_\rho R^z
\;=\;
R^z \Pp_\rho 
\;=\;
\Ff^* \rho (\Ee-z\one)^{-1}\Ff
\;,
\qquad
\Rr^z_{\rho,\alpha}
\;=\;
\langle X\rangle^{-\alpha}
R_\rho^z
\langle X\rangle^{-\alpha}
\;.
\end{equation}
Using a partition of unity, one can then decompose $R^z$ and $\Rr^z$ into a finite number of contributions, stemming from the regular points, the critical points and the degenerate points. These summands are analyzed by different techniques which also lead to different statements. Using Riesz projections, one can, moreover, reduce the study of the regular points and critical points to the scalar case $L=1$ of just one band (see again Section~\ref{sec-Global} for details). Let us first spell out the outcome of the analysis for the regular part in this $1$-band case:

\begin{theo}
\label{theo-RegValues}
Let $d\geq 2$ and $L=1$. Suppose that the support of $\rho$ does not contain a critical point of $\Ee$. Then for $\alpha>\frac{1}{2}$, the limit
\begin{equation}
\label{eq-OpLimitsRegular}
\Rr^{E\pm \imath 0}_{\rho,\alpha}
\;=\;
\lim_{\epsilon\downarrow 0}
\Rr^{E\pm \imath \epsilon}_{\rho,\alpha}
\end{equation}
exists in operator norm.  Moreover, these limit operators are H\"older continuous in $E$ in operator norm:
\begin{equation}
\label{eq-OpHoelderRegular}
\|
\Rr^{E\pm \imath 0}_{\rho,\alpha}
-
\Rr^{E'\pm \imath 0}_{\rho,\alpha}
\|
\;\leq\;
C_\beta\,|E-E'|^\beta
\;,
\end{equation}
with $\beta < 1$ satisfying $0<\beta<\alpha-\frac{1}{2}$.
\end{theo}

Section~\ref{sec-RegPoints} provides a rather elementary proof of Theorem~\ref{theo-RegValues}, based merely on the implicit function theorem and the coarea formula (actually an explicit version of the Sobolev trace theorem). This is similar to the proof for translation invariant differential operators in Chapter 2 of \cite{Yaf2}. 

\vspace{.2cm}

Let us now come to the critical points for which it is again possible to focus on the scalar case $L=1$. By assumption, all critical points are generic (Morse), namely the Hessian of $\Ee$ at these points is non-degenerate. Then a critical point is called definite if all eigenvalues of the Hessian at this point have the same sign, otherwise it is called indefinite.

\begin{theo}
\label{theo-CritValues}
Let $d\geq 3$ and $L=1$. Suppose that $\rho$ is a smooth positive function supported in a sufficiently small neighborhood  $U$ of a critical point $k^*$. Then for $E\in\Ee(U)$ the limit $\langle n| R^{E\pm \imath 0}_\rho|m\rangle=\lim_{\epsilon\downarrow 0}\langle n| R^{E\pm \imath \epsilon}_\rho|m\rangle$ exists and satisfies
$$
|\langle n| R^{E\pm \imath 0}_\rho|m\rangle |
\;\leq\,
\left\{
\begin{array}{cc}
C\,\langle n-m\rangle^{-\frac{d-2}{2}} \;,
& E\in\Ee(U)\;,
\\
C\,\langle n-m\rangle^{-(d-2)} \;,
& E=\Ee(k^*)\;\,\mbox{definite }
\;.
\end{array}
\right.
$$
\end{theo}

As already mentioned above, for definite critical points the claim of Theorem~\ref{theo-CritValues} has appeared in various contexts. For the special case of the Laplacian on $\RM^d$ for $d\geq 3$, it can be obtained from explicit formulas for the Green function in terms of Hankel functions, see Chapters~1 and 7 of \cite{Yaf2} for a history on the numerous contributions. For generic periodic differential operators on $\RM^d$, there are the contributions \cite{MT,KR,KKR}, all dealing with band edges corresponding to definite critical points. Other critical points are analyzed in \cite{Ger}, albeit with exponential rather than powerlaw weights. Furthermore, the proofs in \cite{Ger} appeal to many deep prior results on oscillatory integrals with analytic phase factors, and are therefore not easily accessible for a broader audience. For periodic operators on lattice Hilbert spaces as considered here, there are the works \cite{Mar,BSB} and also the afore mentioned contribution \cite{KSM} for the special case of the discrete Laplacian. Furthermore, in probability theory the same object (Green function at a band edge) is connected to the hitting probability of transient random walks and it thus has been analyzed in great detail, see Chapter~6 in \cite{Spi} and Chapter~4 in \cite{LL}, providing similar results to Theorem~\ref{theo-CritValues}. The novel contribution in Theorem~\ref{theo-CritValues} hence concerns the case of indefinite critical points.

\vspace{.2cm}

Next let us state the main technical statement on Weyl points which is proved in Section~\ref{sec-Weyl}. This is of relevance for Theorem~\ref{theo-3dMain}, but not Theorem~\ref{theo-OpNorm}.

\begin{theo}
\label{theo-WeylMain}
Let $d=3$ and suppose that {\rm Hypothesis~\ref{hyp-Wdisc}} to {\rm \ref{hyp-WeylType}} hold. Further let  $\rho$ be a smooth positive function supported in a sufficiently small neighborhood  $U$ of a Weyl point $k^W$. Then for $E\in\Ee(U)$ the limit $\langle n| R^{E\pm \imath 0}_\rho|m\rangle=\lim_{\epsilon\downarrow 0}\langle n| R^{E\pm \imath \epsilon}_\rho|m\rangle$ exists and satisfies
$$
\|\langle n| R^{E\pm \imath 0}_\rho|m\rangle \|
\;\leq\,
C\,\langle n-m\rangle^{-1}\,\log(\langle n-m\rangle) 
\;.
$$
Furthermore, for $0<\beta< 1$ and  all $E,E'\in\Ee(U)$, 
$$
\|\langle n| R^{E\pm \imath 0}_\rho|m\rangle-\langle n| R^{E'\pm \imath 0}_\rho|m\rangle \|
\;\leq\;
C_\beta\,|z-z'|^\beta\,\langle n-m\rangle^{-1+\beta}\,\log(\langle n-m\rangle) 
\;.
$$
\end{theo}

As to dimension $d=2$, Theorem~\ref{theo-RegValues} covers all regular points, but neither critical points are covered by Theorem~\ref{theo-CritValues} nor band touching is covered by Theorem~\ref{theo-WeylMain}. This latter point is not a limitation for generic operators because again by the theorem of von Neumann and Wigner \cite{NW} there is generically no band touching at all in dimension $d=2$. If one imposes a chiral symmetry, however, band touching in so-called Dirac points is robust within the class of chiral Hamiltonians. While not carried out in detail, we believe that the techniques of Section~\ref{sec-Weyl} transpose to the study such Dirac points. As to critical values in $d=2$, there is no equivalent of Theorem~\ref{theo-CritValues} because there are well-known so-called van Hove singularities. In fact, just as for the Laplacian on $\RM^2$ \cite{Yaf2}, even the diagonal matrix elements of the resolvent explode logarithmically at critical values (more precisely, the real part at definite critical points and the imaginary part at indefinite ones \cite{Eco}). By the techniques of this work, it is possible to prove that for all $E\in\RM$ in a pointed neighborhood of a critical value $E^*$ and suitable $\alpha>0$, the operator norm limits $\Rr^{E\pm \imath 0}_\alpha$ in \eqref{eq-OpLimits} exist and satisfy $\|\Rr^{E\pm \imath 0}_\alpha\| \leq C\,|\log(|E-E^*|)|$. This will be further analyzed in a forthcoming work.

\vspace{.2cm}

\noindent {\bf Notations:} $\langle x,y\rangle$ denotes the euclidean scalar product in $\RM^d$ and $|x|=\sqrt{\langle x,x\rangle}$ the euclidean length. The open upper half-plane is denoted by $\HM=\{z\in\CM\,:\,\Im m(z)>0\}$, its closure by $\overline{\HM}$.  Furthermore $\imath=\sqrt{-1}$. Throughout the text, $C$ denotes changing constants that may depend on parameters. At some instances, this dependence is made explicit to facilitate the exposition.

\section{Limit absorption away from critical points}
\label{sec-RegPoints}

This section is about the proof of Theorem~\ref{theo-RegValues} which addresses regular points $k \in \mathbb{T}^d$ of $\Ee$ at which $\nabla \mathcal{E}(k) \ne 0$. Hence let $\rho$ be a smooth function with support containing only regular points.  Moreover, the focus will here only be on the special case $L=1$ for which $\Ee:\TM^d\to\RM$ is simply a scalar function. Later on in Section~\ref{sec-Global} it will be relatively easy to show that this implies the case of general $L$. Let us outline the strategy of proof. First of all, after the unitary Fourier transform $\Ff$, all arguments will be carried out in Fourier space $L^2(\TM^d)$.  Wave functions will be denoted by $\phi,\psi \in L^2(\TM^d)$. As the statement of Theorem~\ref{theo-RegValues} involves $\Rr^z_{\rho,\alpha}$ and hence the damping factors $\langle X\rangle^{-\alpha}$, it is useful to introduce the notation
$$
M_{\alpha}
\;=\;
\Ff \langle X\rangle^{\alpha}\Ff^*
\;,
$$
and the associated Sobolev space
$$
L^2_\alpha(\TM^d)
\;=\;
\big\{\phi\in L^2(\TM^d)\,:\, M_{\alpha}\phi\in L^2(\TM^d)\big\}
\;,
$$
equipped with the norm $\|\phi\|_{\alpha}=\|M_{\alpha}\phi\|$. Several arguments below will use the fact that the trigonometric polynomials (namely, the image under the Fourier transform of compact supported sequences) form a dense subset in $L^2_\alpha(\TM^d)$. The proof of Theorem~\ref{theo-RegValues} uses the quadratic form 
\begin{equation}
\label{eq-DefQForm}
\Qq^z(u,v)
\;=\;
\langle u | \langle X\rangle^{-\alpha}  R^z_\rho \langle X \rangle^{-\alpha}v \rangle=\langle \Ff\langle X\rangle^{-\alpha} u| \Ff R^z_\rho\Ff^* \Ff\langle X\rangle^{-\alpha} v\rangle
\;,
\qquad
u,v\in \ell^2(\mathbb{Z}^d)\;.
\end{equation}
It will be shown that this form extends (from the compact supported sequences) to a bounded quadratic form on $\ell^2(\mathbb{Z}^d)$ that remains bounded as $\Im m(z)\to 0$. After the change of variable $\phi = \Ff \langle X\rangle^{-\alpha} u=M_{-\alpha}\Ff u$, (note that $\|\phi\|_\alpha = \|u\|_{l^2})$ one hence has to see that 
\begin{equation}
\label{eq-DefQForm2}
\langle\phi| \Ff R^z_\rho\Ff^* \psi\rangle
\;,
\end{equation}
extends (from the trigonometric polynomials) to a quadratic form on $L^2_\alpha(\TM^d)$ which remains bounded  as $\Im m(z)\to 0$. For this it is sufficient to suppose that $\phi$ and $\psi$ are trigonometric polynomials and show that \eqref{eq-DefQForm2} is bounded by their norm in $L^2_\alpha(\TM^d)$. In order to show that, let us first note that $\Ff R^z_\rho\Ff^*$ is a multiplication operator by the function $(\Ee(k)-z)^{-1} \rho(k)$, namely
$$
\langle\phi| \Ff R^z_\rho\Ff^* \psi\rangle
\;=\;
\int_{\mathbb{T}^d}  dk\;\frac{1}{\Ee(k)-z}\,\overline{ \phi(k)}\psi(k)  \rho(k)
\;,
$$
where $dk$ is the Lebesgue measure on $\TM^d$. The integral over $dk$ will be computed via the coarea formula for the energy band function $\Ee$. Hence for $E\in\RM$ let us introduce the level sets by
$$
\Sigma^{E}
\;=\;
\{k\in \mathbb{T}^d: \Ee(k)=E\}
\;.
$$
This set is equipped with the Hausdorff measure $\nu^E$ induced by the Lebesgue measure $dk$. Then the coarea formula reads
\begin{equation}
\label{eq-QForm3}
\langle\phi| \Ff R^z_\rho\Ff^* \psi\rangle
\;=\;
\int_{\mathbb{R}}de\; \frac{1}{e-z} \int_{\Sigma^e}\nu^e(dk) \, \frac{\overline{\phi(k)} \psi(k)}{|\nabla\Ee(k)|}  \rho(k)  
\;.
\end{equation}
In the following it will be shown that the inner integral of  $\Sigma^e$ is bounded w.r.t. the norms of $\phi,\psi$ in $L^2_\alpha(\TM^d)$ and H\"older continuous in $e$. This follows from the Sobolev trace theorem, but is carried out explicitly below. 

\begin{proposi}
\label{pro-cot-lev}
Let $\alpha>\frac{1}{2}$. Then there exists $C > 0$ such that for any trigonometric polynomial
\begin{equation}
\label{equcot}
\left|\int_{\Sigma^E} \nu^E(dk) \;\frac{\overline{\phi(k)} \psi(k)}{|\nabla\Ee(k)|} \,
\rho(k)  \,\right|
\;\leq\; 
C\|\phi\|_{\alpha}\|\psi\|_{\alpha}
\;,
\end{equation}
where $C$ depends on $\rho$ and $\alpha$, but not on $\phi,\psi$ nor on $E$. Furthermore, the l.h.s. of \eqref{equcot} is H\"older continuous in $E$:
\begin{equation}
\label{eqcuta}
\left|\int_{\Sigma^E} \nu^E(dk) \;\frac{\overline{\phi(k)} \psi(k)}{|\nabla\Ee(k)|} \,
\rho(k)  
-
\int_{\Sigma^{E'}} \nu^{E'}(dk) \;\frac{\overline{\phi(k)} \psi(k)}{|\nabla\Ee(k)|} \,
\rho(k)
\,\right|
\;\leq\; 
C'\,|E-E'|^\beta\|\phi\|_{\alpha}\|\psi\|_{\alpha}
\;,
\end{equation}
where $\beta<\min\{\alpha-\tfrac{1}{2},1\}$ and the constant $C'$ depends on $\rho$, $\alpha$ and $\beta$,  but not on $\phi,\psi$ nor on $E,E'$. 
\end{proposi}

With these informations at hand the Plemelj-Privalov theorem \cite{Ple,Pri} applied to  \eqref{eq-QForm3} readily allows to complete the proof of Theorem~\ref{theo-RegValues}. This is carried out at the end of the section. Before focussing on the proof of Proposition~\ref{pro-cot-lev}, let us explain the connections with the Sobolev trace theorem.

\begin{rem}
{\rm
Proposition~\ref{pro-cot-lev} can be restated. Defining  $\Upsilon^E_\rho\,:\,L^2_\alpha(\TM^d)\to L^2(\Sigma^E \cap \supp(\rho),\nu^E)$ by
$$
(\Upsilon^E_\rho\phi)(k)\;=\;
\Big(\frac{\rho(k)}{|\nabla\Ee(k)|}\Big)^{\frac{1}{2}}\,\phi(k)\;,
\qquad
k\in\Sigma^E
\;,
$$
one has the bounds 
$$
\|\Upsilon^E_\rho \phi\| \;\leq\;C\,\|\phi\|_{\alpha}
$$
and
$$
\big|\|\Upsilon^E_\rho\phi\|^2-\|\Upsilon^{E'}_\rho \phi\|^2\big|\;\leq\;C'\,|E-E'|^\beta\,\|\phi\|_{\alpha}^2
\;.
$$
These are statements about the Sobolev trace operator.
}
\hfill $\diamond$
\end{rem}

Let us start the proof of Proposition~\ref{pro-cot-lev} with several preparatory lemmata.

\begin{lemma}
\label{lem-cot-cur}
Let $\alpha>\frac{1}{2}$. Then, there exists a constant $C_\alpha$ such that for any measurable function $G: \mathbb{T}^{d-1}\to \mathbb{T}^1$ and any trigonometric polynomial $\phi$  one has
$$
\int_{\TM^{d-1}} d\hat{k}\;| \phi (G(\hat{k}),\hat{k})|^2 
\; \leq \;
C_\alpha\,\|\phi\|_\alpha^2
\;,
$$
where here $d\hat{k}$ denotes the $(d-1)$-dimensional Lebesque measure.
\end{lemma}

\noindent {\bf Proof.}
Recall that by definition a trigonometric  polynomial $\phi$ has a Fourier transform $\Ff^* \phi$ with finite support. Hence all sums in the following equation are finite:
$$
\phi(\tilde k, \hat{k})
\;= \;
(2\pi)^{-\frac{d}{2}}
\sum_{n\in \mathbb{Z}} e^{\imath  \tilde k n}  \sum_{m\in \mathbb{Z}^{d-1}} e^{\imath  \langle  \hat{k}, m\rangle} (\mathcal{F}^*\phi)(n,m)
\;. 
$$
Using the fact that $\langle n \rangle^{-\alpha}  \in \ell^2(\mathbb{Z},\mathbb{C})$ together with the Cauchy-Schwarz inequality one hence obtains 
$$
|\phi(\tilde k,\hat{k})|^2 
\;\leq \;
(2\pi)^{-d}
\Big(\sum_{n\in \mathbb{Z}} \langle n \rangle^{-2\alpha}\Big) \sum_{n\in \mathbb{Z}} \langle n\rangle^{2\alpha} 
\Big|\sum_{m\in \mathbb{Z}^{d-1}} e^{\imath \langle \hat{k},m\rangle} (\mathcal{F}^*\phi)(n,m)\Big|^2
\;. 
$$
Hence, setting $C_\alpha=(2\pi)^{-1}\sum_{n\in \mathbb{Z}} \langle n \rangle^{-2\alpha}$ and using the monotone convergence theorem,
\begin{align*}
\int_{\TM^{d-1}} d\hat{k}\;| \phi (G(\hat{k}),\hat{k})|^2 
&
\;\leq\;
(2\pi)^{-d+1}C_\alpha
\sum_{n\in \mathbb{Z}} \langle n\rangle^{2\alpha} 
\int_{\TM^{d-1}} d\hat{k}
\Big|\sum_{m\in \mathbb{Z}^{d-1}} e^{\imath \langle \hat{k},m\rangle} (\mathcal{F}^*\phi)(n,m)\Big|^2
\\
&
\;=\;
C_\alpha
\sum_{n\in \mathbb{Z}} \langle n\rangle^{2\alpha} 
\sum_{m\in \mathbb{Z}^{d-1}}| (\mathcal{F}^*\phi)(n,m)|^2
\;,
\end{align*}
where in the last step the unitarity of the discrete Fourier transform in dimension $(d-1)$ was used.  This implies the claim.
\hfill $\Box$

\begin{lemma}
\label{lem-des-cok}
Let $0<\alpha\leq1$.  For all $z,z'\in \mathbb{H}$ and $t\in \mathbb{R}$ one has
$$
 |e^{\imath  zt}-e^{\imath z't}|
\;\leq \;
(e^{-\Im m(z)t}+ e^{-\Im m(z) t})^{1-\alpha}|z-z'|^{\alpha}|t|^\alpha
\;.
$$
\end{lemma}

\noindent {\bf Proof.}
The mean value theorem implies $|e^{\imath zt}-e^{\imath z' t}|\leq |z-z'| |t|.$ Combining this with $|e^{\imath z t}-e^{\imath z' t}|\leq  e^{-t\Im m(z)}+e^{-t\Im m(z')}$, one obtains
$$ 
|e^{\imath  zt}-e^{\imath z't}|
\;=\;
|e^{\imath z t}-e^{\imath z' t}|^{1-\alpha} |e^{\imath z t}-e^{\imath z' t}|^{\alpha}  
\;\leq \;
(e^{-\Im m(z)t}+ e^{-\Im m(z) t})^{1-\alpha}|z-z'|^\alpha |t|^\alpha
\;,
$$
which is the claim.
\hfill $\Box$

\begin{lemma}
\label{lem-des-hol}
Let $\alpha>\frac{1}{2}, \beta \leq 1$ and $0<\beta < \alpha-\frac{1}{2}$. Then, there exists a constant $C_{\alpha,\beta}$ such that for all measurable functions $g,f: \mathbb{T}^{d-1} \to \mathbb{T}^1$ and any trigonometric polynomial $\phi$, one has
$$
\int_{\TM^{d-1}}  d\hat{k} \,|\phi(g(\hat{k}),\hat{k})-\phi(f(\hat{k}),\hat{k})|^2 
\;\leq\; 
C_{\alpha,\beta}\,\sup_{\hat{k}\in \TM^{d-1}} |g(\hat{k})-f(\hat{k})|^{2\beta}  \,\|\phi\|_\alpha^2
\;.
$$
\end{lemma}

\noindent {\bf Proof.}
Using the Fourier representation of $\phi$, one has   
$$
\phi(\tilde k, \hat{k})-\phi(\tilde{k}', \hat{k})
\;=\;
\sum_{n\in \mathbb{Z}} (e^{\imath  \tilde{k}' n}-e^{\imath  \tilde k n})
\sum_{m\in \mathbb{Z}^{d-1}} e^{\imath \langle \hat{k},m\rangle} (\mathcal{F}^*\phi)(n,m)
\;.
$$
Lemma \ref{lem-des-cok} with $t=n$ and  implies that 
$$
\sum_{n\in \mathbb{Z}} | e^{\imath  n \tilde{k}'}-e^{\imath n \tilde k}|^2\langle n\rangle^{-2\alpha}
\;\leq\; 
4\,|\tilde{k}-\tilde{k}'|^{2\beta}\sum_{n\in \mathbb{Z}}\langle n\rangle ^{2(\beta-\alpha)}
\;<\;
\infty
\;.
$$
Together with the Cauchy-Schwarz inequality, one deduces
$$
|\phi(\tilde k, \hat{k})-\phi(\tilde{k}', \hat{k})|^2
\;\leq\; 
C\, |\tilde{k}-\tilde{k}'|^{2\beta}\sum_{n\in \mathbb{Z}} \langle n\rangle^{2\alpha} \left| \sum_{m\in \mathbb{Z}^{d-1}} e^{\imath \langle \hat{k},m\rangle} (\mathcal{F}^*\phi)(n,m)\right|^2
\;.
$$
Choosing $\tilde k =g(k)$ and $\tilde{k}'=f(k)$ and integrating over $\hat{k}$ allows to complete the proof as in Lemma~\ref{lem-cot-cur}.
\hfill $\Box$

\vspace{.2cm}

\noindent {\bf Proof} of Proposition~\ref{pro-cot-lev}. Let us first prove the following statement. For all $E_0\in \spec(H)$ there exists $\delta_{E_0}>0$ and $C_{E_0}\in \mathbb{R}$ such that for any trigonometric polynomial $\phi$ one has 
\begin{equation}
\label{equs-fors-outs}
\int_{\Sigma^E} \nu^E(dk) \, \frac{|\phi(k)|^{2}}{|\nabla\Ee(k)|} \,
\rho(k)  
\;\leq\; 
C_{E_0}\|\phi\|_{\alpha}^2
\;, 
\qquad  |E- E_0|<\delta_{E_0}
\;.
\end{equation} 
Let $E_0\in \spec(H)$. If $\Sigma^{E_0}\cap \supp(\rho)=\emptyset,$ then one can take $\delta_{E_0}>0$ such that for all $E\in (E_0-\delta_{E_0}, E_0+\delta_{E_0})$ one has $\Sigma^E \cap \supp(\rho)=\emptyset$ and $C_{E_0}=0$, so that  \eqref{equs-fors-outs}  trivially holds.  Now, suppose that $\Sigma^{E_0}\cap \supp(\rho) \neq \emptyset$ and consider a point $k_0=(\tilde{k}_{0},\hat{k}_{0})$ in $\Sigma^{E_0}\cap \supp(\rho)$, with $\tilde{k}_{0}\in\TM^1$ and $\hat{k}_{0}\in\TM^{d-1}$. By assumption, $\nabla\Ee(k_0)\neq 0$. W.l.o.g.  one can assume that the partial derivative of $\Ee$ w.r.t. to the first coordinate satisfies $\partial_{k_1}\Ee(k_0)\not=0$. By continuity, there hence exists neighborhoods $\tilde{U}_{k_0}\subset\TM^1$ of $\tilde{k}_0$ and $\hat{U}_{k_0}\subset\TM^{d-1}$ of $\hat{k}_0$ such that for all $k\in U_{k_0} = \tilde{U}_{k_0}\times \hat{U}_{k_0}$ one has
\begin{equation}
\label{eq-as-po}
\left|\partial_{k_1} \Ee (k)\right| \;> \;\tfrac{1}{C_{k_0}}
\end{equation} 
for some $C_{k_0}>0$. Let now $g^E: \hat{U}_{k_0}\to \tilde{U}_{k_0}$ be the implicit function such that $\Ee(g^E(\hat{k}),\hat{k})=E$, and set $G^E(\hat{k})=(g^E(\hat{k}),\hat{k})$. Now one can change of variables and estimate as follows for all $E\in (E_0-\delta_{k_0},E_0+\delta_{k_0})$:
\begin{align*}
\int_{\Sigma^{E} \cap U_{k_0}} \nu^E(dk) \frac{|\phi(k)|^{2}}{|\nabla\Ee(k)|} \rho(k)  
&
\;=\; 
\int_{\hat{U}_{k_0}}  d\hat{k} \, 
\frac{1}{|(\partial_{k_1} \Ee )(G^E(\hat{k}))|}\,
|\phi(G^E(\hat{k}))|^{2} 
\,
\rho(G^E(\hat{k}))
\\
&
\;\leq \;
C_{k_0} \int_{\hat{U}_{k_0}}  d\hat{k}\, |\phi (g^E(\hat{k}),\hat{k})|^{2} 
\\ 
& 
\;\leq\; 
C_{k_0} \|\phi\|^2_\alpha
\,,
\end{align*}
where first  \eqref{eq-as-po} and then  Lemma~\ref{lem-cot-cur} were used. In particular, let us stress that the bound holds uniformly for $E\in (E_0-\delta_{k_0},E_0+\delta_{k_0})$. Next, applying the above argument several times and invoking compactness one can choose a finite open cover $\{U_{k_j} : j=1,\dots,J\}$ of  $\Sigma^{E_0}\cap \supp(\rho)$ such that $U_{k_j}$  satisfies the above properties with an associated $\delta_{k_j}$. Then choose a positive $\delta_{E_0}< \min \delta_{k_j}$ such that, for every $E\in (E_0-\delta_{E_0},E_0+\delta_{E_0})$, $\{U_{k_j} : j=1,\dots,J \}$ is also an open cover of $\Sigma^{E}\cap \supp(\rho)$, and set $C_{E_0} = J\cdot \max \{C_{k_j}: \ j=1,\dots,n\}$. Then one obtains  \eqref{equs-fors-outs}. In order to deduce \eqref{equcot}, one takes a finite subcover of the cover $\{(E-\delta_E,E+\delta_E): E \in \spec(H)\}$ and sets $C_{\rho,\alpha} = \max\{C_{E_j}\}$. Then for all $E\in \spec(H)$, 
$$
\int_{\Sigma^E} \nu^E(dk)  \frac{|\phi(k)|^{2}}{|\nabla\Ee(k)|} 
\rho(k)  
\;\leq \;
C_{\rho,\alpha}\|\phi\|_{\alpha}^2
\;.
$$
Combined with the Cauchy-Schwarz inequality this implies \eqref{equcot}.  

\vspace{.1cm}

For the H\"older continuity, one proceeds in a similar manner by invoking Lemma~\ref{lem-des-hol}. Similar to \eqref{equs-fors-outs}, one first proves a bound that is local in energy, notably one fixes $E_0\in \spec(H)$ and shows that there exists $\delta_{E_0}>0$ and $C_{E_0}\geq 0$ such that for all $E,E'\in (E_0-\delta_{E_0},E_0+\delta_{E_0})$ 
\begin{equation}
\label{equ-cod-dim}
\left| \int_{\Sigma^{E}} \nu^E(dk) \frac{\overline{\phi(k)}\psi(k)}{|\nabla \Ee(k)|} \rho(k) -\int_{\Sigma^{E'}} \nu^{E'}(dk) \frac{\overline{\phi(k)}\psi(k)}{|\nabla \Ee(k)|} \rho(k) \right| 
\;\leq\; 
C_{E_0}|E-E'|^\beta \|\psi\|_{\alpha} \|\phi\|_{\alpha}
\;.
\end{equation} 
If $\Sigma^{E_0}\cap \supp(\rho)= \emptyset$, this holds trivially. Hence let us suppose that $\Sigma^{E_0}\cap \supp(\rho)\neq \emptyset$, and let $k_0\in \Sigma^{E_0} \cap \supp(\rho) $ be such that, w.l.o.g.,  $\partial_{k_1} \Ee (k_0)\neq 0$. In order to estimate the difference in \eqref{equ-cod-dim} locally in $k_0$, let us introduce a compactly supported smooth function $\rho_{k_0}\in C^\infty_0(U_{k_0},[0,1])$. As before, for $\delta_{k_0}>0$ small enough  and $E\in (E_0-\delta_{k_0},E_0+\delta_{k_0})$ one then has the change of variables
$$
\int_{\Sigma^{E} \cap U_{k_0}}\!\! \nu^E(dk) \frac{\overline{\phi(k)}\psi(k)}{|\nabla\Ee(k)|} \rho(k) \rho_{k_0}(k) 
\,=\,
\int_{\hat{U}_{k_0}}  d\hat{k} \, \frac{\overline{\phi(G^E(\hat{k}))}\psi(G^E(\hat{k})) }{|(\partial_{k_1} \Ee )(G^E(\hat{k}))|}
\,  \rho(G^E(\hat{k}))\rho_{k_0}(G^E(\hat{k}))
\,.
$$
This allows to compare the integrals locally in the level sets $\Sigma^E$ and $\Sigma^{E'}$. For the sake of notational convenience, let us introduce 
$$
l^E(\hat{k})
\;=\;
\frac{\rho(G^E(\hat{k}))\rho_{k_0}(G^E(\hat{k}))}{|(\partial_{k_1} \Ee )(G^E(\hat{k}))|}
\;.
$$
Due to \eqref{eq-as-po} one then has $|l^E(\hat{k})|\leq C_{k_0}$. Then for $E,E' \in (E_0-\delta_{k_0},E_0+\delta_{k_0})$
\begin{align*}
\Big| \int_{\Sigma^{E}} & \nu^E(dk) \frac{\overline{\phi(k)}\psi(k)}{|\nabla \Ee(k)|} \rho(k)\rho_{k_0}(k) -\int_{\Sigma^{E'}} \nu^{E'}(dk) \frac{\overline{\phi(k)}\psi(k)}{|\nabla \Ee(k)|} \rho(k)\rho_{k_0}(k) \Big|^2 
\\
\;=\;
&
\Big|
\int_{\hat{U}_{k_0}} d\hat{k}\,
\Big(
\overline{\phi(G^E(\hat{k}))}\,\psi(G^E(\hat{k}))\,l^E(\hat{k})
\,-\,
\overline{\phi(G^{E'}(\hat{k}))}\,\psi(G^{E'}(\hat{k}))\,l^{E'}(\hat{k})
\Big)
\Big|^2
\\
\;\leq\;
&
C_{k_0}^2
\Big(
\int_{\hat{U}_{k_0}} d\hat{k}\,|\phi(G^E(\hat{k}))-\phi(G^{E'}(\hat{k}))|^2
\Big)
\Big(
\int_{\hat{U}_{k_0}} d\hat{k}\,|\psi(G^E(\hat{k}))|^2
\Big)
\\
&
+
C_{k_0}^2
\Big(
\int_{\hat{U}_{k_0}} d\hat{k}\,
 |\phi(G^{E'}(\hat{k}))|^2
\Big)
\Big(
\int_{\hat{U}_{k_0}} d\hat{k}\,|\psi(G^E(\hat{k}))-\psi(G^{E'}(\hat{k}))|^2
\Big)
\\
&
+
\sup_{k\in \tilde U_{k_0}}|l^E(k)-l^{E'}(k)|^2  \Big(
\int_{\hat{U}_{k_0}} d\hat{k}\,|\phi(G^{E'}(\hat{k}))|^2
\Big)
\Big(
\int_{\hat{U}_{k_0}} d\hat{k}\,|\psi(G^{E'}(\hat{k}))|^2
\Big) 
\;.
\end{align*}
Recalling that $G^E(\hat{k})=(g^E(\hat{k}),\hat{k})$, Lemmata~\ref{lem-cot-cur} and \ref{lem-des-hol} imply that
\begin{align*}
\Big(
\int_{\hat{U}_{k_0}} & d\hat{k}\,|\phi(G^E(\hat{k}))-\phi(G^{E'}(\hat{k}))|^2
\Big)
\Big(
\int_{\hat{U}_{k_0}} d\hat{k}\,|\psi(G^E(\hat{k}))|^2
\Big)
\\
&
\;\leq \;
C'_{k_0}\|\psi\|_\alpha^2 \|\phi\|_\alpha^2 \,\sup_{\hat{k}\in \hat{U}_{k_0}} |g^E(\hat{k}) - g^{E'}(\hat{k})|^{2\beta} 
\\
&
\;\leq\; 
C'_{k_0}\, |E-E'|^{2\beta} \|\phi\|_\alpha^2 \|\psi\|_\alpha^2 
\;,
\end{align*}
  where in the last step the continuity of $(E,\hat{k})\in[E_0-\delta_{k_0},E_0+\delta_{k_0}] \times \hat{U}_{k_0}\mapsto \partial_E g^E(\hat{k})$ and the mean value theorem were used. The second summand can be dealt with in the same manner. The third summand satisfies a similar estimate because also $(E,\hat{k})\in[E_0-\delta_{k_0},E_0+\delta_{k_0}] \times \hat{U}_{k_0}\mapsto \partial_E g^E$ is continuous. Together one obtains for all $E,E' \in (E_0-\delta_{k_0},E_0+\delta_{k_0})$
$$
\Big| \int_{\Sigma^{E}} \!\! \nu^E(dk) \frac{\overline{\phi(k)}\psi(k)}{|\nabla \Ee(k)|} \rho(k)\rho_{k_0}(k) -\int_{\Sigma^{E'}} \!\! \nu^{E'}(dk) \frac{\overline{\phi(k)}\psi(k)}{|\nabla \Ee(k)|} \rho(k)\rho_{k_0}(k) \Big| 
\,\leq\,
C_{k_0} |E-E'|^{\beta} \|\phi\|_{\alpha}\|\psi\|_{\alpha}
,
$$
with a constant $C_{k_0}$ that does not depend on $E,E' \in (E_0-\delta_{k_0},E_0+\delta_{k_0})$. Finally let us consider a finite open cover $\{U_{k_j}: j=1,\dots,J\}$ of $\Sigma^{E_0} \cap \supp(\rho)$ with $U_{k_j}$ and associated $\delta_{k_j}$ again satisfying the properties as $k_0$ above. Let $\{\rho_{k_j}: j=1,\dots,J\}$ be a partition of unity subordinated to this cover. Then there exists a $\delta_{E_0}>0$ such that, for every $E\in (E_0-\delta_{E_0},E_0+\delta_{E_0})$, $\{U_{k_j} : j=1,\dots,J \}$ is also an open cover of $\Sigma^{E}\cap \supp(\rho)$. Moreover, one can take $\delta_{E_0}>0$ such that $\delta_{E_0}< \min \delta_{{k_j}}$. Taking $C_{E_0} = J \cdot \max \{C_{k_j}: \ j=1,\dots,J\}$, the above local bound implies  \eqref{equ-cod-dim}, with $C_{E_0}>0$ that does not depend on $E,E' \in (E_0-\delta_{E_0},E_0+\delta_{E_0})$.  In order to obtain the global H\"older continuity, one takes a finite open cover $\{(E_0^i-\delta_{E_0^i}, E_0^i+ \delta_{E_0^i}) : i\in \{1,\dots,I\}\}$ of $\spec(H)$,  and considers  the Lebesgue number $\delta>0$ of this cover. Take $C_{\rho, \beta}=\max\{C_{E_0^i}, \frac{2C_\rho}{\delta^\beta}\}$, where $C_\rho$ is the constant given in   \eqref{equcot}. If $E,E' \in \spec(H)$ satisfy $|E-E'|< \delta$ then  \eqref{eqcuta} is satisfied. If $|E-E'| \geq \delta $ then triangle inequality together with  \eqref{equcot}, and the fact that 
$$
2C_{\rho}
\;=\;
\frac{2C_\rho}{|E-E'|^\beta} |E-E'|^\beta 
\;\leq \;
\frac{2C_\rho}{\delta^\beta} |E-E'|^\beta 
\;\leq\; 
C_{\rho,\beta} |E-E'|^\beta
\;,
$$ 
imply  \eqref{eqcuta}.
\hfill $\Box$

\vspace{.2cm}

\noindent {\bf Proof } of Theorem~\ref{theo-RegValues}.
Let $u, v: \mathbb{Z}^d \to \mathbb{C}$ be compactly supported sequences. Then $\langle X \rangle^{-\alpha} u$, $\langle X\rangle^{-\alpha} v$ are also compactly supported. Proposition~\ref{pro-cot-lev} implies that one can use the Plemelj-Privalov theorem \cite{Ple,Pri} in  \eqref{eq-QForm3} with $\phi = \Ff \langle X\rangle^{-\alpha} u$, $\psi=\Ff \langle X\rangle^{-\alpha} v$,  and obtain that for all $E\in \spec(H)$ the limit (see  \eqref{eq-DefQForm}) 
$$
\Qq^{E\pm \imath 0}(u,v)
\;=\;
\lim_{\epsilon\downarrow 0} \Qq^{E\pm\imath \epsilon}(u,v)
$$ 
exists. Moreover,  \eqref{equcot} and \eqref{eqcuta} imply that
$$
|\Qq^{E\pm \imath 0}(u,v)|
\;\leq\; 
C_\rho \|\phi\|_{\alpha}\|\psi\|_{\alpha} 
\;=\;
C_\rho \| u\|_{\ell^2}\| v \|_{\ell^2}
\;. 
$$ 
Furthermore, \eqref{equcot} and \eqref{eqcuta} imply that for all $E,E'\in \spec(H_0)$ and trigonometric polynomials $\phi,\psi$  one has
\begin{equation}
\label{e}
|\Qq^{E\pm \imath 0}(u,v) - Q^{E'\pm \imath 0} (u,v) |
\;\leq\; 
C_{\rho,\beta}|E-E'|^\beta \|u\|_{\ell^2} \|v\|_{\ell^2}
\;.
\end{equation} 
Therefore, for all $E\in  \spec(H)$, the sesquilinear form $\Qq^{E\pm \imath 0}(\cdot,\cdot)$ is bounded on the set of compactly supported sequences which is a dense subset of $\ell^2(\mathbb{Z}^d)$. Hence one can extend this sesquilinear form to a bounded sesquilinear form $\Qq^{E\pm \imath 0}(\cdot,\cdot):\ell^2(\mathbb{Z}^d)\times \ell^2(\mathbb{Z}^d)\to \mathbb{C}.$ For each $E\in \mathbb{R}$, let us then define $\Rr^{E\pm \imath 0}_{\rho,\alpha}\in \mathbb{B}(\ell^2(\mathbb{Z}^d))$ as the only bounded operator such that 
\begin{equation}\label{eq-de-qu}
\langle  u, \Rr^{E\pm \imath 0}_{\rho,\alpha}  v \rangle 
\;=\; 
\Qq^{E\pm \imath 0}(u,v)
\;, 
\qquad 
u,v \in \ell^2(\mathbb{Z}^d)
\;.
\end{equation} 
By definition and   \eqref{e}, one has  \eqref{eq-OpHoelderRegular}. Next, let us see that the limit in  \eqref{eq-OpLimitsRegular} exists. For   compactly supported sequences $u,v$, one obtains using \eqref{equcot} and \eqref{eqcuta}  that for all $z \in \HM$ and $E\in \spec(H)$ with $|z-E|<1$
$$
|\langle u,(\langle X\rangle^{-\alpha}R^z_\rho \langle X\rangle^{-\alpha} - \mathcal{R}_{\rho,\alpha}^{E+\imath 0})v\rangle|
\;=\;
|\Qq^z(u,v)-\Qq^{E+\imath 0}(u,v)|
\;\leq\; 
C_{\rho,\beta} |z-E|^\beta \|u\|_{\ell^2}\|v\|_{\ell^2}
\,. 
$$
As the compactly supported sequences are dense in $\ell^2(\mathbb{Z}^d)$, this implies that for $z\in \HM$ and $|z-E|<1$ one has  
$$
\| \langle X \rangle^{-\alpha}R^z_\rho\langle X\rangle^{-\alpha} - \mathcal{R}_{\rho,\alpha}^{E+\imath 0}\|
\;\leq\; 
C_{\rho,\beta} |z-E|^\beta
\;,
$$ 
which implies that the limit exists.
\hfill $\Box$

\section{Green function estimates near critical points}
\label{sec-CritPoints}

This section provides the proof of Theorem~\ref{theo-CritValues} which states quantitative bounds on the decay of the Green function for energies at or near a critical energy. These results are based on estimates on oscillatory integrals that we believe to be of independent interest and possibly of wider use, which is why they are given in Appendix~\ref{app-OsciInt}. Furthermore, at the end of the section, it is shown how the estimates on the Green function as given in Theorem~\ref{theo-CritValues} lead to operator norm bounds on the damped resolvent.

\vspace{.2cm}

Again this section is dealing merely the scalar case $L=1$. Let us consider a critical point $k^* \in \mathbb{T}^d$ of the function $\Ee$ and let $\rho\in C_0^\infty(B_{\delta}(k^*))$. Also let us denote $E^* = \Ee(k^*)$. Due to translation invariance, one has $\langle n | R_\rho^z | m\rangle =\langle n-m | R_\rho^z | 0\rangle$. Hence it is sufficient to consider the Green function 
$$
\langle n | R_\rho^z | 0\rangle 
\; =\;  
\langle n | \Ff R_\rho^z \Ff^*| 0\rangle  
\;=\; 
\int_{\mathbb{T}^d} dk\;\frac{e^{\imath \langle k, n\rangle}}{\Ee(k)-z} \,\rho(k)
\;.
$$
This expression is hence the Fourier transform of some function, which does, however, have a singularity as $\Im m(z)\to 0$.  As in \cite{Kos,Mar}, it is useful to replace this singularity by the identity
$$
\frac{1}{\Ee(k)-z} 
\;=\;  
\imath |n| \int_0^\infty  d\eta \ e^{(z-\Ee(k))\imath |n|\eta}
\;.
$$ 
This holds for $\Im m(z)>0$ and $n\neq 0$, and actually the case $n=0$ will not be considered any further (it can readily be dealt with if one uses the integral identity with $|n|=1$). Replacing in the above then leads to
\begin{equation}
 \langle n | R_\rho^z | 0\rangle 
 \;=\; 
\imath |n|\int_{\mathbb{T}^d} dk \int_0^\infty d\eta \ e^{\imath |n| (\langle k, \frac{n}{|n|}\rangle - \eta\Ee(k))} \,e^{\imath z |n|\eta} \, \rho(k)
\;.
\label{eq-Integral}
\end{equation}
This is a oscillatory integral with phase function $k\mapsto \langle k, \frac{n}{|n|}\rangle - \eta\,\Ee(k)$. An essential ingredient for its analysis is the stationary phase method \cite{Dui,DS}. As we could not locate a reference dealing with the particular case needed here, Appendix~\ref{app-OsciInt} provides a detailed analysis of the bounds on these oscillatory integrals.  The special case of a definite critical point of $\Ee$ has been treated in \cite{MT,KR,KKR}. The following result recollects the main facts of Theorem~\ref{theo-mai-res}, where without loss of generality $E^*=\Ee(k^*)=0$. 

\begin{proposi}
\label{prop-CritCollect}
Let $d\geq 3$. Then, for all $E\in \mathbb{R},$ the limit $ \langle n | R_\rho^{E\pm \imath 0}| 0\rangle =\lim_{\epsilon\downarrow 0} \langle n | R_\rho^{E\pm\imath\epsilon} | 0\rangle$ exists. Moreover, there exists a constant $C_\rho\in \mathbb{R}$ and a continuous functions $J_\rho: \overline{\HM} \times [1,\infty)\times \mathbb{S}^{d-1}\to \mathbb{C}$ such that with the notation $ \langle n \rangle = (|n|^2 + 1)^{\frac{1}{2}}  $
$$
\langle n | R_\rho^z| 0\rangle  
\;=\;
\langle n\rangle^{-\frac{d}{2}+1} J_\rho(z,|n|, \tfrac{n}{|n|}) 
\;,
$$ 
with
$$
|J_\rho(z, t , \omega)|\;\leq\; C_\rho
\;, 
\qquad 
\Im m(z)\geq 0\;, \;\; t\geq 1\,,\;\;\omega\in\SM^{d-1}
\;.
$$ 
For most directions $\omega$, a stronger bound can be obtained, namely there exists a constant $D_\rho\geq 0$ such that for all $\omega \in \mathbb{S}^{d-1}$ satisfying
$$
|\langle (\Hess \Ee(k^*))^{-1} \omega, \omega\rangle| 
\;>\; 
D_\rho
\;,
$$
one has 
$$
|J_\rho(z,t,\omega)|
\;\leq\; 
C_{\rho} \,\frac{t^{-\frac{d}{2}+1}}{|\langle (\Hess \Ee(k^*))^{-1} \omega, \omega\rangle|^{\frac{d+1}{2}} }
\;,
$$
uniformly for all $(z,t) \in \overline{\HM}\times [1,\infty)$ with 
$$
0\;\leq\; |z-E^*|\;<\; \tfrac{1}{8} \,\sup\big\{ |\nabla \Ee(k)|^2 : k \in B_{\delta}(k^*)\big\} |\langle (\Hess \Ee(k^*))^{-1} \omega, \omega\rangle|
$$
and 
$$
1 \;\leq \;t\; < \;\frac{\sqrt{|\langle \omega, (\nabla^2 \Ee (k^*))^{-1}\omega\rangle |}}{\sqrt{8|z-E^*|}} 
\;.
$$
The constant $D_\rho$ satisfies $D_\rho \to 0$ as $  \supp(\rho) $  shrinks to $ \{k^*\}$. Further, if $0<\beta\leq 1$ and $\beta <\frac{d-2}{2}$, then for all  $z,z' \in \overline{\HM}$ one has
$$
\big|\langle n | R_\rho^z| 0\rangle -\langle n | R_\rho^{z'}| 0\rangle\big| 
\;\leq \;
C_\beta \langle n\rangle^{-\frac{d}{2}+1+\beta}|z-z'|^{\beta}
\;, 
\qquad  
n \in \mathbb{Z}^d \setminus \{0\}
\;.
$$
\end{proposi} 

In the particular case of a critical point $k^*$ with definite signature, one can bound the quadratic form  $|\langle (\Hess \Ee(k^*))^{-1} \omega, \omega\rangle|>D>0$ uniformly for $\omega \in \mathbb{S}^{d-1}$ with some constant $D$. Then the statement of the proposition considerably strengthened. More precisely, choosing $z=E^*$ leads to the classical result \cite{Spi,LL,MT,KR}
$$
|\langle n | R_\rho^{E^*\pm\imath 0}| 0 \rangle| 
\;\leq\; 
C \langle n\rangle ^{-d+2}
\;.
$$

\noindent {\bf Proof} of Theorem~\ref{theo-CritValues}. This follows directly from Proposition~\ref{prop-CritCollect} and the comment above, combined with the translation invariance of $H$ and the resolvent.
\hfill $\Box$

\vspace{.2cm}

To conclude this section, let us now show that the decay estimates on the Green function allows to show that the damped resolvent exists and is H\"older continuous. Combined with the bounds from Section~\ref{sec-RegPoints} (notably Theorem~\ref{theo-RegValues}) the concludes a proof of Theorem~\ref{theo-OpNorm} in the special case $L=1$.

\begin{proposi}[Limit absorption principle near critical points]
\label{prop-CritLAP}
Let $d\geq 3$ and $\alpha > \frac{d+2}{4}$. Then the limit defining an operator $\Rr^{E\pm \imath 0}_{\rho,\alpha}:\ell^2(\mathbb{Z}^d)\to \ell^2(\mathbb{Z}^d)$ by 
$$
(\Rr^{E\pm \imath 0}_{\rho,\alpha}\phi)(n)
\;=\;
\langle n \rangle^{-\alpha} \sum_{m\in \mathbb{Z}^d} \langle n | R_\rho^{E\pm \imath 0}| m \rangle \langle m\rangle^{-\alpha} \phi(m)
$$
exists for all $E\in \mathbb{R}$, and it satisfies
\begin{equation}
\label{equ-lim-crp}
\lim_{\epsilon \downarrow 0}\|\langle X\rangle^{-\alpha} R^{E\pm\imath\epsilon}_\rho \langle X \rangle^{-\alpha} - \Rr_{\rho,\alpha}^{E\pm\imath 0} \| 
\;=\;
0
\;.
\end{equation} 
Moreover, if  $0<\beta\leq 1$ with  $\beta < \min\{2\alpha- \frac{d+2}{2},\frac{d-2}{2}, \alpha-1\}$ one has 
\begin{equation}
\label{equ-hol-ccp}
\|\Rr_{\rho,\alpha}^{E\pm\imath 0}-\Rr_{\rho,\alpha}^{E'\pm\imath 0}\|
\;\leq\; 
C_\beta |E-E'|^\beta
\;. 
\end{equation}
\end{proposi}

\noindent {\bf Proof.}
Let $\phi \in \ell^2(\ZM^d)$. Using the bound $|\langle n |R^{E\pm\imath 0}_\rho | m \rangle|\leq C \langle n-m\rangle^{-\frac{d}{2}+1}$ from Proposition~\ref{prop-CritCollect} together with 
\begin{equation}\label{equ-ide-prc}
\sum_{m\in\ZM^d} \langle n-m\rangle^{-d'} \langle m\rangle^{-2\alpha}
\;\leq\;C\,\langle n\rangle^{-d'-2\alpha+d}
\;,
\qquad
d'+2\alpha>d
\;,
\end{equation}
one obtains that 
\begin{align*}
\sum_{m\in\ZM^d}  |\langle n|R_\rho^{E\pm\imath 0}|m\rangle|^2 
\langle m\rangle^{-2\alpha} 
&
\;\leq\;
C
\sum_{m\in\ZM^d} \langle n-m\rangle^{-d+2} 
\langle m\rangle^{-2\alpha} 
\;\leq\;
C \langle n\rangle^{2-2 \alpha} 
\;,
\end{align*}
for all $n\in \mathbb{Z}^d$. By the Cauchy-Schwarz inequality, this leads to
$$
|\langle n|R_\rho^{E\pm\imath 0}
\langle X\rangle^{-\alpha}|\phi\rangle|
\;\leq\;
\sum_{m\in\ZM^d}
|\langle n|R_\rho^{E\pm\imath 0}|m\rangle|
\langle m\rangle^{-\alpha}\,|\phi(m)|
\;\leq\; 
C \,\|\phi\|\,\langle n \rangle^{1-\alpha}
\;,
$$
and, moreover,
$$
\|\Rr^{E\pm \imath 0}_{\rho,\alpha}\phi\|^2 
\;\leq\;
\sum_{n\in\ZM^d}
\langle n\rangle^{-2\alpha}
|\langle n|R_\rho^{E\pm\imath 0}
\langle X\rangle^{-\alpha}|\phi\rangle|^2
\;\leq\;
\sum_{n\in\ZM^d}
\langle n\rangle^{-2\alpha}
C \,\|\phi\|^2\langle n \rangle^{2-2\alpha}
\;\leq\;C\,\|\phi\|^2
\;,
$$
where in the last step it was used that $4\alpha-2>d$, which is the part of the hypothesis. This shows that the limit operator exists and is bounded. Next let us prove that \eqref{equ-hol-ccp} holds for $z,z' \in \overline{\HM}$; this then also implies \eqref{equ-lim-crp}. Using the bound  $|\langle n |(R^z_\rho - R^{z'}_\rho )| m \rangle|\leq C |z-z'|^\beta \langle n-m\rangle^{-\frac{d}{2}+1+\beta}$ as given in Proposition~\ref{prop-CritCollect}, one obtains just as above
$$
| \langle n |(R^z_\rho -R^{z'}_\rho)\langle X\rangle^{-\alpha}|\phi\rangle|
\;\leq\; 
C |z-z'|^\beta\|\phi\|\,\langle n \rangle^{1-\alpha+\beta}
\;.
$$ 
Since $4\alpha-2-2\beta> d,$ this implies again as above
$$
\|(\Rr^z_{\rho,\alpha}-\Rr^{z'}_{\rho,\alpha})\phi\|^2 
\;\leq\; 
C |z-z'|^{2\beta} \|\phi\|^2 \sum_{n\in \mathbb{Z}^d} \langle n \rangle^{2+2\beta-4\alpha} 
\;=\;
C_{\alpha,\beta}|z-z'|^{2\beta} \|\phi\|^2
\;,
$$
completing the proof.
\hfill $\Box$

\section{Limit absorption at Weyl points}
\label{sec-Weyl}

This section deals with a smooth selfadjoint function $k\in\RM^3\mapsto \Ee(k)\in\CM^{2\times 2}$ of the form \eqref{eq-2BandDecomp} satisfying Hypothesis~\ref{hyp-WeylType}, and then proves Theorem~\ref{theo-WeylMain}. In Section~\ref{sec-Global} it will be explained how this set-up is obtained by a reduction with Riesz projections in order to complete the proof of Theorem~\ref{theo-3dMain}. To simplify notations in the following, it will be assumed that 
\begin{align}
\label{PH} 
k^W\,=\,0\;, 
\qquad
E^W\,=\,e(k^W)\,=\,0
\;.
\end{align} 
Note that this can readily be realized by a shift in space and $z$ and that then also $h(0)=0$.  Let us start out by laying out the general strategy, and then give details for a toy model, before giving full technical details  for the proof of Theorem~\ref{theo-WeylMain}. Based on the decomposition \eqref{eq-2BandDecomp},  one has 
$$
\det(\Ee(k)-z\,\one)
\;=\;
\big(e(k)-z\big)^2\,-\,|h(k)|^2
\;,
$$
so that
\begin{align*}
(\Ee (k)-z\,\one)^{-1}
&
\;=\;
\frac{1}{(e(k)-z)^2-|h(k)|^2}\;\sigma_2(\Ee(k)-z\,\one)^T \sigma_2
\end{align*}
Now let $\rho$ to be a smooth function compactly supported on a ball $B_{\tau}(0)$ around $k^W=0$ with $\tau>0$ chosen such that the properties of Hypothesis~\ref{hyp-WeylType} hold uniformly in $B_{\epsilon}(0)$. Then  one has
\begin{align*}
\langle n|R^z_\rho|0\rangle&
\;=\; 
\int_{\mathbb{R}^3} dk \
\frac{e^{\imath\langle k,n\rangle }}{(e(k)-z)^2-|h(k)|^2}\,\sigma_2(\Ee(k)-z\,\one)^T \sigma_2 \ \rho(k)
\\
& \;=\;
\int_{\mathbb{R}^3} dk \
\frac{e^{\imath\langle k,n\rangle }}{(e(k)+|h(k)|-z)(e(k)-|h(k)|-z)} (\rho_1(k)+ z\rho_2(k)),
\end{align*}
where 
$$
\rho_1(k)
\;=\; 
\sigma_2 \Ee(k)^T \sigma_2 \rho(k) 
\;,
\qquad
\rho_2(k)\;=\; \rho(k)\,\one_2 
\;,
$$ 
are compactly supported smooth matrix-valued functions. The focus will here be on the limit $ z \to E$ with $E\geq 0$;  the case $E\leq 0$ can be studied in a similar manner. Note that in this case $\Ee_-(k)\leq 0$ and Hypothesis~\ref{hyp-WeylType} implies that for $z$ with $\Re e(z)\geq 0$ one has 
\begin{equation}
\label{eqc}
|\Ee_-(k)-z|
\;=\;
|e(k)-|h(k)|-z|
\;\geq \;
\Re e(z)+|h(k)|- e(k) 
\;\geq\; 
(1-\kappa)|h(k)|
\;,
\end{equation}
with $\kappa<1$ as in Hypothesis~\ref{hyp-WeylType}, namely such that $|e(h^{-1}(x))-e(k^W)| \leq \kappa |x|$. Now, one can use the fact that for $\Im m(z)>0$ 
$$
\frac{1}{e(k)+|h(k)|-z} 
\;=\; \imath t  \int_0^\infty d \eta \ e^{\imath zt\eta}\,e^{-\imath \eta t( e(k)+|h(k)|)}
\;,
$$
to obtain, with $\omega =\frac{n}{|n|}$ and $t=|n|$,
\begin{align}
\langle n|R^z_\rho|0\rangle 
&
\;=\; 
\imath t \int_{\mathbb{R}^3} dk \
\frac{e^{\imath\langle k,n\rangle }}{e(k)-|h(k)|-z} (\rho_1(k)+ z\rho_2(k)) \int_0^\infty d \eta \ e^{\imath t z\eta}\,e^{- t\imath  \eta ( e(k)+|h(k)|)} 
\nonumber
\\ 
& 
\;=\;
I(z,\omega,t)
\,+\,  \imath z t \int_0^\infty d \eta \ e^{\imath z t \eta} \int_{\mathbb{R}^3} dk \ \frac{e^{\imath t\langle k,\omega\rangle- \imath t \eta (e(k) +|h(k)|) }}{e(k)-|h(k)|-z} \rho_2(k)
\;,
\label{eq-RWeylRep}
\end{align}
where
\begin{equation}
\label{eq-IzDef}
I(z,\omega,t)\;= \; \imath t \int_0^\infty d \eta \ e^{\imath z t \eta} \int_{\mathbb{R}^3} dk \ \frac{e^{\imath t\langle k,\omega\rangle- \imath t \eta (e(k) +|h(k)|) }}{e(k)-|h(k)|-z} \rho_1(k)
\;. 
\end{equation}
The integrals of the second term in \eqref{eq-RWeylRep} can be analyzed in same way as the ones of $I(z, w, t)$ and therefore only the latter is addressed. It will be proved below that  $ I(z, w, t) $ is uniformly bounded. This implies that the same holds true for the integrals of the second term in \eqref{eq-RWeylRep} so that the factor $z$ implies that the second term tends to  $0$ as $z=\imath\epsilon\to 0$.  Hence let us only focus on  $I(z,\omega,t)$. The above rewriting of the Green function is useful because the denominator can be estimated by \eqref{eqc}, while the remaining oscillatory integral can be controlled by the non-stationary and stationary phase methods. 

\vspace{.2cm}

Let us start out by illustrating this on a toy model of the Weyl Hamiltonian specified by $h(k)=k$ and $e(k)=0$. Then $\Ee_\pm(k)=\pm|k|$. One gets for $z=0+\imath \epsilon$ with $\epsilon>0$
\begin{align*}
I(\imath\epsilon,\omega,t)
&
\;=\; 
\imath t  \int_0^\infty d \eta \ e^{-\epsilon t \eta} \int_{\mathbb{R}^3} dk \ \frac{e^{\imath t\langle k,\omega\rangle-  \imath t \eta |k|}}{-|k|+ \imath\epsilon} \rho_1(k) 
\\
&
\;=\; 
\imath t \int_0^\infty d \eta \ e^{-\epsilon t \eta} \int_0^R dr \,r^2\,\frac{e^{-\imath t \eta r}}{-r+ \imath\epsilon} \int_{\SM^2}
d\theta\,e^{\imath tr\langle \theta,\omega\rangle} \,\rho_1(r\theta)
\;,
\end{align*}
where as above $t=|n|$ and $\omega=\frac{n}{|n|}\in\SM^2$. The integral over the $2$-sphere is a two-dimensional oscillatory integral with phase function $\Phi(\theta)=\langle \omega,\theta\rangle$. The two-dimensional gradient (restricted to the tangent space of $\SM^2$) is $\nabla_\theta\Phi(\theta)=\omega-\langle \omega,\theta\rangle \,\theta$, and the Hessian is $\nabla_\theta^2 \Phi(\theta)(v)=-\langle \omega,v\rangle\,\theta-\langle \omega,\theta\rangle\,v$ for $v\in T_\theta\SM^2$, namely $\langle v,\theta\rangle=0$. The critical points are the solutions of $\nabla_\theta\Phi(\theta)=0$, {\it i.e.} by $\omega=\langle \omega,\theta\rangle\,\theta$ which are given by $\omega$ and $-\omega$. This shows that $\nabla_\theta^2 \Phi(\pm\omega)(v)=\pm\,v$. Hence $\Phi$ is a Morse function and one can apply the method of stationary phase to lead to 
\begin{align*}
I(\imath\epsilon,\omega,t)
&
\;=\; 
\imath  t \int_0^\infty d \eta \, e^{-\epsilon t \eta} \int_0^R dr \,r^2\,\frac{e^{-\imath t \eta r}}{-r+ \imath\epsilon} 
\Big(
\sum_{\sigma=\pm 1}
\frac{2\pi}{tr}\,e^{\imath tr\sigma}\,e^{-\imath \frac{\pi}{4} 2\sigma}\,\rho_1(\sigma \omega r)\,h(tr)\,+\,g(tr)\Big)
\; ,
\end{align*}
where $|h(tr)|\leq C$ and $|g(tr)|\leq C\frac{1}{r}$, and even for the derivatives $|h^{(n)}(tr)|\leq C_n\frac{1}{t^n}$ and $|g^{(n)}(tr)|\leq C_n\frac{1}{rt^N}$ for $n\geq 0$ and any $N\geq 1$. This already shows that all integrals over $r$ are finite (integrable close to $0$), so that $\frac{1}{-r+ \imath\epsilon}\to\frac{-1}{r}$  in the limit $\epsilon\to 0$. However, in this limit the $\eta$-integral diverges, so that one has to use the oscillations in $e^{-\imath t \eta r}$ in the $r$-integral. These oscillations are small for $\eta$ small, so that it is convenient to decompose the $\eta$-integral in $[0,\eta_0]$, $[\eta_0,\eta_1]$ and $[\eta_1,\infty)$, leading to three contributions $I_\leq(z,\omega,t)$, $I_\parallel(z,\omega,t)$ and $I_\geq(z,\omega,t)$. The first is given by 
\begin{align*}
I_\leq(\imath 0,\omega,t)
&
\;=\; 
\sum_{\sigma=\pm 1}
2\pi\,\imath\,e^{-\imath \frac{\pi}{2} \sigma}
\int_0^{\eta_0} d \eta  \int_0^R dr \,r^2\,\frac{e^{-\imath t \eta r}}{-r}\,\frac{1}{r}\, e^{\imath tr\sigma}
\,\rho_1(\sigma\omega r)\,h(tr)\,+\,\Oo(t^{-N})
\;.
\end{align*}
Here the phase is $e^{-\imath r t (\sigma-\eta )}$, which becomes non-oscillatory for $\eta\to 1$ in the case $\sigma=1$ (for $\sigma=-1$ it remains oscillatory). For $\eta_0<1$, one can use the non-stationary phase argument, to lead to
$$
|I_\leq(\imath 0,\omega,t)|
\;\leq\;
\frac{C}{t}\,\log(|1-\eta_0|^{-1})
\;,
$$
or using non-stationary phase twice gives
$$
|I_\leq(\imath 0,\omega,t)|
\;\leq\;
\frac{C}{t^2}\,\frac{1}{|1-\eta_0|}
\;.
$$
For the third summand, let us change variables $\eta\mapsto \frac{1}{\mu}$ giving, with $\mu_1=\frac{1}{\eta_1}$,
\begin{align*}
I_\geq(\imath\epsilon,\omega,t)
&
\;=\; 
\sum_{\sigma=\pm 1}
2\pi\,\imath\,e^{-\imath \frac{\pi}{2} \sigma}
\int_0^{\mu_1} d \mu\,\frac{e^{-t\frac{\epsilon}{\mu}}}{\mu^2}  \int_0^R dr\,\frac{r}{-r+\imath\epsilon}\, e^{\imath \frac{rt}{\mu}(\sigma \mu-1)}
\,\rho_1(\sigma\omega r)\,h(tr)\,+\,\Oo(t^{-N})
\;.
\end{align*}
Using twice the non-stationary phase argument for the $r$-integral, one gets
\begin{align*}
I_\geq(\imath\epsilon,\omega,t)
&
\;=\; 
\sum_{\sigma=\pm 1}
2\pi\,\imath\,e^{-\imath \frac{\pi}{2} \sigma}
\int_0^{\mu_1} d \mu\,\frac{e^{-t\frac{\epsilon}{\mu}}}{\mu^2}  
\,\frac{\mu^2}{t^2(\sigma \mu-1)^2}\,C
\,+\,\Oo(t^{-N})
\;.
\end{align*}
This gives
$$
|I_\geq(\imath 0,\omega,t)|
\;\leq
\frac{C}{t^2}\,\frac{1}{|\mu_1-1|}
\;.
$$
Note that this is the same as the term above. For the central term, one can simply bound the $\eta$-integral by its volume $\eta_1-\eta_0$. This gives 
$$
|I_\parallel(\imath 0,\omega,t)|
\;\leq\;
C\,|\eta_1-\eta_0|
\;.
$$
Choosing $\eta_0=1-\frac{1}{t}$ and $\eta_1=1+\frac{1}{t}$, one concludes altogether that
$$
|\langle n|R^{\imath 0}_\rho|0\rangle|
\;=\;
|I(\imath 0,\omega,t)|
\;\leq\;
C\,\frac{1}{t}
\;.
$$
Note that this is a slower decay as given by the explicit computation of the whole Green function of the Weyl Hamiltonian in \cite{CGL}. Let us note the above formula might still capture the correct behavior because it only accounts for the local contribution at the origin.

\vspace{.2cm}

Let us now come back to the general case. Motivated by the above analysis, let us start by placing the tilted Weyl point in a normal form.

\begin{lemma}[Morse lemma for a Weyl point]
\label{lem-MorseWeyl}
Let 
$$
\Ee_+\,:\,  B_{\tau}(0) \subset \mathbb{R}^3\to \mathbb{R}
\;, 
\qquad 
\Ee_+(k)\,=\,|h(k)|+e(k)
\;,
$$  
with $\tau>0$ still chosen such that the properties of {\rm Hypothesis~\ref{hyp-WeylType}} hold uniformly on $B_\tau(0)$. Then there exists $R>0$ and a homeomorphism on its image $\varphi:B_R(0)\subset \mathbb{R}^3 \to B_\tau(0)$, diffeomorphism away from $0$, such that
$$
\Ee_+(\varphi(x))
\;=\; |x|
\;.
$$
The Jacobian of $\varphi$ satisfies, for $x\not=0$, the uniform bound
$$
|\det(D\varphi(x))|\;\leq\;C
\;.
$$
Moreover, there is a smooth function $s:(-R,R)\times (\mathbb{R}^3\setminus \{0\}) \to \mathbb{R}$ satisfying
$$
\varphi(x) \;=\; h^{-1} (s (|x|, x) \hat x )
\;,  
\qquad  
\text{ where } 
\;\hat{x}\,=\,\tfrac{x}{|x|}
\;,
$$
Moreover, the function $r\mapsto s(r,x)$ is strictly increasing and is radially simetric, namely $s(r,x)=s(r,\hat{x})$ with $s(0,x)=0$. For every multi-index $\alpha$, there is a constant $C_\alpha$ such that 
\begin{align}\label{PH11}
|D^{\alpha}_x s(r, x)|
\;\leq\; 
C_\alpha |r|
\;,  
\end{align} 
for all  $ x \in \mathbb{S}^2  $ and $ r \in [-R, R] $.
\end{lemma}

\noindent {\bf Proof.} The function $h|_{B_{\tau}(0)}$ is by Hypothesis~\ref{hyp-WeylType} a diffeomorphism. Consider its inverse and notice that 
$$
\Ee_+(h^{-1}(x))=e(h^{-1}(x))+|x|
\;, 
\qquad 
x \in h(B_{\tau}(0))
\;.
$$ 
Let us show the existence of the function $s$. Choose $\zeta>0$ such that $\overline{B_{\zeta}(0 )}\subset h(B_{\tau}(0))$. Consider the function $f: [-\zeta,\zeta]\times (\mathbb{R}^3 \setminus\{0\}) \to \mathbb{R}$ defined by $ f(t,x)=e\circ h^{-1}(t\hat{x}) + t,$ and notice that for each $x\in \mathbb{R}^3\setminus \{0\}$ its derivative with respect $t$ is 
$$
\partial_t f(t,x)
\;=\;
\langle \nabla (e\circ h^{-1})(t\hat x),\hat{x}\rangle\,+\,1 
\;>\;-\kappa+1\;>\; 0\;.
$$
Hence for all $x\in \mathbb{R}^3\setminus \{0\},$ the function $[-\zeta,\zeta]\ni t \mapsto f(t,x)$ is invertible (increasing). Let us denote by $s(\cdot,x):[R_{x,-},R_{x,+}] \to \mathbb{R}$ its inverse, with $R_{x,\pm} = f(\pm \zeta,x)=f(\pm \zeta,\hat{x})$. Define the  global function $s:[-R,R] \times (\mathbb{R}^3\setminus\{0\}) \to \mathbb{R},$ with $R=\min_{x\in \mathbb{S}^2} |R_{x,\pm}|>0$. It satisfies by definition $f(s(r,x),r)=r$. Furthermore $r\in[-R,R] \mapsto  s(r,x)$ is by definition increasing  and since $f(0,x)=e(h^{-1}(0))=e(k^W)=0$, one also has $s(0,x)=0$.  Note that, in fact, $s$ is radially symmetric. Indeed, if $\hat x = \hat y$, then 
$$
f(s(r,y),x)
\,=\,
e\circ h^{-1}(s(r,y)\hat{x})+ s(r,y) 
\,=\, 
e\circ h^{-1}(s(r,y)\hat{y})+ s(r,y)
\,=\,
 f(s(r,y),y)
 \,=\,
 r
 \,,
$$  
since $f(\cdot,x)$ is injective one obtains $s(r,x)=s(r,y)$. For all $r\geq 0$, $x\in \mathbb{R}^3\setminus \{0\}$ one has
\begin{equation}
\Ee_+(h^{-1}(s(r,x)\hat{x}))
\;=\;
e\circ h^{-1}(s(r,x)\hat{x})+s(r,x)
\;=\;
 f(s(r,x),x)
 \;=\;
 r
 \;.
\end{equation}
Therefore the function $\varphi(x)= h^{-1}(s(|x|, x)\hat{x})$ satisfies the desired properties. The function $s:(-R,R)\times (\mathbb{R}^3\setminus \{0\})\to \mathbb{R}$ is smooth. Indeed, the function $H: (-\zeta,\zeta)\times (-R,R)\times (\mathbb{R}^n\setminus\{0\}) \to \mathbb{R}$,  $H(t,r,x)=f(t,x)-r$ is smooth and satisfies $\partial_t H(t,r,x)>0$;  hence the implicit function theorem implies that it has a local smooth solution $\tilde{s}:(r_0-\epsilon,r_0+\epsilon)\times U_{x_0} \to \mathbb{R}, \ H(\tilde{s}(r,x),r,x)=0$, namely $f(\tilde{s}(r,x),x)=r$ and, since $f(\cdot,x)$ is injective,  $s|=\tilde{s}$.  Moreover, since $  s(0, x)  =0$, one has $D^{\alpha}_x s(0, x)=0   $ (for all multi- indexes $\alpha$). By the mean value theorem, 
\begin{align}
\label{PH11}
|D^{\alpha}_x s(r, x)|
\;\leq\; C_\alpha |r|,  
\end{align} 
for some constant $C_{\alpha}$,  and all $ x \in \mathbb{S}^2 $ and $ r \in (-R, R) $. Finally let us argue that all first order derivatives  of  $\varphi(x) = h^{-1} (s (|x|, x) \hat x )$ are bounded. By the chain rule, it is enough to prove that the derivatives of $s (|x|, x) \hat x$ are bounded. This is a consequence of \eqref{PH11} and the fact that the derivatives of the functions $ x \to |x|$ and $ x \to \hat x $ are bounded and grow as $\frac{1}{|x|}$ (for small $|x|$) respectively. This growth is tamed using \eqref{PH11}.      
\hfill $\Box$

\vspace{.2cm}

Let us now come back to the analysis of $I(z,\omega,t)$ defined in \eqref{eq-IzDef},  for the general case of a $k\in B_{\tau}(0)$ with $\epsilon>0$ such that Hypothesis~\ref{hyp-WeylType} holds.  \color{black} Let us split the $\eta$-integral into $[0,\eta_0]$, $[\eta_0,\eta_1]$ and $[\eta_1,\infty)$, leading to three contributions  $I_\leq$, $I_\parallel$ and $I_\geq$. Further let $\chi : \RM \to [0,1]$ be a smooth decreasing function supported on $[0,1]$ and equal to $1$ on $[0,\frac{1}{2}]$, set $\chi_\delta(r)=\chi(\frac{1}{\delta} r)$ and finally  $\chi_\delta^c=1-\chi_\delta$. 
\begin{align}
\label{eq-Rleq}
&
I_\leq(z,\omega,t) 
\;=\;
\imath t \int_0^{\eta_0} d \eta   \  e^{\imath z t \eta}   
\Big(
J_\leq^c(z,\eta,\omega,t,\delta)\,+\,J_\leq(z,\eta,\omega,t,\delta)
\Big)
\;,
\end{align}
where, with a free parameter $\delta$ to be chosen later,
\begin{align*}
&
J_\leq(z,\eta,\omega,t,\delta)
\;=\;
\int_{\mathbb{R}^3} dk \ \frac{e^{\imath t\langle k,\omega\rangle- \imath t \eta (e(k) +|h(k)|) }}{e(k)-|h(k)|-z} \rho_1(k) \chi_\delta(|k|)
\;,
\\
&
J_\leq^c(z,\eta,\omega,t,\delta)
\;=\;
\int_{\mathbb{R}^3} dk \ \frac{e^{\imath t\langle k,\omega\rangle- \imath t \eta (e(k) +|h(k)|) }}{e(k)-|h(k)|-z} \rho_1(k)
\chi_\delta^c(|k|)
\;.
\end{align*}
%

\begin{proposi}
\label{prop-Ileq2}
For all $E\geq 0$, the limit 
$$ 
I_\leq(E+\imath 0,\omega, t) 
\;=\; 
\lim_{\epsilon\downarrow 0} I_\leq(E+\imath \epsilon,\omega,t) 
$$ 
exists. Moreover, one has 
$$
\big| I_\leq(E+\imath 0,\omega,t) \big|
\;\leq\;
C |t|^{-1}\log(t)
\;, 
\qquad t\geq 1
\;.
$$ 
Further, if $0<\beta \leq 1$  for all $z,z' \in \overline{\HM}_+=\{z\in \mathbb{C}: \Im m(z)\geq 0, \Re e(z)\geq 0\}$ one has
$$
|I_\leq(z,\omega,t)-I_\leq(z',\omega,t)| 
\;\leq \;
C_\beta |z-z'|^\beta |t|^{-1+\beta} \log(t) 
\;, 
\qquad 
t\geq 1
\;.
$$
\end{proposi}

\noindent {\bf Proof.} 
The first part is a consequence of Lebesgue dominated theorem, because  \eqref{eqc} and \eqref{eq-hCond} imply $|\Ee_-(k)-z|=|e(k)-|h(k)|-z|\geq (1-\kappa)|h(k)| \geq \frac{1-\kappa}{\gamma}|k| $.  By the latter estimate, one also deduces
\begin{equation}
\label{eq-BoundSmall-k}
|J_\leq(E,\eta,\omega,t,\delta)|
\;\leq\;
C\,\int^{\delta}_0 dr\,r^2\,\frac{\gamma}{(1-\kappa)r}\;\leq\;C\delta^2
\;.
\end{equation}
For a bound on $J^c_\leq(E,\eta,\omega,t,\delta)$ the method of non-stationary phase will be applied. Let us introduce the phase function and its gradient by 
$$
\Phi(k)
\;=\;
\langle k,\omega\rangle- \eta \Ee_+(k)
\;,
\qquad
\nabla \Phi(k)
\;=\;
\omega\,-\,
\eta \nabla \Ee_+(k) 
\;.
$$
Using $\Phi$, one has
$$
J^c_\leq(E,\eta,\omega,t,\delta)
\;=\;
\int_{\mathbb{R}^3} dk  \,
\,e^{\imath t\Phi(k)}\,\hat{\rho}_\delta(k)
\;,
\qquad
\hat{\rho}_\delta(k)
\;=\;
\frac{1}{\Ee_-(k)-E}  \, \rho_1(k)\,\chi^c_\delta(|k|) 
\;.
$$
As  $\nabla \Ee_+(k) $ is uniformly bounded, one  obtains that $|\nabla \Phi(k)|\geq |\omega|-\eta_0 C=1-\eta_0 C$ so that the phase is non-stationary for $\eta_0$ sufficiently small. Now let us  introduce 
$$
(\Ll g)(k)
\;= \;
-\sum_{j=1}^{d} \partial_{ k_j}\left( \frac{1}{|\nabla \Phi|^2 } \big(  \partial_{k_j} \Phi\big) g \right)(k)
\;,
\qquad
g\in C^\infty (B_{\tau}(0))
\;.
$$
The formal adjoint satisfies $\Ll^*e^{\imath t\Phi(k)}=\imath t e^{\imath t\Phi(k)}$. Therefore, by partial integration, 
\begin{align} 
J^c_\leq(E,\eta,\omega,t,\delta)
\;=\;  
\frac{1}{\imath t}
\int_{\mathbb{R}^3} dk  \,
\,e^{\imath t\Phi(k)}\,(\Ll \hat{\rho}_\delta)(k) 
\;= \;
\,\frac{1}{ (\imath t)^2}
\int_{\mathbb{R}^3} dk  \,
\,e^{\imath t\Phi(k)}\,(\Ll^2\hat{\rho}_\delta)(k)
\;.     
\label{eq-StatPhaseAgain}
\end{align}
Let us first use merely the first identity. Hence one needs to bound $\Ll \hat{\rho}_\delta$. As $\nabla \Phi$ is uniformly bounded from below, the largest contributions come for the derivatives of $ (\Ee_-(k)-E)^{-1}$. As $|\Ee_-(k)-E|\geq \frac{1-\kappa}{\gamma}|k|$, one obtains from the product rule
$$
|\Ll \hat{\rho}_\delta(k)|
\;\leq\;
C\,\chi^c_\delta(|k|)\,\big(|k|^{-2}\,+\,|k|^{-1}\delta^{-1}\chi_{2\delta}(|k|)\big)
\;,
$$
where it was used that $|\nabla \chi^c_\delta(|k|)|\leq C\delta^{-1}\chi_{2\delta}(|k|)$ which follows by computing the derivative of $k\mapsto \chi^c_\delta(|k|)$. Replacing and using that one only integrates over a compact set due to the support of $\rho_1$ shows
$$
|J^c_\leq(E,\eta,\omega,t,\delta)|
\;\leq\;
\frac{C}{t}
\;,
$$
which then gives $|I_\leq (E+\imath 0, \omega,t)|\leq C$ and hence no decay in $t$. In order to improve the bound, let us use the second identity in \eqref{eq-StatPhaseAgain} so that one needs to bound  $\mathcal{L}^2\hat{\rho}_\delta$. Using similar arguments as above leads to
$$
|\Ll^2\hat{\rho}_\delta(k)|
\;\leq\;
C\,\chi^c_\delta(|k|)\,\big(|k|^{-3}\,+\,|k|^{-2}\delta^{-1}\chi_{2\delta}(|k|)\,+\,|k|^{-1}\delta^{-2}\chi_{2\delta}(|k|)\big)
\;,
$$
Replacing into \eqref{eq-StatPhaseAgain} gives
\begin{align}
|J^c_\leq(E,\eta,\omega,t,\delta)|
\;\leq\;
\frac{C}{ t^2}
\Big(
|\log(\delta)|\,+\,\delta^{-1}\delta\,+\,\delta^{-2}\delta^2
\Big)
\;,
\label{eq-JcBound}
\end{align}
which when combined with the estimate on $|J_\leq(E,\eta,\omega,t,\delta)|$ in \eqref{eq-Rleq} gives
$$
\big|I_\leq (E+\imath 0, \omega, t)\big|
\;\leq\;
C\big(t\delta^2\,+\,t^{-1}|\log(\delta)|\big)
\;\leq\;
C t^{-1}\log(t)
\;.
$$
Choosing $\delta=t^{-1}$, one obtains the desired estimate. It remains to verify the H\"older continuity. Due to \eqref{eq-Rleq}, it is enough to obtain bounds on $e^{\imath tz\eta}-e^{\imath tz'\eta}$, $J_\leq(z,\eta,\omega,t,\delta)-J_\leq(z',\eta,\omega,t,\delta)$ and  $J^c_\leq(z,\eta,\omega,t,\delta)-J^c_\leq(z',\eta,\omega,t,\delta)$. As to the first term, 
\begin{equation}
\label{eq-ExpHoelder}
|e^{\imath tz\eta}-e^{\imath tz'\eta}|
\;\leq\; 
C(t\eta)^{\beta}|z-z'|^\beta 
\;, 
\qquad 
0<\beta\leq 1
\;.
\end{equation}
This adds a $t^\beta$ factor to the estimate. The other two terms are estimated as above using again that $|\Ee_-(k)-E|\geq \frac{1-\kappa}{\gamma}|k|$, but considering the following estimate for $z,z'\in \overline{\HM}_+$
\begin{align}
\left| \frac{1}{(\Ee_-(k)-z)^N}-\frac{1}{(\Ee_-(k)-z')^N} \right|  
&
\;=\;
\left| \frac{1}{(\Ee_-(k)-z)^N}-\frac{1}{(\Ee_-(k)-z')^N} \right|^{\beta+(1-\beta)}
\nonumber
\\
&
\;\leq\; 
\tfrac{\gamma}{1-\kappa}\, |z-z'|^\beta \frac{1}{|k|^{(N+1)\beta}} \frac{1}{|k|^{N(1-\beta)}}
\nonumber
\\
&
\;=\;  
\tfrac{\gamma}{1-\kappa}\, |z-z'|^\beta \frac{1}{|k|^{\beta+N}}  
\;,
\label{eq-es-ho}
\end{align}
where the first and second factors are estimated using the mean value theorem and the triangle inequality respectively.  Then, using a volume estimate as in  \eqref{eq-BoundSmall-k} togehter with  \eqref{eq-es-ho} one obtains for $\beta\leq 1$ and $N = 1$
\begin{equation}
\label{eqeshol} 
|J_\leq(z,\eta,\omega,t,\delta)-J_\leq(z',\eta,\omega,t,\delta)| 
\;\leq\; 
C \delta^{2-\beta} |z-z'|^\beta
\;.
\end{equation}
For the last term one applies non-stationary phase method as before and using the amplitude function 
$$
\hat{\rho}_\delta(k)
\;=\;
\left(\frac{1}{\Ee_-(k)-z}\,-\,\frac{1}{\Ee_-(k)-z'}\right)  \, \rho_1(k)\,\chi^c_\delta(|k|) 
\;.
$$  
Similar arguments as above and  \eqref{eq-es-ho} lead to
$$
|\Ll^2\hat{\rho}_\delta(k)|
\;\leq\;
C|z-z'|^\beta\,\big(|k|^{-3-\beta}\,+\,|k|^{-2-\beta}\delta^{-1}\chi_{2\delta}(|k|)\,+\,|k|^{-1-\beta}\delta^{-2}\chi_{2\delta}(|k|)\big)
\;.
$$
Then one obtains
\begin{align}
|J^c_\leq(z,\eta,\omega,t,\delta)\,-\, J^c_\leq(z',\eta,\omega,t,\delta)|
\;\leq\;
\frac{C_\beta}{ t^2}|z-z|^\beta
\Big(
\delta^{-\beta}\,+\,\delta^{-1}\delta^{1-\beta}\,+\,\delta^{-2}\delta^{2-\beta}
\Big)
\;.
\label{eq-JcBoundhol}
\end{align}
Choosing  $\delta = t^{-1}$ this implies the desired H\"older estimate.
\hfill $\Box$
 
\vspace{.2cm}

For the summand $I_\geq$ let us change variables $\eta\mapsto \mu=\eta^{-1}$ and set $\mu_1=(\eta_1)^{-1}$. Then
\begin{align}
\label{eq-Rgeq}
&I_\geq(z,\omega, t ) 
\;=\;
\imath t
\int^{\mu_1}_0 d \mu\,\frac{1}{\mu^2}   \,  e^{\imath z t \frac{1}{\mu}}   
\Big(
J_\geq^c(z,\mu,\omega,t,\delta)+J_\geq(z,\mu,\omega,t,\delta)
\Big)
\;,
\end{align}
where, again for some $\delta$ to be chosen later,
\begin{align*}
&
J_\geq(z,\mu,\omega,t,\delta)
\;=\;
\int_{\mathbb{R}^3} dk  \,
\frac{e^{\imath \frac{t}{\mu}(\mu\,\langle k,\omega\rangle-  \Ee_+(k))}}{\Ee_-(k)-z}  \, \ \rho_1(k) \,\chi_{\delta}(|k|)
\;,
\\
&
J^c_\geq(z,\mu,\omega,t,\delta)
\;=\;
\int_{\mathbb{R}^3} dk  \,
\frac{e^{\imath \frac{t}{\mu}(\mu\,\langle k,\omega\rangle-  \Ee_+(k))}}{\Ee_-(k)-z}  \, \ \rho_1(k)  \,\chi^c_\delta(|k|)
\;.
\end{align*}

\begin{proposi}
\label{prop-Igeq}
For all $E\geq 0$, the limit 
$$ 
I_\geq(E+\imath 0,\omega, t) 
\;=\; 
\lim_{\epsilon\downarrow 0} I_\geq(E+\imath \epsilon,\omega,t) 
$$ 
exists. Moreover,  one has 
$$
\big| I_\geq(E+\imath 0,\omega,t) \big|
\;\leq\;
C |t|^{-1}\log(t)\;, \qquad t\geq 1 
\;.
$$ 
Further, if $0<\beta< 1$  for all $z,z' \in \overline{\HM}_+$, 
$$
|I_\geq(z,\omega,t)\,-\,I_\geq(z',\omega,t)| 
\;\leq \;
C_\beta |z-z'|^\beta |t|^{-1+\beta}\log(t)
\;, 
\qquad 
t\geq 1
\;.
$$
\end{proposi}

\noindent {\bf Proof.} 
Note that in \eqref{eq-Rgeq} the limit $\epsilon\downarrow 0$ cannot be taken directly because the factor $e^{\imath z t \frac{1}{\mu}}$ for $\epsilon >0$ compensates the singularity $\frac{1}{\mu^2}$, assuring that the $\mu$-integral is finite. All estimates below will be shown to hold uniformly in $\epsilon>0$.  The first integral $J_\geq(z,\mu,\omega,t,\delta)$ can be bounded by a volume estimate exactly as in \eqref{eq-BoundSmall-k}, leading to $|J_\geq(z,\mu,\omega,t,\delta)|\leq C\delta^2$. In the following, we will choose $\delta=\frac{\mu}{t}$ so that replacing in \eqref{eq-Rgeq} will lead to a finite $\mu$-integral and a contribution bounded by $t\frac{1}{t^2}=\frac{1}{t}$. As to $J_\geq^c(\mu,\omega,t,\delta,\epsilon)$, the non-stationary phase analysis will be carried out with the phase function
$$
\Phi(k)
\;=\;
\mu\,\langle k,\omega\rangle-  \Ee_+(k)
\;,
\qquad
\nabla \Phi(k)
\;=\;
\mu\,\omega\,-\,
\nabla \Ee_+(k)
\;.
$$
Using Hypothesis~\ref{hyp-WeylType} and that  $  \Ee_\pm(h^{-1}(x))=e(h^{-1}(x))\pm |x| $, one obtains that $  | \nabla  \Ee_\pm(h^{-1}(x))|   $ is uniformly bounded from below. This implies that, for $\mu_1$ sufficiently small,  $|\nabla \Phi(k)|\geq C$ uniformly for $k\in \supp(\rho_1 \cdot \chi_\delta(|\cdot|))$ and $\mu\in [0,\mu_1]$.  Now one can apply the non-stationary phase argument to $J_\geq^c(z,\mu,\omega,t,\delta)$ twice just as above, leading to exactly the same bound as \eqref{eq-JcBound} with $t$ replaced by $\frac{t}{\mu}$. Replacing this and the above in \eqref{eq-Rgeq} leads to
$$
\big| I_\geq(z,\omega,t) \big|
\;\leq\;
t
\int^{\mu_1}_0 d \mu\,\frac{1}{\mu^2}   \,     
\Big(
C\delta^2
\,+\,
C\frac{\mu^2}{ t^2}
\big(
|\log(\delta)|\,+\,\delta^{-1}\delta\,+\,\delta^{-2}\delta^2
\big)
\Big)
\;\leq\;
\frac{C}{t}\log(t)
\;,
$$
where the last bound follows by choosing $\delta=\frac{\mu}{t}$. This bound is uniform in $\Im m(z)\geq 0$ and $ \Re e(z)\geq 0$. For the H\"older continuity one also proceeds as before. One takes the difference $I(z,t,\omega)-I(z',t,\omega)$ using  \eqref{eq-Rgeq} and one uses the estimates, still with $\delta=\frac{\mu}{t}$,  
$$
\mu^{-2}\,|e^{\imath z\frac{t}{\mu}}- e^{\imath z'\frac{t}{\mu}}|\,|J_\geq(z,\mu,\omega,t,\delta)| 
\;\leq\; 
C\,|z-z'|^\beta \,\mu^{-\beta}\, t^{-2+\beta}
\;,
$$ 
and 
$$
\mu^{-2}\,|e^{\imath z\frac{t}{\mu}}- e^{\imath z'\frac{t}{\mu}}|
\,
|J^c_\geq(z,\mu,\omega,t,\delta)| 
\;\leq\; 
C\,|z-z'|^\beta \,\mu^{-\beta} \,|\log(\mu)|\, t^{-2+\beta}\,\log(t)
\;,
$$ 
and both terms are integrable on $[0,\mu_1]$ provided that $\beta<1$. Using volume estimates and non-stationary phase method, just as in \eqref{eqeshol} and \eqref{eq-JcBoundhol}, together with  \eqref{eq-es-ho} one obtains 
\begin{align*}
&
\mu^{-2}|J_\geq(z,\mu,\omega,t,\delta) - J_\geq(z',\mu,\omega,t,\delta)|
\;\leq \;
C\, |z-z'|^\beta \,\mu^{-\beta}\, t^{2-\beta}
\;,
\\
&
\mu^{-2}|J^c_\geq(z,\mu,\omega,t,\delta) - J^c_\geq(z',\mu,\omega,t,\delta)|
\;\leq\; C \,|z-z'|^\beta \,\mu^{-\beta} t^{2-\beta}
\;.
\end{align*}
Both terms are integrable on $[0,\mu_1]$ provided that $\beta<1$.  
\hfill $\Box$ 

\vspace{.2cm}

Finally let us turn to the summand $I_\parallel(z,\omega,t)$.  Instead of splitting $I(z,\omega,t)$ defined in \eqref{eq-IzDef} using three indicator functions in $\eta$ on $[0,\eta_0]$, $[\eta_0,\eta_1]$ and $[\eta_1,\infty)$, let us rather use a smooth partition of unity so that $I_\parallel(z,\omega,t)$ rather contains a smooth function $\chi \in C_0^\infty ( (\eta_0, \eta_1))$ instead of a the indicator functions. The above arguments on $I_\leq(z,\omega,t)$ and $I_\geq(z,\omega,t)$ readily transpose to this case. As to $I_\parallel(E
+\imath\epsilon,\omega,t)$, the limit $\epsilon\downarrow 0$ can now be taken, leading to 
$$
I_\parallel(E+\imath 0,\omega,t)
\;=\;
\imath t
\int^{\eta_1}_{\eta_0} d \eta \,  e^{\imath E t \eta}   
\int_{\mathbb{R}^3} dk  \,
\frac{e^{\imath t(\langle k,\omega\rangle-  \eta \Ee_+(k))}}{\Ee_-(k)-E}  \rho_1(k)\,\chi(\eta)
\;.
$$
Next let us apply the change of variables $k=\varphi(x)$ as given in Lemma~\ref{lem-MorseWeyl} and note that 
$$
\Ee_-(\varphi(x)) 
\;=\; 
\Ee_+ (\varphi(x)) - 2 |h ( \varphi(x) )|
\;=\;
|x|-2s(|x|, \hat x)
\;.
$$
Then
$$
I_\parallel(E+\imath 0,\omega,t)
\;=\;
\imath t
\int^{\eta_1}_{\eta_0} d \eta \,  e^{\imath E t \eta}   
\int_{\mathbb{R}^3} dx  \,|\det(D\varphi(x))|\,
\frac{e^{\imath t(\langle \varphi(x),\omega\rangle-  \eta |x|)}}{|x|-2  \,s(|x|, \hat x)-E}  \rho_1(\varphi(x))\,\chi(\eta)
\;.
$$
Passing to polar coordinates $x=r\theta$ with $r\geq0$ and $\theta\in\SM^2$ and introducing the smooth function $\tilde{\rho}(r,\theta)=|\det(D\varphi(r\theta))|\rho_1(\varphi(r\theta))$, one  obtains 
\begin{align}
I_\parallel(E+\imath 0,\omega,t)
&
\;=\;\imath t\int_{\eta_0}^{\eta_1} \ d\eta \,  e^{\imath E t \eta}   
\int_0^R  dr\,r^2  \, \int_{\mathbb{S}^2}d\theta\;
\frac{e^{\imath t(\langle \varphi(r\theta),\omega\rangle-  \eta r)}}{r-2 s(r,\theta)-E} \, \tilde \rho(r,\theta)\,\chi(\eta)
\nonumber
\\
&
\;=\;
\imath t
\int_0^R  dr\,r^2 \int_{\eta_0}^{\eta_1} d\eta \,  e^{\imath E t \eta}    \, \int_{\mathbb{S}^2}d\theta\;
\frac{e^{\imath t(\langle \varphi(r\theta),\omega\rangle-  \eta r)}}{r-2 s(r,\theta)-E}\, \tilde \rho(r,\theta)\,\chi(\eta)
\nonumber
\\
&
\;=\;
\imath t
\int_0^R  dr\,r^2 \int_{\mathbb{S}^2}d\theta\;\frac{e^{\imath t \langle \varphi(r\theta),\omega\rangle} }{r-2 s(r, \theta)-E} \, \tilde \rho(r,\theta)
\int_{\eta_0}^{\eta_1}  d\eta \,   
e^{\imath t \eta (E-r)}\,\chi(\eta)
\;.
\label{eq-I=}
\end{align}
An important point is that the $\theta$-integral and the $\eta$-integral are {\it not} coupled and can be dealt with separately. The $\eta$-integral can be computed by partial integrations because $\chi$ does not lead to any boundary contributions as long as $\Re e(z)\not=r$, $z\in \overline{\mathbb{H}}^+$
\begin{equation}
\label{eqcot1}
\left|\int_{\eta_0}^{\eta_1}  d\eta \,   
e^{\imath t \eta (z-r)}\,\chi(\eta)\right|
\;\leq \;
\frac{C_N}{t^N|\Re e(z)-r|^N}
\;,
\end{equation}
for $N$ to be chosen later. Next for $z\in \overline{\mathbb{H}}^+$, $r\geq 0$ and $\omega \in \mathbb{S}^{d-1}$ set
$$
J(z,t,r,\omega)
\;=\;
\int_{\mathbb{S}^2}d\theta\;e^{\imath tr  \langle \frac{1}{r} \varphi(r\theta),\omega\rangle}\,
\frac{\tilde \rho(r,\theta)}{r-2 s(r, \theta)-z} 
\;.
$$

\begin{proposi}
\label{prop-Jestimate}
For any $\omega_0\in \mathbb{S}^{d-1}$  there exists $r_{\omega_0}>0$ and $W_{\omega_0}\subset\mathbb{S}^{d-1}$ with $\omega_0\in W_{\omega_0}$, such that for all $(r,\omega) \in [0,r_{\omega_0})\times W_{\omega_0}$ and $z\in \overline{\mathbb{H}}^+$ one has 
\begin{equation}\label{equa-cota-jota}
|J(z,t,r,\omega)|
\;\leq \;
C_{\omega_0} \,\frac{1}{r^2t}
\;.
\end{equation}
Moreover, for all $z,z' \in \overline{\mathbb{H}}^+$ and $0\leq \beta<1$ one has 
\begin{equation}
\label{eqhoco}
|J(z,t,r,\omega)-J(z',t,r,\omega)| 
\;\leq\; 
C_{\omega_0,\beta} \,\frac{1}{r^{2+\beta} t}\, |z-z'|^\beta
\;.
\end{equation} 
\end{proposi}

\noindent {\bf Proof.} Due to the definition of $J(z,t ,r,\omega)$,  let us introduce the phase function 
$$
\Phi_{\omega,r}(\theta)
\;=\;
\Phi(\theta,\omega,r)
\;=\; 
\langle \tfrac{1}{r}  \varphi(r\theta),\omega\rangle
\;.
$$
Note that the function $r\mapsto \frac{1}{r} \varphi(r\theta)=\frac{1}{r}h^{-1}(s(r,\theta)\theta)$ is indeed smooth because $s(0,\theta)=0$, $h^{-1}(0)=0$ and $h$ is a local diffeomorophism. Moreover, one has 
\begin{equation}
\label{equ.fun.fas}
\Phi_{\omega,0}(\theta)
\;=\;
\lim_{r\to 0}\Phi_{\omega,r}(\theta)
\;=\; 
\partial_r s(0,\theta)\,\langle  \theta, D h^{-1}(0)^T \omega \rangle
\;,
\end{equation}
where $D h^{-1}(0)$ denotes the Jacobian of $h^{-1}$ at $0$. Let us recall from the proof of Lemma~\ref{lem-MorseWeyl} that $s(r,\theta)$ is the solution to the equation 
$$
e\circ h^{-1}(s(r,\theta)\theta )+ s(r,\theta)
\;=\;r
\;.
$$ 
Taking derivative with respect $r$ shows
$$ 
\langle \nabla (e\circ h^{-1})( s(r,\theta)\theta), \partial_r s(r,\theta) \theta \rangle + \partial
_r s(r,\theta)
\;=\;
1
\;.
$$ 
Then evaluating the above equation at $r=0$ and recalling $  s(0, \theta) = 0$, one concludes that
$$
\partial
_r s(0, \theta)\big(\langle \nabla (e\circ h^{-1})( 0), \theta\rangle + 1 \big)
\;=\;
1
\;,
$$ 
which implies the following expression for $\partial_rs(0,\theta)$, 
\begin{equation}
\label{equa-fors-salv}
\partial_r s(0, \theta)
\;=\;
\frac{1}{  \langle \nabla (e\circ h^{-1})( 0), \theta\rangle + 1 \ }
\;.
\end{equation} 
The two cases $\nabla e(0)=0$ or $\nabla e(0)\neq 0$ have to be distinguished. If $\nabla e(0)=0$, then \eqref{equ.fun.fas} and \eqref{equa-fors-salv} imply that
$$
\Phi_{\omega,0}(\theta)
\;=\; 
\langle \theta, D h^{-1}(0)^T \omega \rangle
\;.
$$ 
This is clearly a Morse function on the sphere, and therefore one can apply the stationary phase method. Let $ \Cc(\Phi_{\omega_0,0})=\{\theta_1,\theta_2\}$   be the  critical points of the function $\Phi_{\omega_0,0}$ which by definition satisfy $\nabla_\theta\Phi_{\omega_0,0}(\theta_j)=0$ where $\nabla_\theta$ denotes the gradient on $\mathbb{S}^2$. Then $\theta_1 = \frac{1}{| D h^{-1}(0)^T \omega  |}D h^{-1}(0)^T \omega$ and $\theta_2 = - \theta_1 $. Let us first estimate the integral $J(z, t, r,\omega_0)$ restricted to an open neighborhood $V_j$ of a critical point $\theta_j$. Let $\theta_j \in \Cc(\Phi_{\omega_0,0})$ and $\psi_j:U_j \subset \mathbb{R}^2 \to V_j \subset \mathbb{S}^2$ a local chart near $\theta_j$ with $\psi_j(0)=\theta_j$.  By definition, the function 
$$
F: U_{\omega_0, j}\times \mathbb{S}^2 \times [0,R)\,\to\, \mathbb{R}^2
\;, 
\qquad 
F(x,\omega,r) \,=\, \nabla(\Phi_{\omega,r}\circ    \psi_j)(x)
\;,
$$ 
is smooth and such that $F(0,\omega_0,0)=0$, and it has non-degenerated derivative w.r.t. $x$. Therefore, the implicit function theorem implies that there exists $W_{\omega_0,j}\subset \mathbb{S}^2$, $r_{\omega_0,j}>0$ and a smooth function $\Theta_j: W_{\omega_0,j}\times [0,r_{\omega_0,j}) \to U_{\omega_0, j}$ such that $ \nabla (\Phi_{\omega,r}\circ\psi_j)(\Theta_j(\omega,r))=0$. Moreover, the parametrized Morse lemma (see Lemma~\ref{prop-mors-theo} in Appendix~\ref{app-OsciInt}) implies that there exists a smooth function $\xi_j: U_{\omega_0,j} \times W_{\omega_0,j}\times [0,r_{\omega_0,j})\to  B_a(0)\subset \mathbb{R}^2$ with the map  $x\in U_{\omega_0,j}\mapsto  \xi_{j, \omega, r}(x) =  \xi_j(x, \omega,r)$ being diffeomorphism on its image, such that $\xi_j(\Theta(\omega,r),\omega,r)=0$ and 
$$
\Phi_{\omega,r}(\psi_j(x))
\;=\;
\tfrac{1}{2}\langle \Hess (\Phi_{\omega, r}\circ \psi_j)(\Theta_j(\omega,r)) \xi_j(x,\omega,r),\xi_j(x,\omega,r) \rangle + \Phi_{\omega,r}\circ \psi_j(\Theta_j(\omega,r))
\;.
$$ 
Then, for a compactly supported function $\rho_j$ whose support is contained in   $V_{\omega_0,j} = \psi_j(U_{\omega_0,j})$,  using the change of variable and Parserval's identity (see Proposition \ref{prop-case-crit}), one obtains for all $(\omega,r)\in W_{\theta_0,j}\times [0,r_{\omega_0,j})$ that
$$
\left|\int_{\mathbb{S}^2}d\theta\;e^{\imath tr  \Phi_{\omega,r}(\theta)}\,
\frac{\tilde \rho(r,\theta)}{r-2s(r,\theta)-z} 
\; \rho_j(\theta) \right| 
\;\leq \;
\frac{C_j}{tr} \sup_{k\in B_a(0),\,|\alpha|\leq 3}  \left| D^\alpha_k \frac{\hat\rho(k,r,\omega)}{r-2s(r,\phi_{\omega, r} (k))-z} \right|
\;,
$$
where $\hat \rho$ is a smooth function such that $\supp (\hat \rho(\cdot, r,\omega)) \subset \xi_{j, \omega, r}(U_{\omega_0,j})$ and $\phi_{r,\omega}= \psi_j \circ \xi_{j,\omega,r}^{-1}$.  Now recall from Lemma \ref{lem-MorseWeyl} that there is a constant $  C_{\alpha} $  such that    $|D^{\alpha}_x{ s (r,  x) }| \leq r C_{\alpha} $ uniformly for $x \in \mathbb{S}^2$ and $r \in (-R, R)$.  This, in particular, implies $|D^\alpha_k s(r,\phi_{r,\omega})| \leq r C_\alpha$, for some constant $  C_{\alpha} $. \color{black} Furthermore, for $\Re e (z) \geq 0$ and due to $s(r, \theta ) \geq 0$,
$$
|r-2s(r,\theta)-z|
\;\geq \;
\Re e(z)+2s(r,\theta)-r
\;\geq\; 
\frac{1-\kappa}{1+\kappa} r
\;,
$$ 
where last inequality follows from $\frac{r}{1-\kappa}\geq s(r,\theta)\geq \frac{r}{1+\kappa}$ which in turn is a consequence of $e(h^{-1}(s(r,\theta))+s(r,\theta)=r$ and $|e(h^{-1}(x))|\leq \kappa |x|$. Hence one concludes
\begin{align}\label{PH5}
\left|D^\alpha_k \frac{1}{r-2s(r,\phi_{r,\omega}(k))-z}\right| 
\;\leq\; 
C_\alpha \frac{1}{r}
\;,
\end{align}
which combined with the above leads to
\begin{equation}
\label{equcotuno}
\left|\int_{\mathbb{S}^2}d\theta\;e^{\imath tr  \Phi_{\omega,r}(\theta)}\,
\frac{\tilde \rho(r,\theta)}{r-2s(r,\theta)-z} 
\; \rho_j(\theta) \right| 
\;\leq\; 
\frac{C_j}{tr^2}
\;.
\end{equation}
In the same manner one obtains, for $z,z' \in  \overline{\mathbb{H}}^+$ and $0\leq\beta< 1$,
\begin{equation}
\label{equcotunoh}
\left|\int_{\mathbb{S}^2}d\theta\;e^{\imath tr  \Phi_{\omega,r}(\theta)}\,
A(r,\theta,z,z') \rho_j(\theta) \right|  
\;\leq\; 
\frac{C_{j,\beta} |z-z'|^\beta}{tr^{2+\beta} }
\;,
\end{equation}
where 
\begin{align}\label{A}
A(r,\theta,z,z')
\;=\;
\frac{1}{r-2s(r,\theta)-z}\,-\,\frac{1}{r-2s(r,\theta)-z'}
\;,
\end{align}
which satisfies the bound 
$$
|D^\alpha_k A(r,\phi_{r,\omega}(k),z,z')| 
\;\leq \; 
\frac{C_{\alpha,\beta}}{r^{1+\beta}}|z-z'|^\beta
\;.
$$
Indeed, 
$$
|D_k^\alpha A(r,\phi_{r,\omega}(k),z,z')|
\;=\;
|D_k^\alpha A(r,\phi_{r,\omega}(k),z,z')|^\beta|D_k^\alpha A(r,\phi_{r,\omega}(k),z,z')|^{1-\beta}
\;,
$$
where the first factor is bounded by adding the fractions and using \eqref{PH5} by a $r^{-2\beta}|z- z'|^\beta$ factor, the other factor without adding fractions and using \eqref{PH5}  is bounded by a $r^{\beta-1}$ factor.
Next  let us study the non-stationary part for $\nabla e(0)=0$, namely the integral $J(z,t,r,\omega_0)$ restricted to $\mathbb{S}^2 \setminus \bigcup_j V_j$. For each $\theta_0 \in \mathbb{S}^2 \setminus \bigcup_j V_j,$ one has $ \nabla_\theta \Phi(\theta_0,\omega_0,0) \neq 0$. Since the function $\mathbb{S}^2 \times \mathbb{S}^2\times [0,R) \ni (\theta, \omega,r) \mapsto \nabla_\theta \Phi(\theta,\omega,r)$ is continuous, one can take open sets $V_{\theta_0,\omega_0}$ and $  W_{\theta_0, \omega_0}$ with $(\theta_0,\omega_0)\in V_{\theta_0,\omega_0} \times W_{\theta_0, \omega_0}$, and $r_{\theta_0,\omega_0}>0$ and a constant $C_{\theta_0, \omega_0}>0$ such that 
\begin{equation}
\label{equ-bou-fsf}
|\nabla_\theta \Phi_{\omega,r}(\theta)|
\;>\; 
C_{\theta_0,\omega_0}, \ (\theta,\omega,r)\in  V_{\theta_0,\omega_0}\times W_{\theta_0, \omega_0} \times [0,r_{\theta_0,\omega_0})
\;.
\end{equation} 
Let $\rho_{\theta_0}$ be a smooth function with compact support in $V_{\theta_0,\omega_0}$ and let us take a  local chart $\psi_{\theta_0}:U_{\theta_0,\omega_0} \subset \mathbb{R}^2 \to V_{\theta_0,\omega_0} \subset \mathbb{S}^2$ with $\psi_{\theta_0}(0)=\theta_0$. Then 
$$
\int_{\mathbb{S}^2}d\theta\;e^{\imath tr  \Phi_{\omega,r}(\theta)}\,
\frac{\tilde \rho(r,\theta)}{r-2s(r,\theta)-z} 
\; \rho_{\theta_0}(\theta)
\;=\; 
\int_{\mathbb{R}^2} dk \ e^{\imath tr  \Phi_{\omega,r}(\psi_{\theta_0}(k))}
\;\hat \rho(k,r,z) 
\;,
$$ 
where 
$$
\hat \rho(k,r,z)  
\;=\; 
\frac{\tilde \rho(r,\psi_{\theta_0}(k))}{r-2s(r,\psi_{\theta_0}(k))-z} 
\; \rho_{\theta_0}(\psi_{\theta_0}(k)) |\det((D \psi_{\theta_0})_k)|
\;.
$$ 
Let us introduce the operator 
$$
(\Ll g)(k)
\;= \;
-\sum_{j=1}^{2} \partial_{ k_j}\left( \frac{1}{|\nabla (\Phi_{\omega,r}\circ \psi_{\theta_0})|^2 } \big(  \partial_{k_j} (\Phi_{\omega,r}\circ \psi_{\theta_0}) \big) g \right)(k)
\;.
$$ 
Note that $|\nabla (\Phi_{\omega,r}\circ \psi_{\theta_0})(k)| =| D \psi_{\theta_0}(k)^T \nabla_{\theta} \Phi_{\omega,r}(\psi_{\theta}(k)) | \geq C_{\theta_0,\omega_0}$, that follows from the fact that $D\psi_{\theta_0}(k)^T: T_k \mathbb{S}^2 \to \mathbb{R}^2$ is bounded from below combined with  \eqref{equ-bou-fsf}. Furthermore one has $\left| D^\alpha_k \frac{1}{r-2s(r,\psi_{\theta_0}(k))-z} \right| \leq \frac{1}{r} C_{\theta_0,\omega_0}.$ Thus
$$ 
|\Ll  \hat \rho(\cdot,r,z)| 
\;\leq \;
\frac{1}{r} \,C_{\theta_0,\omega_0}
\;.
$$ 
Hence
\begin{equation}
\label{equcotdos}
\left| \int_{\mathbb{R}^2} dk \ e^{\imath tr  \Phi_{\omega,r}(\psi_{\theta_0}(k))}\,
\hat \rho(k,r,z) \right| 
\;=\; 
\frac{1}{rt} \left|  \int_{\mathbb{R}^2} dk \ e^{\imath tr  \Phi_{\omega,r}(\psi_{\theta_0}(k))}\,
(\Ll \hat \rho(\cdot,r,z))(k)  \right| 
\;\leq \;
\frac{C_{\theta_0,\omega_0} }{tr^2}
\;,
\end{equation}  
uniformly for $(\omega,r)\in  W_{\theta_0, \omega_0} \times [0,r_{\theta_0,\omega_0})$.  Using the  estimate  
$$
|\Ll  A(\psi_{\theta_0},r,z,z')| 
\;\leq\; 
\frac{C_{\beta,\theta_0}}{r^{1+\beta}}|z-z'|^\beta
$$ 
which holds for $z,z'\in \overline{\mathbb{H}}_+$ and $0\leq \beta< 1$ and follows for \eqref{A}, one obtains in the similar manner
\begin{equation}
\label{equcotdosh}
\left| \int_{\mathbb{R}^2} d\theta \ e^{\imath tr  \Phi_{\omega,r}(\theta))}\,
A(\theta,r,z,z') \rho_{\theta_0}(\theta) \right|  
\;\leq\; 
\frac{C_{\theta_0,\omega_0} }{tr^{2+\beta}} |z-z'|^\beta
\;.
\end{equation}  
Finally, taking a finite cover $\{V_j\}\cup\{V_{\theta}\}$  of $\mathbb{S}^2$ and a partition of unity $\{\rho_j\} \cup \{\rho_{\theta_l}\}$ associated to this cover, setting $C_{\omega_0}=\max_{l,j} (C_j, C_{\theta_l,\omega_0})$, $r_{\omega_0}=\min_{l,j} (r_{\omega_0,j},r_{\omega_0,\theta_l})$ and $W_{\omega_0}=\bigcap_j W_{\omega_0,j} \cap \bigcap_l W_{\omega_0,\theta_l}$ and using \eqref{equcotuno} and \eqref{equcotdos}, one obtains \eqref{equa-cota-jota}, and using  \eqref{equcotunoh} and  \eqref{equcotdosh} one obtains \eqref{eqhoco}.

\vspace{.1cm}

Finally let us address the case $\nabla e(0)\neq 0$. Let $U$ be a rotation matrix with $\det(U)=1$ such that 
\begin{align}
\label{U}
UD  h^{-1}(0)^T \nabla e(0)
\;=\; 
|D h^{-1}(0)^T \nabla e(0)|\,e_1
\;,
\end{align}
where $e_1=(1,0,0)$. The change of variable $\theta \mapsto U\theta$ leads to
$$
J(z,t,r,\omega)
\;=\;
\int_{\mathbb{S}^2} d\theta\;e^{\imath tr  \langle \frac{1}{r} \varphi(rU\theta),\omega\rangle}\,
\frac{\tilde \rho(r,U\theta)}{\Ee_-(\varphi(rU\theta))-E} 
\;.$$ 
Hence one is confronted  with an oscillatory integral with phase function 
$$
\Phi^U_{\omega,r}(\theta)
\;=\; 
\langle \tfrac{1}{r} \varphi(rU\theta),\omega\rangle
\;.
$$  
As in \eqref{equ.fun.fas},  
$$
\Phi^U_{\omega,0}(\theta)
\;=\; 
\partial_r s(0,U\theta) \langle U\theta, D h^{-1}(0)^T \omega \rangle
\;=\;
\partial_r s(0,U\theta) \langle \theta, U D h^{-1}(0)^T \omega \rangle 
\;,
$$ 
so that combined with \eqref{equa-fors-salv} and \eqref{U} one obtains 
$$
\partial_r s(0,U\theta)
\;=\;  
\frac{1}{\langle 
UD h^{-1}(0)^T \nabla e(0), \theta\rangle + 1 }
\;=\; 
\frac{1}{|D h^{-1}(0)^T \nabla e(0)|\,\theta_1+1}
\;,
$$
where $1>\kappa\geq |D h^{-1}(0)^T \nabla e(0)|\neq 0$. Namely, one has
$$
\Phi_{\omega,0}(\theta)
\;=\;
\frac{1}{|D h^{-1}(0)^T \nabla e(0)|\,\theta_1+1} \, \langle \theta, U D h^{-1}(0)^T \omega \rangle
\;.
$$ 
Lemma \ref{lem-mor-fun} below implies that $\Phi_{\omega,0}$ is a Morse function, then one can apply the stationary phase method as above. This allows to conclude the proof.
\hfill $\Box$

\begin{lemma}
\label{lem-mor-fun}
For fixed $ 0\leq \mu  <1$ and $\omega\in \mathbb{S}^{2}$, the function $f_\omega:\mathbb{S}^2\to \mathbb{R}$ given by 
$$
f_\omega(\theta)
\;=\; 
\frac{1}{\mu  \,\theta_1+1}\, 
\langle \theta, \omega\rangle
$$  
is a Morse function.
\end{lemma}

\noindent {\bf Proof.}
After applying a two-dimensional rotation in the second and third coordinates, one can assume that $\omega=(\omega_1,\omega_2,0)$.  The gradient of $f_\omega$ at $\theta\in \mathbb{S}^2$ is given by
$$
\nabla_\theta f_\omega(\theta)
\;=\; 
\nabla f_\omega(\theta)-\langle \nabla f_\omega(\theta),\theta\rangle \theta
\;,
$$ 
where $\nabla f_\omega$ denotes the gradient of a local smooth extension $f_\omega : U\subset \mathbb{R}^3 \to \mathbb{R}$. This can  be explicitly computed
$$ 
\nabla f_\omega(x)
\;=\; 
\langle x, \omega\rangle 
\,\tfrac{1}{(\mu x_1+1)^2}
\begin{pmatrix}
-\mu \\ 0 \\ 0
\end{pmatrix}
\, +\, \tfrac{1}{\mu  x_1+1} \omega 
\;.
$$
Therefore one readily checks
\begin{align*}
\nabla_\theta f_\omega(\theta)
&
\;=\;
\tfrac{1}{(\mu  \theta_1+1)^2}\big(-\mu  \langle \theta, \omega\rangle  e_1 + (\mu   \theta_1+1) \omega -  \langle \theta, \omega\rangle \theta\big)
\;.
\end{align*}
Then $\nabla_\theta f_\omega(\theta)=0$ if and only if 
\begin{equation}
\label{equatio}
(\mu   \theta_1+1) \omega -\mu   \langle \theta,\omega\rangle e_1 
\;=\;
\theta \langle \theta,\omega\rangle
\;.
\end{equation}
Note that $\langle \theta,\omega\rangle \neq 0$, otherwise $(\mu  \theta_1+1)\omega =0$ which is not possible. If $\omega=e_!$ one obtains that the critical points are given by $\{\pm e_1\}$. If one takes the local coordinates  $\psi(x,y)=(\sqrt{1-x^2-y^2},x,y)$ with $\psi(0)=e_1$, then $\nabla^2 (f_{\omega}\circ \psi)(0)= -\frac{1}{(\mu  +1)^2}\one_2$, which is non-singular. Let us hence assume that $\omega\neq e_1$ and consider a critical point $\theta=(\theta_1,\theta_2,\theta_3)$ of $f_\omega$.  Identity \eqref{equatio} implies that $\theta \in \mbox{\rm span}\{e_1,\omega\} = \mathbb{S}^2 \cap \{(x_1,x_2,0): x\in \mathbb{R}^3\}$, then $\theta_3=0$. Note that $\theta\neq e_1$ (otherwise  \eqref{equatio} would imply  $\omega=e_1$), so that $|\theta_1|<1$ and $\theta_2\neq 0$. Then, there exists $U\subset \mathbb{R}^2$ with $(\theta_1,\theta_2,0 )\in  U$ and  such that $x^2+y^2<1$ for all $(x,y)\in U$. Consider the local chart $\psi:U \subset \mathbb{R}^2 \to \mathbb{S}^2$ given by $\psi(x,y)=(x,\sqrt{1-x^2-y^2},y)$ with $\psi(\theta_1,0)=\theta$. Let us compute the Hessian of $f_\omega \circ \psi$ at the critical point $\binom{\theta_1}{0}$. With the notations $f_\omega^\psi=f_\omega \circ \psi$ and  $h(x,y)=\sqrt{1-x^2-y^2}$, one has 
\begin{align*}
f_\omega^\psi(x,y)
&
\;=\; 
\tfrac{1}{\mu  x+1}(\omega_1 x + \omega_2 h(x,y))
\;,
\\
\nabla f_\omega^\psi(x,y)
&
\;=\; 
\tfrac{1}{\mu  x+1}
\begin{pmatrix}
\frac{-\mu  (\omega_1 x + \omega_2 h(x,y))}{\mu  x+1}+(\omega_1+\omega_2 \partial_x h(x,y))
\\
\omega_2 \partial_y h(x,y)
\end{pmatrix}
\;,
\\
\Hess f_\omega^\psi(x,y)
&
\;=\; 
\tfrac{1}{\mu  x+1}
\begin{pmatrix}
-2\mu   \partial_x f_\omega^\psi (x,y)+ \omega_2  \partial_x^2h(x,y) 
& 
-\mu   \partial_y f_\omega^\psi  (x,y)+ \omega_2 \partial_x\partial_y h (x,y) 
\\ 
-\mu  \partial_y f_\omega^\psi (x,y)+ \omega_2 \partial_x\partial_y h (x,y) 
&  
\omega_2\partial_y^2 h (x,y) 
\end{pmatrix}
\;.
\end{align*}
The fact that $\nabla f_\omega^\psi(\theta_1,0)=0$ and $\partial_x\partial_y h(\theta_1,0)=0$ imply 
$$
\Hess f_\omega^\psi(\theta_1,0)
\;=\; 
\tfrac{\omega_2}{\mu  \theta_1+1}  \begin{pmatrix}
 \partial_x^2 h(\theta_1,0) & 0 \\ 0 &   \partial^2_y h(\theta_1,0) 
\end{pmatrix}
\;=\; 
\tfrac{\omega_2}{\mu  \theta_1+1} 
\begin{pmatrix}
 -\frac{1}{\theta_2^3} & 0 \\ 0 &  -\frac{1}{\theta_2}
\end{pmatrix}
\;,
$$
showing that $\Hess f_\omega^\psi(\theta_1,0)$ is not singular so that $f_\omega$ is a Morse function also for $\omega\not=e_1$.
\hfill $\Box$

\vspace{.2cm}

Using a compactness argument over the sphere $\SM^2$, Proposition~\ref{prop-Jestimate} implies that
$$
J(z,t,r,\omega)
\;\leq\;
\frac{C}{tr^2}
\;,
$$
uniformly in $z\in\overline{\HM}_+$ and $\omega$. Replacing this and \eqref{eqcot1} into \eqref{eq-I=} shows
\begin{align*}
|I_\parallel(E+\imath 0,\omega,t)|
&
\;\leq\;
 t
\int_0^R dr\,r^2 
\,
\frac{C}{tr^2}
\,
\frac{C_N}{t^N(E-r)^N}
\;,
\end{align*}
for any $N\geq 0$. Of course, the singularity $\frac{1}{(E-r)^N}$ is not integrable at $E$ for $N\geq 1$. Hence one cuts the $r$-integral into three parts $[0,E-\delta]$, $[E-\delta,E+\delta]$ and $[E+\delta,R]$, and actually only two intervals in the special case $E=0$. On the first and third, one uses the above estimates with $N=1$, and this gives a contribution of the order
$$
t
\int_0^{E-\delta} dr\,r^2 
\,
\frac{C}{tr^2}
\,
\frac{C_1}{t(E-r)}
\;\leq \;
\frac{C}{t }\,\log(\delta^{-1})
$$
and similarly for the third term. In the middle term, one does not use the oscillations in $\eta$ and bounds the $\eta$-integral simply by $\eta_1-\eta_0$, leading to a contribution
$$
t
\int_{E-\delta}^{E+\delta} dr\,r^2 
\,
\frac{C}{tr^2}
\,
(\eta_1-\eta_0)
\;\leq\;
C  \,\delta\,(\eta_1-\eta_0)
$$
Choosing $\delta=\frac{1}{t}$, one then gets a decay of the type $t^{-1}\log(t)$. Hence one concludes:

\begin{proposi}
\label{prop-Iparallel}
For all $E\geq 0$,  one has 
$$
\big| I_\parallel (E,\omega,t) \big|
\;\leq\;
C t^{-1}\log(t)\;, \qquad t\geq 1
\;.
$$ 
 Further, if $0\leq \beta< 1$  for all $z,z' \in \overline{\HM}_+$, 
$$
|I_\parallel(z,\omega,t)\,-\,I_\parallel(z',\omega,t)| 
\;\leq \;
C_\beta |z-z'|^\beta t^{-1+\beta}\log(t)
\;, 
\qquad 
t\geq 1
\;.
$$
\end{proposi}

\noindent {\bf Proof}. The first statement was already proved above so that only remains to check the H\"older continuity in $z$. Due to \eqref{eq-I=} with $E+\imath 0$ replaced by $z$, there are two dependences on $z$ contributions. Let us start out from, for $z,z' \in  \overline{\mathbb{H}}_+$,
\begin{align*}
I_\parallel(z,\omega,t)-I_\parallel(z',\omega,t)
\;=\; 
&
\imath t \int_0^R dr \ r^2 \big(J(z,t,r,\omega)-J(z',t,r,\omega)\big) \tilde J(z,t,r) 
\\ 
&+  \imath t \int_0^R dr \ r^2 J(z',t,r,\omega)\big(\tilde J(z,t,r)-\tilde J(z',t,r)\big)
\;, 
\end{align*} 
where 
$$
\tilde J(z,t,r) 
\;= \;
\int_{\eta_0}^{\eta_1}  d\eta \,   
e^{\imath t \eta (z-r)}\,\chi(\eta)
\;.
$$
For the first summand and $E=\Re e(z)\not=0$, one cuts the $r$-integral into three parts $[0,(\Re e(z)-\delta)^{\frac{1}{1-\beta}}]$, $[(\Re e(z)-\delta)^{\frac{1}{1-\beta}},(\Re e(z)+\delta)^{\frac{1}{1-\beta}}]$, $[(\Re e(z)+\delta)^{\frac{1}{1-\beta}},R]$. Using Eqs. \eqref{eqhoco} as well as \eqref{eqcot1} for $N=1$, one can bound the integral over the interval  $[0,(\Re e(z)-\delta)^{\frac{1}{1-\beta}}] $ by
$$ 
C_{\beta} t  \int_0^{(\Re e(z)-\delta)^{\frac{1}{1-\beta}}} dr \, r^2 \,\frac{1}{r^{2+\beta}t}\,|z-z'|^\beta \,\frac{1}{ t |\Re e(z)-r|} 
\;\leq\; 
C_{\beta} |z-z'|^{\beta} t^{-1} \log(\delta^{-1})
\;.
$$ 
In a similar manner, one obtains the same estimate for the integral over $[(\Re e(z)+\delta)^{\frac{1}{1-\beta}},R]$. For the integral over the central part  $[(\Re e(z)-\delta)^{\frac{1}{1-\beta}},(\Re e(z)+\delta)^{\frac{1}{1-\beta}}]$, one uses again \eqref{eqhoco}, but bounds $\tilde J$ simply by a constant, so that one obtains an upper bound by
$$ 
C_{\beta} t  \int_{(\Re e(z)-\delta)^{\frac{1}{1-\beta}}}^{(\Re e(z)+\delta)^{\frac{1}{1-\beta}}} dr \, r^2\, \frac{1}{r^{2+\beta}t}\,|z-z'|^\beta 
\;\leq\; 
C_{\beta} |z-z'|^{\beta}\, \delta
\;.
$$ 
Choosing $\delta=\frac{1}{t}$, these bound result in the H\"older continuity of the first summand as claimed. For $E=0$, one splits the $r$-integral in merely two intervals $[0,(\Re e(z)+\delta)^{\frac{1}{1-\beta}}]$ and $[(\Re e(z)+\delta)^{\frac{1}{1-\beta}},R]$, but otherwise proceeds in the same manner.

\vspace{.1cm}

For the second summand in $I_\parallel(z,\omega,t)-I_\parallel(z',\omega,t)$, let us start out with the bound 
\begin{align*}
\big|
\tilde J(z,t,r) 
-
\tilde J(z',t,r) 
\big|
&
\;=\;
\big|
\tilde J(z,t,r) 
-
\tilde J(z',t,r) 
\big|^{\beta+(1-\beta)}
\nonumber
\\ 
&
\;\leq \; 
C_\beta t^\beta |z-z'|^\beta  \Big(\Big|\int_{\eta_0}^{\eta_1}  d\eta \,   
e^{-t\eta \Im m(z)} e^{\imath t\eta (\Re e(z) -r)}\,\chi(\eta) \Big|^{1-\beta}
\nonumber
\\
&
\hspace{3cm}
+  \Big|\int_{\eta_0}^{\eta_1}  d\eta \,   
e^{-t\eta \Im m(z')} e^{\imath t\eta (\Re e(z') -r)}\,\chi(\eta) \Big|^{1-\beta} \Big)
\;,
\end{align*}
which follows from \eqref{eq-ExpHoelder}. Using this and \eqref{equa-cota-jota},  the second summand is bounded above by
\begin{align*}
 & C_{\beta} t \int_0^R dr \ r^2 \frac{1}{r^2 t} t^{\beta}|z-z'|^{\beta}\left|\int_{\eta_0}^{\eta_1}  d\eta \,   
e^{-t\eta \Im m(z)} e^{it\eta (\Re e(z) -r)}\,\chi(\eta) \right|^{1-\beta}\\& +C_{\beta} t \int_0^R dr \ r^2 \frac{1}{r^2 t} t^{\beta} |z-z'|^\beta\left|\int_{\eta_0}^{\eta_1}  d\eta \,   
e^{-t\eta \Im m(z')} e^{it\eta (\Re e(z') -r)}\,\chi(\eta) \right|^{1-\beta} 
\;.
\end{align*} 
These summands are essentially the same, so let us focus on the first.  The $r$-integral is now split into the intervals  $[0,\Re e(z)-\delta], [\Re e(z)-\delta, \Re e(z)+\delta], [\Re e(z)+\delta, R]$. In the intervals  $[0,\Re e(z)-\delta]$ and $[\Re e(z)+\delta,R]$ one uses \eqref{eqcot1} for $N=1$, while in $[\Re e(z)-\delta, \Re e(z) + \delta]$ one just bounds the $\eta$-integral by a constant. Choosing $\delta=\frac{1}{t}$ and proceeding similar as in the above, one obtains in all three cases the H\"older continuity bound as claimed.
\hfill $\Box$

\vspace{.2cm}

\noindent {\bf Proof} of Theorem~\ref{theo-WeylMain}. Based on the representation \eqref{eq-RWeylRep} of the (translation invariant) resolvent as in oscillatory integral $I$ which is decomposed into the three parts $I_\leq$, $I_\parallel$ and $I_\geq$, one obtains both claimed bounds by combining Propositions~\ref{prop-Ileq2}, \ref{prop-Igeq} and \ref{prop-Iparallel} because the second summand in \eqref{eq-RWeylRep} is subdominant due to the prefactor $t$.
\label{prop-Ileq}
\hfill $\Box$

\section{Global limit absorption}
\label{sec-Global}

The preceding sections dealt with the limit absorption in regular points (Section~\ref{sec-RegPoints}), at and near critical points (Section~\ref{sec-CritPoints}) and at Weyl points (Section~\ref{sec-Weyl}). For this purpose, smooth indicator functions $\rho$ were inserted into the resolvent that allow to separate the various contributions, see ~\eqref{eq-LocResol}. Furthermore, all these technical sections only dealt with the one-band or two-band situation, namely $L=1$ for regular and critical points and $L=2$ for Weyl points. In this section, it will be shown how to combine the local results in order to provide a proof of Theorem~\ref{theo-3dMain}. This is based on two standard techniques, namely the use of Riesz projections and partitions of unity.

\vspace{.2cm}

Hence let be given a periodic Hamiltonian with smooth symbol $k\in\TM^d\mapsto \Ee(k)=\Ee(k)^*\in\CM^{L\times L}$ having all the properties stated in the Hypothesis stated in the introduction. Let us start by isolating the finite set $k^W_1,\ldots,k^W_J$ of Weyl points which due to the Hypothesis make up the subset $\{k\in\TM^d\,:\,\Ee(k) \,\mbox{\rm has degenerate eigenvalues}\}$. At each Weyl point, there is a band touching of two bands (in principle, there could be several band touching over each Weyl point, but this is non-generic and simply excluded for sake of notational simplicity). Around each $k^W_j$, let us isolate a ball $B_{2r}(k^W_j)$ of radius to be chosen later. Then set $A=\TM^d\setminus \cup_{j}\overline{B_r(k^W_j)}$ so that $A,B_{2r}(k^W_1),\ldots,B_{2r}(k^W_J)$ is an open cover of $\TM^d$ and let $\rho^{\mbox{\rm\tiny 1b}},(\rho^W_1)^3,\ldots,(\rho^W_J)^3$ be a subordinate smooth partition of unity of $\TM^d$. It will be become apparent below why it is convenient to insert third powers. Then
$$
R^z
\;=\;
\Ff^*\,\rho^{\mbox{\rm\tiny 1b}}\,(\Ee-z\,\one)^{-1}\Ff
\;+\;
\sum_{j=1}^J
\Ff^*\,(\rho^W_j)^3\,(\Ee-z\,\one)^{-1}\Ff
\;.
$$
Let us first focus on the first summand corresponding to non-intersecting bands. For each such band $k\in A\mapsto\Ee_l(k)$ there is an associated Riesz projection of rank $1$ given by
\begin{equation}
\label{eq-Riesz1b}
P_l(k)
\;=\;
\oint_{\Gamma_l} \frac{dz}{2\pi\imath}\,(z\,\one-\Ee(k))^{-1}
\;,
\end{equation}
where $\Gamma_l$ is a positively oriented loop in the complex plane encircling only $\Ee_l(k)$ and no other eigenvalue of $\Ee(k)$.  Then $\sum_{j=1}^JP_l(k)=\one_L$ for $k\in A$, $[P_l(k),\Ee(k)]=0$ and $P_l(k)P_{l'}(k)=0$ for $l\not=l'$. It is well-known \cite{Kat} that $k\in A\mapsto P_l(k)$ is real analytic for all $l=1,\ldots,L$. Similarly, let us introduce the Riesz projection onto the two bands touching over a Weyl point $k^W_j$:
$$
P^W_j(k)
\;=\;
\oint_{\Gamma_j} \frac{dz}{2\pi\imath}\,(z\,\one-\Ee(k))^{-1}
\;,
$$
where $\Gamma_j$ is a positively oriented loop in the complex plane encircling the two eigenvalues involved in the band touching. Again $k\mapsto P^W_j(k)$ is real analytic on some ball $B_{2r}(k^W_j)$ where $r$ is chosen sufficiently small such that this analyticity statement holds for all $j=1,\ldots,J$ \cite{Kat}. Then $P^W_j(k)$ is of rank $2$, and it is completed by $L-2$ Riesz projections $P_{j,1}(k),\ldots,P_{j,L-2}(k)$ of rank $1$ given as in \eqref{eq-Riesz1b}, namely one has $P^W_j(k)+\sum_{l=1}^{L-2}P_{j,l}(k)=\one_L$. Next let us recall the following

\vspace{.2cm}

\noindent {\bf Fact} \cite{Nen} {\it Let $U\subset\RM^d$ be a simply connected open set. For every real analytic family of projections $k\in U\mapsto P(k)$ of rank $r$, one can construct a real analytic $r$-frame $k\in U\mapsto \Phi(k)\in\CM^{L\times r}$, namely a function of matrices satisfying $\Phi(k)^*\Phi(k)=\one_r$ and $P(k)=\Phi(k)\Phi(k)^*$.}

\vspace{.2cm}

\noindent This can be immediately applied on the sets $B_{2r}(k^W_j)$, providing frames $\Phi^W_j(k)\in\CM^{L\times 2}$ and $\Phi_{j,l}(k)\in\CM^{L\times 1}$ for $P^W_j(k)$ and $P_{j,l}(k)$. In order to apply this also on $A$, let us cover $A$ by simply connected open set $A_1,\ldots,A_I$. Let $(\rho_1)^3,\ldots,(\rho_I)^3$ be an associated subordinate smooth partition of unity. For each $l=1,\ldots, L$ and $i=1,\ldots,I$, there are then real analytic $1$-frames $k\in A_i\mapsto \Phi_{l,i}(k)$ for $P_l(k)$. One can now insert all these partitions in order to decompose the resolvent:
\begin{align*}
R^z
\;=\;
&
\sum_{l=1}^L\sum_{i=1}^I
\big(\Ff^*\rho_i\,\Phi_{l,i}\Ff\big)
\big(\Ff^*\rho^{\mbox{\rm\tiny 1b}}\rho_i (\Phi_{l,i})^* (\Ee-z\,\one)^{-1} \Phi_{l,i}\Ff\big)
\big(\Ff^*\rho_i (\Phi_{l,i})^*\Ff\big)
\\
&
+\,
\sum_{l=1}^{L-2}\sum_{j=1}^J
\big(\Ff^*\rho^W_j\,\Phi_{j,l}\Ff\big)
\big(\Ff^*\rho_j^W (\Phi_{j,l})^* (\Ee-z\,\one)^{-1} \Phi_{j,l}\Ff\big)
\big(\Ff^*\rho_j^W (\Phi_{j,l})^*\Ff\big)
\\
&
+\,
\sum_{j=1}^J
\big(\Ff^*\rho^W_j\,\Phi^W_{j}\Ff\big)
\big(\Ff^*\rho_j^W (\Phi^W_{j})^* (\Ee-z\,\one)^{-1} \Phi^W_{j}\Ff\big)
\big(\Ff^*\rho_j^W (\Phi^W_{j})^*\Ff\big)
\;.
\end{align*}
Now in the first sum $(\Phi_{l,i})^* (\Ee-z\,\one)^{-1} \Phi_{l,i}=(\Ee_l-z)^{-1}$ is a scalar function, notably of type $L=1$, and similarly in the second sum, while in the third sum one has $2\times 2$ matrices $\Ee^W_j$. After simplification and furthermore introducing the weights as in \eqref{eq-Ralpha}, one then gets
\begin{align}
\Rr^z_\alpha
\;=\;
&
\sum_{l=1}^L\sum_{i=1}^I
M_{l,i}
\langle X\rangle^{-\alpha} 
\big(\Ff^*\rho^{\mbox{\rm\tiny 1b}}\rho_i (\Ee_l-z)^{-1}\Ff\big)
\langle X\rangle^{-\alpha} 
(M_{l,i})^*
\label{eq-1band}
\\
&
+\,
\sum_{l=1}^{L-2}\sum_{j=1}^J
M_{j,l}
\langle X\rangle^{-\alpha} 
\big(\Ff^*\rho_j^W  (\Ee_j-z)^{-1} \Ff\big)
\langle X\rangle^{-\alpha} 
(M_{j,l})^*
\label{eq-1bandWeyl}
\\
&
+\,
\sum_{j=1}^J
M^W_j
\langle X\rangle^{-\alpha} 
\big(\Ff^*\rho_j^W (\Ee_j^W-z\,\one_2)^{-1} \Ff\big)
\langle X\rangle^{-\alpha} 
(M^W_j)^*
\;,
\label{eq-2bandWeyl}
\end{align}
where
$$
M_{l,i}
\;=\;
\langle X\rangle^{-\alpha} 
\Ff^*\rho_i\,\Phi_{l,i}\Ff
\langle X\rangle^{\alpha} 
\;,
\qquad
M_{j,l}
\;=\;
\langle X\rangle^{-\alpha} 
\Ff^*\rho^W_j\,\Phi_{j,l}\Ff
\langle X\rangle^{\alpha} 
\;,
$$
and 
$$
M^W_j
\;=\;
\langle X\rangle^{-\alpha} 
\Ff^*\rho^W_j\,\Phi^W_{j}\Ff
\langle X\rangle^{\alpha} 
\;.
$$

\begin{lemma}
\label{lem-Mbounds}
$M_{l,i}$, $M_{j,l}$ and $M^W_{j}$ are bounded operators for all $\alpha\in\RM$.
\end{lemma}

\noindent {\bf Proof.} By construction, the entries of $\rho_i\,\Phi_{l,i}$ are smooth compactly supported functions. Hence their Fourier transform $\Ff^*\rho_i\,\Phi_{l,i}\Ff$ is a convolution operator with off-diagonal decay that is faster than any power. Hence the commutators $[\Ff^*\rho_i\,\Phi_{l,i}\Ff,\langle X\rangle^{-\alpha} ]$ are bounded operators for any $\alpha\in\RM$, and this implies the claim for $M_{l,i}$. For $M_{j,l}$ and $M^W_{j}$ one argues in the same manner.
\hfill $\Box$

\vspace{.2cm}

\noindent {\bf Proof} of Theorem~\ref{theo-3dMain}. The sums in \eqref{eq-1band}, \eqref{eq-1bandWeyl} and \eqref{eq-2bandWeyl} are finite. Due to Lemma~\ref{lem-Mbounds}, it is hence sufficient to prove that the central factors in \eqref{eq-1band}, \eqref{eq-1bandWeyl} and \eqref{eq-2bandWeyl} converge to bounded operators as $\Im m(z)\to 0$. In \eqref{eq-1band}  this factor is of scalar $L=1$ type. One still has to partition the support into balls around the finite number of critical points as well as the regular points. For the regular points, one can directly apply Theorem~\ref{theo-RegValues} to obtain the existence of the limit operators and their H\"older continuity. For the critical points, Proposition~\ref{prop-CritLAP} also shows that the decay estimates of Theorem~\ref{theo-CritValues} imply the existence and H\"older continuity of the limit operators. Finally, the proof of Proposition~\ref{prop-CritLAP} transposes directly to the Weyl points for which Theorem~\ref{theo-WeylMain} provides the needed estimates on the resolvent (which are stronger than the estimates at the critical values given by Theorem~\ref{theo-CritValues}).
\hfill $\Box$

\vspace{.2cm}

\noindent {\bf Proof} of Theorem~\ref{theo-OpNorm}. Due to the hypothesis of the theorem, there is no band touching and hence only the $1$-band contributions \eqref{eq-1band} remain, simply because \eqref{eq-1bandWeyl} and \eqref{eq-2bandWeyl} vanish. For the $1$-band case $L=1$, Theorem~\ref{theo-RegValues} deals with the regular values and Proposition~\ref{prop-CritLAP} with the neighborhoods of critical values. Together this concludes the proof.
\hfill $\Box$

\appendix

\section{Bounds on an oscillatory integral}
\label{app-OsciInt}

This appendix provides asymptotic estimates for an oscillatory integral of the type
\begin{equation}
\label{equa-inte-osci}
I(z, t , \omega) 
\;= \;
\int_{\mathbb{R}^d}  dx \int_0^{\infty} d\eta\; e^{\imath z t \eta} e^{\imath t(\langle x,\omega\rangle-\eta f(x))} \rho(x)  
\;,
\end{equation}
in the regime $t\to\infty$ and $\Im m(z)\downarrow 0$ for $z\in\CM$, uniformly in $\omega \in \mathbb{S}^{d-1}$. Here $\langle x,\omega\rangle$ denotes the Euclidean scalar product in $\RM^d$ and the phase function $f:B_R(x_0)\to \RM$ is supposed to have a unique critical point $x_0$ in the ball $B_R(x_0)=\{x\in \mathbb{R}^d: |x-x_0|<R\}$, and $\rho: B_R (x_0)\to\RM$ is smooth and has compact support in an open subset $U \subset B_R(x_0)$ containing $x_0$. The set $U$ will be chosen shortly. The latter property immediately implies that, for $z=E+\imath\epsilon$ with $\epsilon>0$, one has
$$
|I(E+\imath\epsilon,t,\omega)|
\;\leq\; C\,\frac{1}{\epsilon t}
\;,
$$
for $t\geq 1$ and $\epsilon>0$. Although this proves that integral \eqref{equa-inte-osci} exists, the bound blows up as $\epsilon\to 0$. In this appendix, it is shown how to obtain better estimates by exploring the oscillatory behavior of the integrand. This combines techniques available in the literature \cite{Dui,DS,MT,KR,KKR}.

\vspace{.2cm}

In order to state the main result, some notations need to be introduced. Let $\nabla f:B_R(x_0)\to\RM^d$ and $\Hess f:B_R(x_0)\to\RM^{d\times d}$ denote the gradient and (symmetric) Hessian of $f$. By assumption, the equation $\nabla f(x)=0$ hence has a single solution $x_0$ in $B_R(x_0)$.  To simplify notations, let us also assume that $f(x_0)=0$.  Furthermore, it will be assumed that $f$ is a Morse function, namely that the Hessian is non-singular $\det(\Hess f(x_0))\not=0$. In particular, this implies that for some open  $U \subset B_R(x_0)\subset \mathbb{R}^d$ with $x_0\in U$, the map $\nabla f : U \to B_{r}(0) \subset \mathbb{R}^d$ is a diffeomorphism for some $r\leq \frac{1}{2}$. Consider the smooth function $\theta: [0,r) \times \mathbb{S}^{d-1} \to U,$
\begin{equation}
\label{equ-def-the}
\theta(\mu,\omega)
\;=\; 
(\nabla f)^{-1}(\mu \omega).
\end{equation} 
By definition, for all $(\mu,\omega)\in [0,r) \times \mathbb{S}^{d-1}$, the equation
\begin{equation}
\label{equ-pun-cri}
\nabla f(x) - \mu \omega
\;=\;
0
\;,
\end{equation}
has exactly one solution $x\in U$ and is given by $\theta(\mu,\omega)$. For $\mu=0$, this solution is $\theta(0,\omega)=x_0$. Note that the choice of $U$ implies 
\begin{equation}
\label{equ-ass-gra}
\|\nabla f(x)\| 
\;<\; r \;\leq \;\tfrac{1}{2}
\;, 
\qquad 
x\in U
\;.
\end{equation}
By choosing $R$ sufficiently small, one knows that the signature $\sig(\Hess f(x))\in\{d,d-2,\dots,-d\}$, defined as the difference of the number of positive  eigenvalues minus the number of negative eigenvalues, is constant on $U$. The main result of this appendix is the following.

\begin{theo}
\label{theo-mai-res} 
Suppose that $d\geq 3$. Then, there exists a continuous function $M: \overline{\HM} \times [1,\infty) \times \mathbb{S}^{d-1}\to \mathbb{C}$   such that for all $(t,\omega) \in [1,\infty)\times \mathbb{S}^{d-1}$ 
\begin{equation}
\label{equa-cota-boul}
I(z,t,\omega)
\;=\;
t^{-\frac{d}{2}}M(z,t,\omega)
\;, 
\qquad 
\Im m(z)\geq 0
\;,
\end{equation} 
and the following holds: there is a constant  $C$ that does not depend on $(t,\omega)\in [1,\infty) \times \mathbb{S}^{d-1}$ and $z$ such that 
$$
|M(z,t,\omega)|
\;\leq\; C
\;.
$$ 
Moreover, there exists a constant $C_U\geq 0$ such that, for all $\omega \in \mathbb{S}^{d-1}$ satisfying
$$
|\langle  \omega,(\nabla^2 f(x_0))^{-1}\omega\rangle|
\;>\;
C_U
\;,
$$
one has
$$
|M(E,t,\omega)|
\;\leq\;
\frac{C\,t^{-\frac{d}{2}+1}}{|\langle \omega,(\nabla^2 f(x_0))^{-1}\omega\rangle|^{\frac{d+1}{2}}}
\;,
$$
uniformly for  
$$
0
\;\leq \;
|E| 
\;< \;
\tfrac{1}{8}\, r^2 \,
|\langle  \omega,(\nabla^2 f(x_0))^{-1}\omega\rangle|
\;, 
\qquad 
1 \;\leq\; t\;< \;\frac{\sqrt{|\langle \omega,(\nabla^2 f(x_0))^{-1}\omega\rangle|}}{\sqrt{8E}}\;,
$$
where $r=r_U$ is specified by \eqref{equ-ass-gra}. Finally, one has the H\"older continuity in $z$ with exponent $0<\beta\leq 1$ satisfying $\beta<\frac{d-2}{2}$, namely 
$$
|M(z,t,\omega)-M(z',t,\omega)|
\;\leq \;
C_\beta t^{\beta}|z-z'|^{\beta} 
\;, 
\qquad
\omega \in \mathbb{S}^{d-1}
\;. 
$$
\end{theo}

The remainder of this appendix is dedicated to the proof of Theorem~\ref{theo-mai-res}.  Let us start by splitting the integral \eqref{equa-inte-osci} in the following manner
\begin{equation}
\label{equat-split-integ}
I(z,t,\omega) 
\;=\; 
I_1(z,t,\omega) + I_2(z,t,\omega)
\;,
\end{equation}
where 
\begin{align*}
&
I_1(z,t,\omega)
\;=\;
\int_{\mathbb{R}^d}dx \int_0^1 d\eta \;e^{\imath z t \eta} e^{\imath t( \langle x,\omega\rangle-\eta f(x))} \rho(x) 
\;,
\\
&
I_2(z,t,\omega)
\;= \;
\int_{\mathbb{R}^d} dx \int_1^{\infty}d\eta\; e^{\imath z t \eta} e^{\imath t(\langle x,\omega\rangle-\eta f(x))} \rho(x) 
\;.
\end{align*} 
%

\begin{proposi} 
\label{prop-meto-nstp}
For all integers $j\geq 0$ the limit $\partial_E^j I_1(E+\imath 0,t,\omega)=\lim_{\epsilon\downarrow 0} \partial_z^j I_1(z, t,\omega)$ exists for all $\omega \in \mathbb{S}^{d-1}$ and $E\in \mathbb{R}$, and it is given by
\begin{equation}
\label{equ-lim-in1}
\partial_E^j I_1(E+\imath 0,t,\omega)
\;=\;
(\imath t)^j \int_0^1 d\eta  \ \eta^j e^{\imath E t\eta}\, \int_{\mathbb{R}^d}dx \;e^{\imath t(\langle x,\omega\rangle-\eta f(x))} \, \rho(x)
\;.
\end{equation}
Moreover, for all $N\in \mathbb{N}$, there exists a constant $C_{N,j}$  such that, uniformly in $\omega \in \mathbb{S}^{d-1}$ and $z\in\overline{\HM}$,
\begin{equation}
\label{equa-cota-pare}
|\partial_z^j I_1(z,t,\omega)| 
\;\leq \;
C_{N,j} t^{-N}
\;.
\end{equation}
\end{proposi}

\noindent {\bf Proof.} Let us first focus on the case $j=0$. The first claim \eqref{equ-lim-in1} is an immediate consequence of the Lebesgue dominated convergence theorem and Fubini's theorem (note that the integrand is  bounded by $1$ and $\rho$ has compact support). The second claim \eqref{equa-cota-pare} is an application of the method of non-stationary phase \cite{DS} to the oscillatory integral 
$$
J_1(t,\eta,\omega)
\;=\;
\int_{\mathbb{R}^d}dx\;  e^{\imath t(\langle x,\omega\rangle-\eta f(x))}  \ \rho(x) 
\;.
$$
Hence let us introduce the phase function
$$
\Phi_1(x,\omega,\eta)
\;=\;
\langle x,\omega\rangle-\eta f(x)
\;.
$$
 \eqref{equ-ass-gra} implies that if $\eta\leq 1$, then $\eta\|\nabla f(x)\|<\tfrac{1}{2}$ for all $x\in U$ and $\omega,\eta$ fixed, which leads to, for $\nabla=\nabla_x$,
\begin{equation}
\label{eq-co-wd}
\|\nabla \Phi_1(x,\omega,\eta)\|
\;=\;
\|\omega-\eta \nabla f(x) \| 
\;\geq \;
\|\omega \|-\eta\|\nabla f(x)\| 
\;\geq \;
\tfrac{1}{2}\;,  
\end{equation}
still for $x\in B_r(x_0)$. For a function $g\in C^\infty(B_r(x_0))$  one can hence define 
$$
(\Ll g)(x)
\;= \;
-\sum_{j=1}^{d} \partial_{ x_j}\left( \frac{1}{\|\nabla \Phi_1\|^2 } \big(  \partial_{x_j} \Phi_1\big) g \right)(x)
\;.
$$
Then one can readily check that its (formal) adjoint satisfies
\begin{equation}
\label{eqaul1}
(\Ll^*e^{\imath t \Phi_1})(x)
\;=\; 
\imath t \,e^{\imath t  \Phi_1}
\;.
\end{equation} 
Furthermore,  if $h \in C^\infty_0(B_r(x_0),\mathbb{C})$ and $g\in C^\infty(B_r(x_0),\mathbb{C})$ integration by parts implies 
$$
\int_{\mathbb{R}^d} dx\;(\Ll h)(x) \,g(x) 
\; =\; 
\int_{\mathbb{R}^d} dx\; h(x)\,(\Ll^*g)(x) 
\;.
$$
Applying this $N$ times and using  \eqref{eqaul1}, one obtains 
$$
J_1(t,\eta,\omega)
\; =\; 
(\imath t)^{-N}\int_{\mathbb{R}^d}dx\;  e^{\imath t\Phi_1(x,\omega,\eta)}\, (\Ll^N\rho)(x)
\;.
$$
 \eqref{eq-co-wd} implies that for $\eta \leq 1$, there exists a constant $C_{\rho,N}$ that does not depend on $\omega\in \mathbb{S}^{d-1}$ and $E\in \mathbb{R}$ such that 
$$
\|(\Ll^N \rho )(x)\| 
\;\leq\; 
C_{\rho,N} \chi_{B_r(x_0)}(x)
\;.
$$
Using these last two equations shows that $J_1(t,\eta,\omega)\leq C_N\,t^{-N}$. When replacing this bound in \eqref{equ-lim-in1}, one obtains the claimed bound by an $L^2$-estimate of the integral over $\eta$. For $j>0$, one proceeds in the same manner, possibly requiring a larger constant.  
\hfill $\Box$

\vspace{.2cm}

Let us stress that the (weak) oscillations of the factor $e^{\imath E t\eta}$ in \eqref{equ-lim-in1} were simply discarded for the estimate \eqref{equa-cota-pare} on $I_1(z,t,\omega)$. Clearly, in $I_2(z,t,\omega)$ the oscillations of the factor $e^{\imath z t \eta}$ are much larger and will actually play a crucial role in the last step of the argument leading to Theorem~\ref{theo-mai-res}. In order to analyze $I_2(z,t,\omega)$, it is convenient change variables to $\mu=\frac{1}{\eta}\in [0,1]$. Again combined with Fubini's theorem, one obtains
\begin{equation}
\label{equa-osci-int2}
I_2(z,t,\omega)
\;=\; 
\int_{0}^1 d\mu\;e^{\frac{\imath z t }{\mu}} \frac{1}{\mu^2} \int_{\mathbb{R}^d} dx\; e^{\imath t\frac{1}{\mu}(\mu \langle x,\omega \rangle)-f(x)} \rho(x)
\;.
\end{equation}
Hence one is lead to study the oscillatory integral 
\begin{equation}
\label{equa-osci-inte}
J_2(t,\mu,\omega) 
\;=\; 
\int_{\mathbb{R}^d}dx\; e^{\imath t\frac{1}{\mu} (\mu \langle x, \omega \rangle)-f(x))} \rho(x) 
\;.
\end{equation}
It is then again convenient to introduce a phase function  $\Phi_2 : U\times \mathbb{R}_\geq \times \mathbb{S}^{d-1} \to \mathbb{R}$ as well as fiberwise restriction $\Phi_{2,\mu,\omega} : U \to \mathbb{R}$ for fixed $\mu,\omega$ by
\begin{equation}
\label{eq-PhiPhaseDef}
\Phi_2(x,\mu,\omega) 
\;=\; 
\mu \langle x, \omega \rangle-f(x)
\;,
\qquad
\Phi_{2 ,\mu,\omega}(x) 
\;=\; 
\Phi_2(x,\mu,\omega) 
\;.
\end{equation} 
Note that the gradient and Hessian of $\Phi_{2,\mu,\omega}$ are, for $x\in U$, given by 
\begin{equation}\label{equ-cer-nab}
\nabla \Phi_{2,\mu,\omega}(x)
\;=\;  \mu \omega - \nabla f(x)
\;, 
\qquad
\nabla^2 \Phi_{2,\mu,\omega}(x)\;=\; -\,\nabla^2 f(x)
\;.
\end{equation} 
For each $(\mu, \omega)\in [0,r)\times \mathbb{S}^{d-1}$, Eqs. \eqref{equ-cer-nab}, \eqref{equ-pun-cri} and \eqref{equ-def-the} imply that   $\Phi_{2,\mu,\omega}$ has only one critical point  in $U$ which is given by $\theta(\mu,\omega)$. On the other hand, ~\eqref{equ-ass-gra} implies that for $(\mu, \omega)\in [r,1]\times \mathbb{S}^{d-1}$, $\Phi_{2,\mu,\omega}$ has no critical points on $U$. Now let us proceed to study the oscillatory integral \eqref{equa-osci-inte}, beginning with the case of non-stationary points.

\begin{proposi}
\label{prop-case-nonc}
Let $(\mu_0,\omega_0)\in [r,1]\times \mathbb{S}^{d-1}$. Then, there exist open sets $V_{\mu_0}\subset \mathbb{R}, \ W_{\omega_0} \subset \mathbb{S}^{d-1}$, with $\mu_0 \in V_{\mu_0}$ and $\omega_0\in W_{\omega_0}$, with the following property: for all $N\in \mathbb{N}$, there exists a smooth function $g_{N,\mu_0,\omega_0}:(0,\infty)\times V_{\mu_0}\times W_{\omega_0}\to \mathbb{C}$ such that for all $(\mu,\omega) \in V_{\mu_0}\times W_{\omega_0}$ one has for any $N\in\NM$
\begin{equation}
\label{equ-jot-pun}
J_2(t,\mu,\omega) 
\;=\; 
\left(\frac{\mu}{t}\right)^N g_{N,\mu_0,\omega_0}(t,\mu,\omega)
\;,
\end{equation}
where $|g_{N,\mu_0,\omega_0}(t,\mu,\omega)|\leq C_{N,\mu_0,\omega_0}$ uniformly for $(t,\mu,\omega) \in [1,\infty)\times V_{\mu_0}\times W_{\omega_0}$.
\end{proposi}

\noindent {\bf Proof.}
The claim follows from a standard application of the method of non-stationary phase, for example the second part of  Proposition  \ref{prop-meto-nstp}, can be adapted. Let us just point out which modifications are needed. Since $\Phi_{2,\mu_0,\omega_0}$ has no critical points in $U$,  for all $x\in \supp (\rho)\subset U$, there are open sets  $U_x\subset U,$ $V_{\mu_0,x}\subset \mathbb{R}, \ W_{\omega_0,x} \subset \mathbb{S}^{d-1}$ with $(x,\mu_0,\omega_0) \in U_x \times V_{\mu_0,x}\times W_{\omega_0,x}$ and such that for all $(y,\mu,\omega)\in U_x \times V_{\mu_0,x}\times W_{\omega_0,x}$, one has
$$
\|\nabla \Phi_{2,\mu,\omega}(y)\| 
\;\geq \;
C_{x,\mu_0,\omega_0}
\;>\;0
\;.
$$ 
Since the support of $\rho$ is compact, it admits a finite open cover $(U_{x_j})_{j = 1,\ldots,n_j}$, where the sets $U_{x_j}$ are as above. Set
$$
V_{\mu_0} 
\;=\; 
\bigcap_{j} V_{\mu_0,x_j}
\;, 
\qquad  
W_{\omega_0} 
\;= \;
\bigcap_{j} W_{\omega_0,x_j}
\;, 
\qquad 
C_{\mu_0,\omega_0}
\;=\; 
\min_j \{C_{x_j,\mu_0,\omega_0}\}
\;.
$$ 
Then, for all $(\mu,\omega) \in V_{\mu_0}\times W_{\omega_0}$ and $x\in \supp (\rho)$,
$$
\|\nabla \Phi_{2,\mu,\omega}(x)\|
\;\geq\; 
C_{\mu_0,\omega_0}
\;>\;
0
\;, 
\qquad 
\left| \partial_{x_j}\Phi_{2,\mu,\omega} (x) \right|
\;=\;
 \left|\partial_{x_j}f (x) + \mu \omega_j \right|  
 \;\leq \;
 C
 \;.
 $$
For fixed $(\mu,\omega)\in V_{\mu_0}\times W_{\omega_0}$, the method of non-stationary phase hence applies and implies the claim.
\hfill $\Box$

\vspace{.2cm}

As a preparation for the study of the stationary points, one has to study a set-up in which a critical point and its second derivative depend on some parameters. One can nevertheless attain a quadratic form for each parameter, via a suitable basis change.

\begin{proposi}[Parametrized version of Morse lemma \cite{Dui}]\label{prop-mors-theo}
Let $\omega_0\in \mathbb{S}^{d-1}$, and $\mu_0 \in [0,r)$.  Then, there exist open sets $W_{\omega_0} \subset \mathbb{S}^{d-1}$, $V_{\mu_0} \subset [0,r)$,  $U_{\mu_0\omega_0}\subset U\subset\RM^d$ with  $\omega_0 \in W_{\omega_0}, \mu_0 \in V_{\mu_0}$, $\theta(\mu_0,\omega_0) \in U_{\mu_0\omega_0}$ and a smooth map  
$$
\xi_{\mu_0,\omega_0}
\,:\, 
U_{\mu_0\omega_0} \times V_{\mu_0} \times W_{\omega_0} 
\;\to\; 
\mathbb{R}^d
\;,
$$ 
such that for all $(\mu, \omega)\in V_{\mu_0}\times W_{\omega_0}$ the map  
$$
\xi_{\mu_0,\omega_0}(\cdot,\mu,\omega)
\,:\,
U_{\mu_0\omega_0}
\,\to\, \mathbb{R}^d
\;,
$$ 
is a diffeomorphism satisfying $\xi_{\mu_0,\omega_0}(\theta(\mu,\omega),\mu,\omega)=0$ and, for all $x  \in   U_{\mu_0\omega_0}$, 
\begin{equation}\label{equ-phi-rep}
\Phi_2(x,\mu,\omega)
\;=\;
-\,\frac{1}{2}\,
\langle  \xi_{\mu_0,\omega_0}(x,\mu,\omega),\Hess f (\theta(\mu,\omega)) \xi_{\mu_0,\omega_0}(x,\mu,\omega)\rangle
\,+\, 
\Phi_2(\theta(\mu,\omega),\mu,\omega)
\;.
\end{equation}
\end{proposi}

\noindent {\bf Proof.} As \cite{Dui} only contains a sketch of the argument, let us here provide some details. Choose an open convex set $\tilde U \subset U$  such that $\theta(\mu_0,\omega_0) \in \tilde U$. For $y\in \tilde U$ and $(\mu,\omega)\in \theta^{-1}(\tilde U)$,  consider the function
$$
t\in (1-\epsilon,1+\epsilon)\; \mapsto\; \Gamma(t)\,=\,\Phi_{2,\mu,\omega}((1-t)\theta(\mu,\omega)+ty)
\;.
$$
As $\nabla \Phi_{2,\mu,\omega}(\theta(\mu,\omega))=0$ (see \eqref{equ-def-the},  \eqref{equ-pun-cri},  \eqref{equ-cer-nab}),
a second order Taylor expansion with integral remainder gives
$$ 
\Phi_{2,\mu,\omega}(y)
\;=\;
\Phi_{2,\mu,\omega}(\theta(\mu,\omega))
\,+\,
\int_0^1 dt\; (1-t) \Gamma''(t) \;.
$$ 
Computing $\Gamma''$ explicitly using  \eqref{equ-cer-nab} gives 
$$
\Gamma''(t)
\;=\; 
-\,\langle y -\theta(\mu,\omega),\Hess f( (1-t) \theta(\mu,\omega) + ty) (y- \theta(\mu,\omega))\rangle
\;.
$$
Then 
\begin{equation}
\label{equ-exp-pao}
\Phi_{2,\mu,\omega}(y) 
\;=\; 
\Phi_2(\theta(\mu,\omega),\mu,\omega) 
\,-\, \frac{1}{2}\langle y-\theta(\mu,\omega),F(y,\mu,\omega) (y-\theta(\mu,\omega))\rangle
\;,
\end{equation}	 
with 
$$
F(y,\mu,\omega)
\;=\;
2\int_0^1 dt\;(1-t)\Hess f((1-t)\theta(\mu,\omega)+ty)
\;.
$$ 
In particular, note that $F$ is a smooth function, with $F(y,\mu,\omega)=F(y,\mu,\omega)^T$ being symmetric. Let us denote by $\Sym(d,\mathbb{R})\subset\RM^{d\times d}$ the vector space of symmetric matrices and then consider the function $G:\RM^{d\times d}\times \tilde U \times \theta^{-1}( \tilde U)\to \Sym(d,\mathbb{R})$ defined by 
$$
G(R,y,\mu,\omega)
\;=\;
R^T\Hess f(\theta(\mu,\omega))R-F(y,\mu,\omega)
\;.
$$ 
If $\one\in\RM^{d\times d}$ denotes the identity matrix, then  
$$
G(\one,\theta(\mu_0,\omega_0),\mu_0,\omega_0)
\;=\;
\Hess f(\theta(\mu_0,\omega_0))-\Hess f(\theta(\mu_0,\omega_0))
\;=\;	
0\;.
$$
Moreover, the derivative of the function  $ R \mapsto G(R,\theta(\mu_0,\omega_0),\mu_0,\omega_0)$ at $R=\one$ is given by 
$$
D_R G(R, \theta(\mu_0,\omega_0),\mu_0,\omega_0)|_{R=\one}(S)
\;=\;
S^T \Hess f(\theta(\mu_0,\omega_0))
\,+\, 
\Hess f(\theta(\mu_0,\omega_0))S
\;,
$$ 
and it is surjective because $C\in \Sym(d,\mathbb{R})$ is attained for $S=\frac{1}{2} \Hess f(\theta(\mu_0,\omega_0))^{-1}C$. Therefore, the implicit function theorem implies that there exists an open set $V \times U_{\mu_0\omega_0} \times V_{\mu_0}\times W_{\omega_0}\subset \GL(d,\mathbb{R})\times \tilde U \times \theta^{-1}(\tilde U)$ with $(\one,\theta(\mu_0,\omega_0),\mu_0,\omega_0)\in V \times U_{\mu_0 \omega_0} \times V_{\mu_0}\times W_{\omega_0}$,  and a smooth map $ R:U_{\mu_0\omega_0} \times V_{\mu_0}\times W_{\omega_0} \to V\subset \GL(d,\mathbb{R})$ such that 
$$
0
\;=\;
G(R(y,\mu,\omega),y,\mu,\omega)
\;=\;
R(y,\mu,\omega)^T\Hess f (\theta (\mu,\omega)) R(y,\mu,\omega)- F(y,\mu,\omega)
\;.
$$
Let us define $\xi_{\mu_0,\omega_0}:U_{\mu_0\omega_0} \times V_{\mu_0}\times W_{\omega_0} \to \mathbb{R}^d$ by 
$$
\xi_{\mu_0,\omega_0}(y,\mu,\omega)
\;=\;
R(y,\mu,\omega)(y-\theta(\mu,\omega))
\;.
$$ 
Combining this with  \eqref{equ-exp-pao}, one obtains  \eqref{equ-phi-rep}. Finally, since  $\nabla_y\xi_{\mu_0,\omega_0}(y, \mu_0,\omega_0)|_{y=\theta(\mu_0,\omega_0)}$ can readily be verified to be an invertible matrix, it follows that $y\mapsto \xi_{\mu_0,\omega_0}(y, \mu,\omega)$ is a local diffeomorphism. 
\hfill $\Box$

\begin{proposi}
\label{prop-case-crit}
 Let $(\mu_0,\omega_0) \in [0,r) \times \mathbb{S}^{d-1}$. There exist open sets $V_{\mu_0} \subset [0,r)$, $W_{\omega_0} \subset \mathbb{S}^{d-1}$, with $\mu_0 \in V_{\mu_0}$ and $\omega_0 \in W_{\omega_0}$, with the following property: there are smooth functions $h, r: V_{\mu_0}\times W_{\omega_0} \times (0,\infty) \to \mathbb{C}$ such that for all $(\mu,\omega) \in V_{\mu_0}\times W_{\omega_0}$ one has 
\begin{equation}
\label{equ-jot-puc}
J_2(t,\mu,\omega)
\;=\; 
\left(\frac{\mu}{t}\right)^{\frac{d}{2}}  e^{\imath t\frac{1}{\mu} \Phi(\theta(\mu,\omega),\mu,\omega)} h(\mu,\omega,t) 
\;+\;
\left(\frac{\mu}{t}\right)^d r(\mu,\omega,t)
\;,
\qquad  t> 0\;, 
\end{equation}
where  for $n \in \mathbb{N}$,  $\ |\partial_\mu^n  h(\mu,\omega,t)|\leq C_{n,\mu_0,\omega_0}$ and $|r(\mu,\omega,t)|\leq C_{\mu_0,\omega_0},$ where $C_{n,\mu_0,\omega_0}, C_{\mu_0,\omega_0}$ do not depend on $(\mu,\omega,t)\in V_{\mu_0}\times W_{\omega_0}\times [1,\infty)$. 
\end{proposi}

\noindent {\bf Proof.}
Proposition~\ref{prop-mors-theo} provides open sets  $W_{\omega_0} \subset \mathbb{S}^{d-1}$, $V_{\mu_0} \subset [0,r)$, $U_{\mu_0\omega_0} \subset U$, with  $\omega_0 \in W_{\omega_0}, \mu_0 \in V_{\mu_0}$, $\theta(\mu_0,\omega_0) \in U_{\mu_0\omega_0}$ and a smooth map $\xi_{\mu_0,\omega_0}: U_{\mu_0\omega_0} \times V_{\mu_0} \times W_{\omega_0} \to \mathbb{R}^d$ such that for all $(y,\mu,\omega) \in U_{\mu_0\omega_0}\times V_{\mu_0} \times W_{\omega_0}$ one can express $\Phi_2(y,\mu,\omega)$ as in  \eqref{equ-phi-rep}. Furthermore, one may assume that $\xi_{\mu_0,\omega_0}$ is bounded by $s=s_{\mu_0\omega_0}$. Let us choose an open set $O \subset U$  with $\theta(\mu_0,\omega_0) \in O$ and $\overline{O}\subset U_{\mu_0\omega_0}$ and then consider the open set $U_{\reg}= U\setminus  \overline{O}$. Furthermore, let $\{\rho_{\mu_0\omega_0}, \rho_{\reg}\}$ be a smooth partition of unity subordinated to the open cover $\{U_{\mu_0\omega_0}, U_{\reg}\}$.  Note that the function $\Phi_{2,\mu_0,\omega_0}$ has no critical points on $U_{\reg}$, so that one can apply the arguments of the proof  of Proposition~\ref{prop-case-nonc} with $N=d$. Without restrictions, one can assume $V_{\mu_0}$, $W_{\omega_0}$ be the same as that proof and thus conclude that for all $(\mu,\omega) \in V_{\mu_0}\times W_{\omega_0}$
\begin{equation}
\label{equa-part-nonc}
\int_{\mathbb{R}^d} dx\;e^{\imath t\frac{1}{\mu}( \mu \langle x, \omega\rangle- f(x))} \rho(x) \cdot \rho_{\reg}(x) 
\;=\;  
\left(\frac{\mu}{t}\right)^d r(\mu,\omega,t)
\;,\qquad 
t>0\;,
\end{equation}
with $|r(\mu,\omega,t)|\leq C_{\mu_0,\omega_0}$, where $C_{\mu_0,\omega_0}$ does not depend on $\mu,\omega,t \in V_{\mu_0}\times W_{\omega_0}\times [1,\infty)$. Now, using  \eqref{equ-phi-rep} and the change of variable theorem we can write for $(\mu,\omega)\in V_{\omega_0}\times W_{\omega_0}$, the integral as follows
$$
\int_{\mathbb{R}^d} dx\,e^{\imath t\frac{1}{\mu}(\mu\langle x, \omega \rangle-f(x))} \rho(x) \, \rho_{\mu_0\omega_0}(x)  
\;=\;  
e^{\imath t\frac{1}{\mu} \Phi_2(\theta(\mu,\omega),\mu,\omega)} \int_{\mathbb{R}^d}dy\, e^{-\imath t\frac{1}{2\mu} \langle  y, \Hess f(\theta(\mu,\omega)) y \rangle} \tilde{\rho}(y,\mu,\omega) \,, 
$$
where $\tilde{\rho}(\xi_{\mu_0,\omega_0}(x,\mu,\omega),\mu,\omega)= \rho(x)\rho_{\mu_0\omega_0}(x)|\det(D_x \xi_{\mu_0,\omega_0}(x,\mu,\omega))|^{-1}$, for $y=\xi_{\mu_0,\omega_0}(x,\mu,\omega)$ with $ y\in \xi_{\mu_0,\omega_0}(U_{\mu_0\omega_0},\mu,\omega)$ and is extended to $0$ for $(y,\mu,\omega) \in \xi_{\mu_0,\omega_0}(U_{\mu_0\omega_0},\mu,\omega)^c \times \{(\mu,\omega)\}$.  In particular, $\tilde \rho$ is smooth with all derivates bounded in $B_s(0) \times V_{\mu_0}\times W_{\omega_0}$ and $\supp(\tilde \rho(\cdot, \mu,\omega)) \subset B_s(0)$. Then Parseval's identity and an explicit computation of the Fourier transform $\Ff_x$ in the variable $x$ (just as in \cite{DS}) leads to
\begin{align*}
&
\int_{\mathbb{R}^d}  dy  \,e^{\imath t\frac{1}{2\mu} \langle  y, \Hess f(\theta(\mu,\omega)) y \rangle} \tilde{\rho}_{\mu,\omega}(y) 
\\
&
= \;\left(\frac{2\pi \mu}{t}\right)^{\frac{d}{2}} \frac{ e^{\imath  \frac{\pi}{4} \mbox{\rm\tiny sig} (\Hess f(\theta(\mu,\omega)))}}{|\det(\Hess f(\theta(\mu,\omega)))|^{\frac{1}{2}}} \int_{\mathbb{R}^d} dk\,e^{-\imath \mu \frac{1}{2t} \langle  k, \Hess f(\theta(\mu,\omega))^{-1}k \rangle }  
(\Ff_x\tilde{\rho}) (k,\mu,\omega)
\\
&
= \; \left(\frac{2\pi \mu}{t}\right)^{\frac{d}{2}} 
\frac{ e^{\imath  \frac{\pi}{4} \mbox{\rm\tiny sig} (\Hess f(\theta(\mu,\omega)))}}{|\det(\Hess f(\theta(\mu,\omega)))|^{\frac{1}{2}}} 
\int_{\mathbb{R}^d} dk\,
\sum_{j\geq 0}\frac{1}{j!}\,\frac{(-\imath \mu \langle  k, \Hess f(\theta(\mu,\omega))^{-1}k \rangle)^j}{(2t)^j}
(\Ff_x\tilde{\rho})(k,\mu,\omega)
\\
&
= \;\left(\frac{2\pi \mu}{t}\right)^{\frac{d}{2}} \frac{ e^{\imath  \frac{\pi}{4} \mbox{\rm\tiny sig} (\Hess f(\theta(\mu,\omega)))}}{|\det(\Hess f(\theta(\mu,\omega)))|^{\frac{1}{2}}} 
\sum_{j\geq 0}\frac{1}{j!}\,\frac{(-\imath \mu )^j}{(2t)^j}
\int_{\mathbb{R}^d} dk\,(\Ff_x  (P_{\mu,\omega})^j \tilde{\rho})(k, \mu,\omega)
\\
&
= \;\left(\frac{2\pi \mu}{t}\right)^{\frac{d}{2}} \frac{ e^{\imath  \frac{\pi}{4} \mbox{\rm\tiny sig} (\Hess f(\theta(\mu,\omega)))}}{|\det(\Hess f(\theta(\mu,\omega)))|^{\frac{1}{2}}} 
\sum_{j\geq 0}\frac{1}{j!}\,\frac{(-\imath \mu )^j}{(2t)^j}
( (P_{\mu,\omega})^j \tilde{\rho})(0, \mu,\omega)
\;,
\end{align*}
where $(P_{\mu,\omega} g)(k)= (\langle \nabla^2 f(\theta(\mu,\omega))^{-1} \nabla, \nabla\rangle g)(k)$ for a smooth function $g$. Now one can check that $|( (P_{\mu,\omega})^j \tilde{\rho})(0, \mu,\omega)|\leq C^j$ uniformly in $\mu$ and $\omega$ for some constant $C$ that only depends on $\mu_0,\omega_0$. The same holds for derivatives  $\partial_\mu^N( (P_{\mu,\omega})^j \tilde{\rho})$. As the sum over $j$ is summable, this implies the claim.
\hfill $\Box$

\begin{rem}
\label{rem-Covering}
{\rm
Propositions~\ref{prop-case-nonc} and \ref{prop-case-crit} imply that for any $(\mu_0,\omega_0)\in [0,1]\times \mathbb{S}^{d-1}$, there are open sets $V_{\mu_0}\subset [0,1]$ and $ W_{\omega_0}\subset \mathbb{S}^{d-1}$ such that for all $(\mu,\omega)\in V_{\mu_0}\times W_{\omega_0}$ one can write $J_2(t,\mu,\omega)=f_{\mu_0,\omega_0}(t,\mu,\omega)$ with a function $f_{\mu_0,\omega_0}$ that satisfies \eqref{equ-jot-pun} or \eqref{equ-jot-puc} (with $N= d$). Now choose a finite subcover of the open cover $\{V_{\mu_0}\times W_{\omega_0}: (\mu_0,\omega_0) \in [0,1]\times \mathbb{S}^{d-1}\}$ and let us denote it simply by $ \{V_{\mu_j} \times W_{\omega_j}\}$. Further let  $\{\rho_j\}$ be a subordinate partition of unity for this open cover. Then for all $(\mu,\omega)\in V_{\mu_0}\times W_{\omega_0}$ one has $J_2(t,\mu,\omega)=\sum_j f_j(t,\mu,\omega) \rho_j(\mu,\omega)$.  Therefore there exist smooth functions $h,r : [0,1]\times \mathbb{S}^{d-1} \times (0,\infty) \to \mathbb{C}$ such that for all $(\mu,\omega)\in [0,1]\times \mathbb{S}^{d-1}$  one has
\begin{align}\label{PH8}
J_2(t,\mu,\omega)
\;=\; 
\left(\frac{\mu}{t}\right)^{\frac{d}{2}}  e^{\imath t\frac{1}{\mu} \Phi_2(\theta(\mu,\omega),\mu,\omega)} h(\mu,\omega,t) 
\;+\;
\left(\frac{\mu}{t}\right)^d r(\mu,\omega,t)
\;, 
\qquad   t>0\,, 
\end{align}
where for $n\in \mathbb{N}, |\partial^n_\mu h(\mu,\omega,t)|\leq C_n$ and $ |r(\mu,\omega,t)|\leq C$ with constants $C,C_n$ that do not depend on $(\mu,\omega,t) \in [0,1]\times \mathbb{S}^{d-1} \times [1,\infty)$. In particular, if $\mu_j< r$, then  $ \supp (\rho_j) \subset V_{\mu_j}\times W_{\omega_j} \subset [0,r)\times \mathbb{S}^{d-1}$, which implies that $\supp(h) \subset [0,r) \times \mathbb{S}^{d-1}\times (0,\infty)$.
\hfill $\diamond$  
}
\end{rem}

Now all is prepared for the study of the asymptotics of $I_2(z,t,\omega)$.

\begin{proposi}
\label{prop-cota-parc}
Let us assume that $d\geq 3$. Then, there exists a continuous function  $M_2: \overline{\HM} \times (0,\infty) \times \mathbb{S}^{d-1}\to \mathbb{C}$  such that for all $(t,\omega) \in (1,\infty)\times \mathbb{S}^{d-1}$ one has
$$
I_2(z,t,\omega)
\;=\;
t^{-\frac{d}{2}} \,M_2(z,t,\omega)
\;, 
\qquad 
\Im m(z)\geq 0
\;.
$$
There exists a constant $C\in \mathbb{R}$ independent of $(t,\omega) \in [1,\infty)\times \mathbb{S}^{d-1} $ such that, uniformly in $\Im m(z)\geq 0$ and $t\geq 1$, 
\begin{equation}
\label{equ-cot-inc}
|M_2(z,t,\omega)|
\;\leq \; C
\;. 
\end{equation}
Moreover, there exists a constant $C_U\geq 0$ such that, for all $\omega \in \mathbb{S}^{d-1}$ satisfying
\begin{equation}
\label{eq-OmegaCond}
|\langle (\nabla^2 f(x_0))^{-1} \omega,\omega\rangle|
\;>\;
C_U
\;,
\end{equation} 
one has
\begin{equation}
\label{equ-cot-inc2}
|M_2(E,t,\omega)|
\;\leq \;
\frac{C\,t^{-\frac{d}{2}+1}}{|\langle (\nabla^2 f(x_0))^{-1}\omega,\omega\rangle|^{\frac{d+1}{2}}}
\;,
\end{equation}
uniformly for  
$$
0
\;\leq \;
|E| 
\;< \;
\tfrac{1}{8}\, r^2 \,
|\langle \omega,(\nabla^2 f(x_0))^{-1} \omega\rangle|
\;, 
\qquad 
1 \;\leq\; t\;< \;\frac{\sqrt{|\langle (\nabla^2 f(x_0)) ^{-1}\omega,\omega\rangle|}}{\sqrt{8E}}\;,
$$
where $r=r_U$ is specified by \eqref{equ-ass-gra}. Finally, if $0<\beta\leq 1$ with $\beta <\frac{d-2}{2}$, then for all  $z,z' \in \overline{\HM}$ one has
$$
|M_2(z,t,\omega)-M_2(z',t,\omega)| 
\;\leq\; 
C_\beta t^{\beta}|z-z'|^{\beta}
\;.
$$
\end{proposi}

Let us stress that for a definite critical point $x_0$, the condition \eqref{eq-OmegaCond} holds for all $\omega\in\SM^d$ and then the bound \eqref{equ-cot-inc2} reduces to what is known to hold for the Laplacian, see \cite[p. 78]{Yaf2}.

\vspace{.2cm}

\noindent {\bf Proof} of Proposition~\ref{prop-cota-parc}.  Using \eqref{PH8}, \eqref{equa-osci-int2} and \eqref{equa-osci-inte}, it follows that 
\begin{equation}
\label{equ-in2-rer}
I_2(z,t,\omega) 
\;=\; 
t^{-\frac{d}{2}}  \int_0^1  d\mu \ e^{\imath t\frac{1}{\mu}(\Phi_2(\theta(\mu,\omega),\mu,\omega),\mu,\omega)+z)} \mu^{\frac{d-4}{2}} h(\mu,\omega,t)
\;+\; 
t^{-d} \int_0^1  d\mu  \  e^{\frac{\imath zt}{\mu}} r(\mu,\omega,t) \mu^{d-2} 
\;.
\end{equation}
The first summand can be bounded by 
\begin{equation}
\label{eq-FinalOscillation}
\left| \int_0^1 d\mu\; e^{\imath t\frac{1}{\mu}(\Phi_2(\theta(\mu,\omega),\mu,\omega)+z)} \mu^{\frac{d-4}{2}} h(\mu,\omega,t) \right| 
\;\leq\; 
C \int_0^1 d\mu\;\mu^{\frac{d-4}{2}} 
\;\leq\; C
\;, 
\qquad  t\in [1,\infty)
\;.
\end{equation} 
As to the second summand, one gets in the same manner
$$
\left|\int_0^1 d\mu \;e^{\frac{\imath zt}{\mu}} r(\mu,\omega,t) \mu^{d-2} \right| 
\;\leq \;
C  \int_0^1 d\mu\;\mu^{d-2} 
\; \leq\; 
C \,, 
\qquad   t\in [1,\infty)
\;. 
$$
This proves \eqref{equ-cot-inc}, but not yet \eqref{equ-cot-inc2} for which one has to carry out an analysis of the oscillations in $\mu$ in the first summand. As just shown, the limits $I_2(E+\imath 0,t,\omega) $ exists so that it remains to bound the oscillatory integral 
$$
K(E,t,\omega)
\;=\;
\int_0^1 d\mu\;e^{\imath t\frac{1}{\mu}(E+\Phi_2(\theta(\mu,\omega),\mu,\omega))} \mu^{\frac{d-4}{2}} h(\mu,\omega,t)
\;.
$$
Hence let us introduce the phase functions
$$
\Psi_{E,\omega}(\mu)
\;=\;
\tfrac{1}{\mu}(\Phi_2(\theta(\mu,\omega),\mu,\omega)+E)
\;=\;
\tfrac{1}{\mu} (E-f(\theta(\mu,\omega))) \,+\, \langle \omega, \theta(\mu,\omega)\rangle
\;,
$$ 
for fixed $\omega$ and $E$, and where $\Phi_2$ is as in  \eqref{eq-PhiPhaseDef}. Now recall that $\theta(\mu, \omega) $ is a critical point of $\Phi_2$ which by \eqref{equ-cer-nab} implies that $  \nabla f(\theta (\mu, \omega) )  = \mu \omega   $. Based on \eqref{eq-PhiPhaseDef}, one can then check that 
\begin{align}\label{PH9} \notag
\partial_\mu
\Psi_{E,\omega}(\mu)
&
\;=\;
-\tfrac{1}{\mu^2}(E-f(\theta(\mu,\omega)))- \tfrac{1}{\mu}\langle \nabla f(\theta(\mu,\omega)), \partial_\mu \theta(\mu,\omega)\rangle +\langle \partial_\mu\theta(\mu,\omega),\omega\rangle
\\
&
\;=\;
\tfrac{1}{\mu^2}\big(f(\theta(\mu,\omega))-E\big)
\;.
\end{align}
This shows that the derivative of the function $\Psi_{E,\omega}$ may vanish. In fact, if $\Hess f$ is indefinite, then for some direction $\omega\in \mathbb{S}^{d-1}$ and some $E\in \mathbb{R}$, the function $\Psi_{E,\omega}$ may even be constant. Consequently, the function $K(E,t,\omega)$, does not have further decay in $t$ uniformly in $\omega$. However, one can exclude those directions for which the derivative of $\Psi_{E,\omega}$ vanishes. Let $\omega\in \mathbb{S}^{d-1}$, using Taylor's formula  with integral remainder with the function $\mu\in (-r,r) \mapsto f((\nabla f)^{-1}(\mu \omega))$  around $0$, one obtains
\begin{equation}
\label{equ-tay-the}
f((\nabla f)^{-1}(\mu \omega))
\;=\; 
\tfrac{\mu^2}{2}\langle \omega, (\Hess f(x_0))^{-1} \omega\rangle + \tfrac{1}{6}\int_0^\mu d\nu\;(\partial^3_\nu f ((\nabla f)^{-1}(\nu \omega))) \nu^2 \;,
\end{equation} 
where an explicit computation of the first and second derivative is used, and the fact that $(\nabla f)^{-1}(0)=x_0$ as $\nabla f (x_0) = 0$, see \eqref{equ-def-the} and the text above it). Let us set 
$$
D_U
\;= \;
\sup 
\big\{ |\partial^3_\nu f ((\nabla f)^{-1}(\nu \omega))|: \nu \in [0,r], \omega \in \mathbb{S}^{d-1}
\big\}
\;<\;
\infty
\;.
$$ 
Recall that  $r \geq \sup\{|\nabla f(x)| : x\in U\}$ by \eqref{equ-ass-gra}, and consider $\omega \in \mathbb{S}^{d-1}$ such that 
$$
|\langle (\nabla^2 f (x_0))^{-1} \omega,\omega\rangle| 
\;>\; 
C_U
\;,
\qquad
\mbox{\rm where }
C_U
\;=\;
\tfrac{2}{9} D_U \sup\{|\nabla f(x)| : x\in U\}
\;.
$$
For such $\omega$, one has for all $\mu \in [0,r)$  
\begin{equation}
\label{ome-des-com}
\tfrac{1}{4} \mu^2 |\langle (\nabla^2 f (x_0))^{-1} \omega,\omega\rangle| 
\;\geq\; 
\tfrac{D_U}{18} r \mu^2 
\;\geq\; 
\tfrac{D_U}{18} \mu^3
\;. 
\end{equation}
Now take $E\in \mathbb{R}$ such that
\begin{equation}
\label{eq-mu-co}
r 
\;>\;
\mu 
\;\geq \;
L_E\;,
\qquad
\mbox{\rm where }L_E\;=\;\frac{\sqrt{8|E|}}{\sqrt{|\langle \omega, (\nabla^2 f(x_0))^{-1}\omega\rangle|}}
\;.
\end{equation}
Then combining Eqs. \eqref{equ-tay-the}, \eqref{ome-des-com}, \eqref{eq-mu-co} one obtains
\begin{align*}
\tfrac{1}{\mu^2}|f(\theta(\mu,\omega))-E|
&
\;\geq \;
\tfrac{1}{\mu^2}|f(\theta(\mu,\omega))| - \tfrac{1}{\mu^2}|E|
\\
&
\;\geq \;
\tfrac{1}{2}|\langle \omega, (\nabla^2 f(x_0))^{-1}\omega\rangle| - \tfrac{1}{4} |\langle \omega, (\nabla^2 f(x_0))^{-1}\omega\rangle| - \tfrac{1}{8} |\langle \omega, (\nabla^2 f(x_0))^{-1}\omega\rangle| 
\\ 
&
\;\geq\; 
\tfrac{1}{8}|\langle \omega, (\nabla^2 f(x_0))^{-1}\omega \rangle|
\;.
\end{align*}
Therefore, if one takes $\omega\in \mathbb{S}^{d-1}$ satisfying  \eqref{eq-OmegaCond}, the function $\Psi_{E,\omega}$ has no critical points in $[L_E,r)$. Moreover, for $\mu \in [L_E,r)$ one has due to \eqref{PH9}
\begin{equation}\label{equ-der-peo}
|\partial_\mu
\Psi_{E,\omega}(\mu)| 
\;\geq \;
\tfrac{1}{8}|\langle \omega, (\nabla^2 f(x_0))^{-1}\omega \rangle|
\;. 
\end{equation} 
For $E\in \mathbb{R}$ with $L_E<r$, let us now consider $1 \leq t<\frac{1}{L_E}$. The integral $K(E,t,\omega)$ will be split into two contributions $\mu\in[0,\frac{1}{t}]$ and $\mu\in[\frac{1}{t},1]$. For the first contribution one proceeds as in \eqref{eq-FinalOscillation}: 
\begin{align*}
\left|\int^{\frac{1}{t}}_0 d\mu\;e^{\imath t \Psi_{E,\omega}(\mu) } \,\mu^{\frac{d-4}{2}} h(\mu,\omega,t) \right|
&
\;\leq\;
C\int_0^{\frac{1}{t}} d\mu\; \,\mu^{\frac{d-4}{2}} 
\;=\;
C
\,t^{-\frac{d}{2}+1}
\;.
\end{align*}
For the second, let us integrate by:
\begin{align*}
\int_{\frac{1}{t}}^1d\mu\; e^{\imath t\Psi_{E,\omega}(\mu)} \mu^{\frac{d-4}{2}} h(\mu,\omega,t)
&
\;=\;
\tfrac{1}{\imath t}
\Big[-
\int_{\frac{1}{t}}^1 d\mu\; e^{\imath t\Psi_{E,\omega}(\mu)} \partial_\mu \big((\partial_\mu\Psi_{E,\omega}(\mu))^{-1}\mu^{\frac{d-4}{2}} h(\mu,\omega,t)\big) 
\\
&
\hspace{1.4cm}
 +  e^{\imath t\Psi_{E,\omega}(\mu)} 
(\partial_\mu\Psi_{E,\omega}(\mu))^{-1} \mu^{\frac{d-4}{2}} h(\mu,\omega,t)\big|^1_{\frac{1}{t}}
\Big]
\end{align*} 
Since $\supp (h(\cdot, \omega,t)) \subset [0,r)$ and  $\frac{1}{t}> L_E$, then using  \eqref{equ-der-peo} one obtains 
$$
\Big|  e^{\imath t\Psi_{E,\omega}(\mu)}  (\partial_\mu\Psi_{E,\omega}(\mu))^{-1} \mu^{\frac{d-4}{2}} h(\mu,\omega,t) \big|^1_{\frac{1}{t}}
\Big| 
\;\leq\;   
\frac{C t^{-\frac{d}{2}+2}}{|\langle \omega, (\nabla^2 f(x_0))^{-1}\omega \rangle|} 
\;.
$$ 
Note that, for $N\in \mathbb{N}$, one has
$$
\partial_\mu \big((\partial_\mu\Psi_{E,\omega}(\mu))^{-M}\mu^N\big)
\,=\,  
\mu^{N-1}((\partial_\mu \Psi_{E,\omega}(\mu))^{-M}  c_1(\mu,\omega,E)
+ 
(\partial_\mu \Psi_{E,\omega}(\mu))^{-(M+1)}  c_2(\mu,\omega,E)) 
\,,
$$ 
where $c_j(\mu,\omega,E)$ are smooth functions in $(0,r)$  uniformly bounded with respect  $(\mu,E,\omega)$.  Then one applies integration by parts $\frac{d-1}{2}$ times for $d$ odd and $\frac{d-2}{2}$ times for $d$ even, in such a way that one obtains  (let us say, for $d$ odd)
\begin{align*}
\int_{\frac{1}{t}}^1d\mu\; e^{\imath t\Psi_{E,\omega}(\mu)} \mu^{\frac{d-4}{2}} h(\mu,\omega,t) 
\;=\;
& 
(\imath t)^{-\frac{d}{2}+\frac{1}{2}}
\int_{\frac{1}{t}}^1 d\mu\; e^{\imath t\Psi_{E,\omega}(\mu)} \mu^{-\frac{3}{2}} \tilde{\rho}_1(\mu,\omega,E,t)  
\\
&
\;-\;
t^{-\frac{d}{2}+1} \tilde{\rho}_2(\mu,\omega,E,t)
\;,
\end{align*}
where $\tilde{\rho}_j(\cdot, E,\omega,t)$ are smooth functions with compact support on $(0,r)$ satisfying 
$$
|\tilde{\rho}_j(\mu,\omega,E,t)| 
\;\leq\; 
\frac{C}{|\langle \omega, (\nabla f(x_0))^{-1}\omega \rangle |^{\frac{d+1}{2}}}
\;,
$$ 
with a constant $C\in \mathbb{R}$ that does not depend on $\mu,\omega, E,t$. Finally performing the last integral, one obtains 
\begin{align*}
\left|\int_{\frac{1}{t}}^1 d\mu\; e^{\imath t\Psi_{E,\omega}(\mu)} \mu^{-\frac{3}{2}} \tilde{\rho}_1(\mu,\omega,E,t)  \right|
& 
\;\leq \;
\frac{C}{|\langle \omega, (\nabla f(x_0))^{-1}\omega \rangle |^{\frac{d+1}{2}}}\int_{\frac{1}{t}}^1 d\mu \ \mu^{-\frac{3}{2}} 
\\
&
\;\leq \;
\frac{C}{|\langle \omega, (\nabla f(x_0))^{-1}\omega \rangle |^{\frac{d+1}{2}}} (t^{\frac{1}{2}}-1)
\;.
\end{align*}
For the H\"older continuity, let us consider  $\beta <\frac{d-2}{2}$ and $\beta\leq 1$  using Lemma \ref{lem-des-cok} and  \eqref{equ-in2-rer} one obtains for $z,z' \in \overline{\HM}$ 
$$
|t^{\frac{d}{2}}I_2(z,t,\omega)-t^{\frac{d}{2}} I_2(z',t,\omega)| 
\;\leq\; 
C |z-z'|^\beta t^{\beta} \int_0^1 d\mu \ (\mu^{\frac{d-4}{2}}+\mu^{d-2}) \mu^{-\beta}
\;.
$$
Since $\frac{d-4}{2}-\beta>-1$, the last integral is bounded. 
\hfill $\Box$

\begin{rem}
{\rm
The proof of Proposition~\ref{prop-cota-parc} shows that for $\omega$ and $E$ satisfying \eqref{eq-mu-co}, one has $|M_2(E\pm\imath 0,t,\omega)|\leq C\,t^{-\frac{d}{2}+1}$ also if the $E$ is near a critical point of indefinite signature. In particular, if $E=f(x_0)=0$ is the critical value, then the condition \eqref{eq-mu-co} is valid for all $\omega$ not lying in the isotropic subspace of the quadratic form $\nabla^2 f(x_0) $. Furthermore, if the form $\nabla^2 f(x_0)$ is definite, then \eqref{eq-mu-co} is merely a condition on $E$ and one obtains an estimate that is similar to (1.2.31) in \cite{Yaf2}. Finally, if $\nabla^2 f(x_0)$ is definite {\it and} $E=0$, one recovers  the bounds in \cite{MT,LL,KR,KKR}.
}
\hfill $\diamond$
\end{rem}

Finally let us combing all bounds to obtain the estimates of the integral  \eqref{equa-inte-osci}.

\vspace{.2cm}

\noindent {\bf Proof} of Theorem~\ref{theo-mai-res}.
One splits the integral using  \eqref{equat-split-integ} in two parts $I_1$ and $I_2$. Then Proposition~\ref{prop-meto-nstp} allows to bound  $I_1$ and its derivatives. For $I_2$ one applies  Proposition~\ref{prop-cota-parc}. 
\hfill $\Box$

\vspace{.2cm}

\noindent {\bf Acknowledgements:} The work of M. Ballesteros and G. Franco C\'ordova was supported by CONACYT, FORDECYT-PRONACES 429825/2020 and PAPIIT-DGAPA-UNAM  IN114925. Furthermore, G. Franco C\'ordova received funding from the DAAD. The work of H.~S-B. was supported by the DFG grant SCHU 1358/8-1. Data sharing not applicable to this article as no datasets were generated or analyzed during the current study. The authors have no competing interests to declare that are relevant to the content of this article.


\begin{thebibliography}{99} 

\bibitem{AMV} N.~P.~Armitage, E.~J.~Mele, A.~Vishwanath, {\sl Weyl and Dirac semimetals in three-dimensional solids}, Rev. Mod. Phys. {\bf 90}, 015001 (2018).



\bibitem{BSB} J.~Bellissard, H.~Schulz-Baldes, {\sl Scattering theory for lattice operators in dimension $d\geq 3$}, Rev. Math. Phys. (2012).

\bibitem{CGL} A.~Carey, F.~Gesztesy, G.~Levitina, R.~Nichols, F.~Sukochev, D.~Zanin, {\sl The limiting absorption principle for massless Dirac operators, properties of spectral shift functions, and an application to the Witten index of non-Fredholm operators},  (Memoires EMS, Berlin, 2023).


\bibitem{DS} M.~Dimassi, J.~Sjostrand,  {\sl Spectral asymptotics in the semi-classical limit}, (Cambridge Univ. Press, 1999).


\bibitem{Dui} J.~J.~Duistermaat, {\sl Fourier Integral Operators}, Reprint of 1996 edition, (Birkh\"auser, Boston, 2011).

\bibitem{Eco} E.~N. Economou, {\sl Green's functions in quantum physics}, 3rd Edition (Springer, New York, 2005).

\bibitem{Ger} C.~G\'erard,  {\sl Resonance theory for periodic Schr\"odinger operators},  Bull. Soc. Math. France {\bf 118},  27-54 (1990).

\bibitem{Kat} T. Kato, {\sl Perturbation theory for linear operators}, 2nd edition, (Springer, Berlin, 2012).


\bibitem{KKR} M.~Kha, P.~Kuchment, A.~Raich, {\sl Green's function asymptotics near the internal edges of spectra of periodic elliptic operators. Spectral gap interior},  J. Spectral Theory {\bf 7}, 1171-1233 (2017). 
 

\bibitem{KSM} E.~L.~Korotyaev, J.~Schach M{\o}ller, {\sl Weighted estimates for the Laplacian on the cubic lattice}, Ark. Mat. {\bf 57}, 397-428 (2019).

\bibitem{Kos} G.~F.~Koster, {\sl Theory of scattering in solids}, Phys. Rev. {\bf 95}, 1436-1443  (1954).

\bibitem{KR} P.~Kuchment, A.~Raich, {\sl Green's function asymptotics near the internal edges of spectra of periodic elliptic operators. Spectral edge case}, Math. Nachrichten {\bf 285}, 1880-1894 (2012).

\bibitem{Kur} S.~T.~Kuroda, {\sl An introduction to scattering theory},  Aarhus Lecture Series  51, (Aarhus universitet, Matematisk institut,1978)

\bibitem{LL} G.~Lawler, V.~Limic, {\sl Random Walk: A Modern Introduction}, (Cambridge University Press, Cambridge, 2010).

\bibitem{Mar} P.~A.~Martin, {\sl Discrete scattering theory: Green's function for a square lattice}, Wave Motion {\bf 43}, 619-629 (2006).

\bibitem{MT} M.~Murata, T.~Tsuchida, {\sl Asymptotics of Green functions and the limiting absorption principle for elliptic operators with periodic coefficients}, J. Math. Kyoto Univ. {\bf 46}, 713-754 (2006).


\bibitem{Nen} G.~Nenciu, {\sl Existence of the exponentially localised Wannier functions}, Commun. Math. Phys. {\bf 91}, 81-85 (1983).

\bibitem{NW} J.~von Neumann, E.~Wigner, {\sl \"Uber das Verhalten von Eigenwerten bei adiabatischen Prozessen}, Phys. Zeitschrift {\bf 30}, 467-470 (1929).

\bibitem{Nic} L.~I. Nicolaescu, {\sl An invitation to Morse theory}, (Springer, New York, 2007).

\bibitem{Par} L.~Parnovski, {\sl Bethe-Sommerfeld conjecture}, Ann. Henri Poincar\'e {\bf 9}, 457-508 (2008).

\bibitem{PaS} L.~Parnovski,  A.~V.~Sobolev, {\sl Bethe-Sommerfeld conjecture for periodic operators with strong perturbations}, Invent. Math. {\bf 181}, 467-540  (2010).


\bibitem{Ple} J.~Plemelj, {\sl Ein Erg\"anzungssatz zur Cauchy'schen Integraldarstellung analytischer Funktionen, Randwerte betreffend}, Monatshefte Math. Phys. {\bf 19}, 205-210 (1908).

\bibitem{Pri} I.~Priwaloff, {\sl Sur les functions conjugu\'ees}, Bull. Soc. Math. France {\bf 44}, 100- 103 (1916). 

\bibitem{RS3} M.~Reed, B.~Simon, {\sl Methods of modern mathematical physics III: Scattering theory}, (Academic Press, New York, 1979).


\bibitem{SSt} H.~Schulz-Baldes, T.~Stoiber, {\sl  Spectral localization for semimetals and Callias operators}, J. Math. Phys.
{\bf 64}, 081901 (2023).


\bibitem{Spi} F.~Spitzer, {\sl Principles of Random Walk}, (Springer-Verlag, Heidelberg, 1976).



\bibitem{Yaf} D.~R.~Yafaev, {\sl Mathematical Scattering Theory: General Theory}, (AMS, Providence,
1992).

\bibitem{Yaf2} D.~R.~Yafaev, {\sl   Mathematical Scattering Theory: Analytic Theory}, Mathematical Surveys and Monographs {\bf 158}, (AMS, Providence,  2010).

\end{thebibliography}
\end{document}